\documentclass[10pt]{article}

\usepackage[]{amsmath, amssymb, theorem}
\usepackage{amsfonts}

\usepackage{theorem}
{\theorembodyfont{} \newtheorem{proposition}{Proposition}[section]}
{\theorembodyfont{} \newtheorem{theorem}[proposition]{Theorem}}
{\theorembodyfont{} }
{\theorembodyfont{\rm} \newtheorem{definition}[proposition]{Definition}}
{\theorembodyfont{} }
{\theorembodyfont{\rm} \newtheorem{example}[proposition]{Example}}
{\theorembodyfont{\rm} }
{\theorembodyfont{\rm} }
{\theorembodyfont{} }
{\theorembodyfont{} }
{\theorembodyfont{} }
{\theorembodyfont{} }

\newcommand{\Pomega}{{P\hspace*{-1pt}\omega}}
\newcommand{\bfSig}{\mathbf{\Sigma}}
\newcommand{\bfPi}{\mathbf{\Pi}}
\newcommand{\bfDelta}{\mathbf{\Delta}}
\newcommand{\bfGamma}{\mathbf{\Gamma}}
\newcommand{\dom}{\mathit{dom}}
\newcommand{\rng}{\mathit{rng}}

\newcommand{\calC}{\mathcal{C}}

\newcommand{\calN}{\mathcal{N}}

\newcommand{\IQ}{\mathbb{Q}}
\newcommand{\IS}{\mathbb{S}}
\newcommand{\IR}{\mathbb{R}}
\newcommand{\RR}{\mathbb{R}}

\title{Well Quasiorders and Hierarchy Theory}

\author{Victor Selivanov%\thanks{The work was funded by the subsidy allocated to Kazan Federal University
\\A.P. Ershov
Institute of Informatics
Systems SB RAS\\
and Kazan Federal University
}

\date{}

\begin{document}

\maketitle

\begin{abstract}
We discuss some applications of WQOs to several fields were hierarchies and reducibilities are the principal classification tools, notably to Descriptive Set Theory, Computability theory and Automata Theory. While the classical hierarchies of sets usually degenerate to structures very close to ordinals, the extension of them to functions requires more complicated WQOs, and the same  applies to reducibilities. We survey some results obtained so far and discuss  open problems and possible research directions.

\end{abstract}

%%%%%%%%%%%%%%%%%%%%%%%%%%%%%%%%%%%%%%%%%%%%%%%%%%%%%%%%%%%%%%
%
%
%%%%%%%%%%%%%%%%%%%%%%%%%%%%%%%%%%%%%%%%%%%%%%%%%%%%%%%%%%%%%%

\section{Introduction}\label{in}

WQO-theory is an important part of combinatorics with deep connections and applications to several parts of mathematics (proof theory, reverse mathematics, descriptive set theory, graph theory) and computer science (verification of infinite-state systems, combinatorics on words and formal languages, automata theory).

In this paper we discuss some applications of WQOs to several fields where hierarchies and reducibilities are the principal classification tools, notably to descriptive set theory (DST), computability theory and automata theory. 
The starting point of our discussion are  three important parts of  DST:

\begin{enumerate}\itemsep-1mm
\item The classical Borel, Luzin, and Hausdorff hierarchies in Polish spaces, which are defined  using set-theoretic operations.
\item  The Wadge hierarchy which is  non-classical in the sense that it is based on a notion of reducibility that was not recognized in the classical DST, and on using ingenious versions of Gale-Stewart games rather than the properties of set-theoretic operations.
\item  The classification of  Borel equivalence relations by means of  Borel reducibility, which  uses deep  analytical tools.
 \end{enumerate}

J. Addison \cite{ad59,ad62,ad62a,ad65} suggested to develop a general hierarchy theory, in order to have precise notions and tools to study analogies between the classical hierarchies and some hierarchies that appeared later in logic and computability theory. In particular, he suggested a general notion of a hierarchy of sets. The hierarchy theory was continued  in a series of the author's papers (see e.g. \cite{s95,s08,s12} and references therein) where new general notions and techniques of  hierarchy theory were suggested, in particular the notion of a hierarchy of $k$-partitions and of a reducibility that fits a given hierarchy.  

While the classical hierarchies of sets usually degenerate to structures very close to ordinals, the attempt to extend them to $k$-partitions requires more complicated WQOs (namely, the so called $h$-quasiorder on labeled forests), and the same  applies to reducibilities. This  was the original author's motivation for a systematic study of relationships between WQO-theory and hierarchy theory. 

In this paper we survey some results obtained in this direction so far. The general theory of hierarchies and reducibilities based on WQO-theory seems already matured and homogeneous, including the extension to $k$-partition. In contrast, several attempts to include in the  theory more general functions and Borel equivalence relations is still in the beginning, and the role of WQOs in such further generalizations is not yet clear.  For this reason we  mention several open questions which seem interesting for such generalizations. We decided not to include proofs (which are sometimes technical and long), instead concentrating on the formulations of basic results and discussions of the main tools.

In the next section we recall some well-known notions and  facts, but we also mention some less-known facts about the Wadge hierarchy and the extension of the classical hierarchies to the so called quasi-Polish spaces which are of interest to computer science. In Section \ref{wbqo} we recall the basic notions of WQO and BQO and provide examples which are important for the sequel. In Section \ref{wlike} we discuss several versions and extensions of Wadge reducibility which are based on the $h$-quasiorder. 
In Section \ref{redfun} we discuss some reducibilities on objects more complex than $k$-partitions, notably for equivalence relations and functions on the Baire space. In Section \ref{redfun} we recall some notions of the general hierarchy theory needed to unify terminology. In Sections \ref{comput} and  \ref{auto} we discuss some
hierarchies in computability theory and automata theory respectively, trying to relate this to WQO-theory and the $h$-quasiorder. We conclude in Section \ref{con} with comments on a recent preprint and some open questions.

As is well known, outside the Borel sets in DST or the hyperarithmetical sets in computability theory, some properties of hierarchies and reducibilities depend on set-theoretic axioms. Although the axiomatic issues are important and interesting, we decided to avoid the foundational discussions and to include only  results  provable in the widely accepted axiomatic system ZFC. As a result, we stay mainly within the Borel sets, although many facts may be extended far beyond the Borel sets under suitable set-theoretic assumptions.

%%%%%%%%%%%%%%%%%%%%%%%%%%%%%%%%%%%%%%%%%%%%%%%%%%%%%%%%%%%%%%%%
%
%
%%%%%%%%%%%%%%%%%%%%%%%%%%%%%%%%%%%%%%%%%%%%%%%%%%%%%%%%%%%%%%%%

%%%%%%%%%%%%%%%%%%%%%%%%%%%%%%%%%%%%%%%%%%%%%%%%%%%%%%%%%%%%%%%%
%
%
%%%%%%%%%%%%%%%%%%%%%%%%%%%%%%%%%%%%%%%%%%%%%%%%%%%%%%%%%%%%%%%%

%%%%%%%%%%%%%%%%%%%%%%%%%%%%%%%%%%%%%%%%%%%%%%%%%%%%%%%%%%%%%%%%
%
%
%%%%%%%%%%%%%%%%%%%%%%%%%%%%%%%%%%%%%%%%%%%%%%%%%%%%%%%%%%%%%%%%

\section{Preliminaries}\label{prelim}

In this section we recall some notation, notions and results used in
the subsequent sections.
We  use the standard set-theoretic notation like
$\dom(f)$ and $\rng(f)$  for the domain and range  of
a function $f$, respectively, $X\times Y$ for the Cartesian
product, $X \oplus Y$ for the disjoint union of sets $X$
and $Y$, $Y^X$ for the set of functions $f \colon X\to Y$, and $P(X)$ for the set of all subsets of $X$.
For $A\subseteq X$, $\overline{A}$ denotes the complement
$X\setminus A$ of $A$ in $X$. 
The notation $f:X\to Y$
means that $f$ is a (total) function from a set $X$ to a set $Y$.

\subsection{Ordinals}\label{ordinal}

We assume the reader to be acquainted with the notion of an  ordinal
 (see e.g.  \cite{km67}). Ordinals are important for
the hierarchy theory because levels of  hierarchies of sets are 
(almost) well ordered by inclusion. This opens the possibility to
estimate the complexity of sets (and other objects) by ordinals.

Ordinals are denoted by $\alpha$, $\beta$, $\gamma$, $\ldots$. The
successor $\alpha+1$ of an ordinal $\alpha$ is defined by
$\alpha+1=\alpha\cup\{\alpha\}$. Every ordinal $\alpha$ is the set
of all smaller ordinals, in particular $ k=\{0,1,\ldots,k-1\}$ for each $k<\omega$, and $ \omega=\{0,1,2,\ldots\}.$ Ordinals may
be considered as the order types of well orders (see the next subsection).

We use some well-known facts about the ordinal arithmetic. As usual,
$\alpha+\beta$, $\alpha\cdot\beta$ and $\alpha^{\beta}$ denote the
ordinal addition, multiplication and exponentiation of $\alpha$ and
$\beta$, respectively. The context will help to distinguish the
ordinal exponentiation from the set exponentiation denoted in the
same way but having a quite different meaning.

Below we will  mention the ordinals $\omega$, $\omega^2$,
$\omega^3,\ldots$ and $\omega^\omega$. The last ordinal  is the
order type of finite sequences $(k_1,\ldots,k_n)$ of natural numbers
$k_1\geq\cdots\geq k_n$, ordered lexicographically. Any non-zero
ordinal $\alpha<\omega^\omega$ is uniquely representable in the form
$\alpha=\omega^{k_1}+\cdots+\omega^{k_n}$ with
$\omega>k_1\geq\ldots\geq k_n$.
We will also use the bigger ordinal $\varepsilon_0=sup\{\omega
,\omega ^\omega , \omega ^{(\omega ^\omega)},\ldots\}$. Any non-zero ordinal $\alpha<\varepsilon_0$ is
uniquely representable  in the form $\alpha=\omega^{\gamma_0}+\cdots
+\omega^{\gamma_k}$  for a finite sequence
$\gamma_0\geq\cdots\geq\gamma_k$ of ordinals $<\alpha$. The ordinal
$\varepsilon_0$ is the smallest solution of the ordinal equation
$\omega^\varkappa=\varkappa$.

All concrete ordinals mentioned above are computable, i.e. they are
order types of computable well orders on computable subsets of
$\omega$. The first non-computable ordinal $\omega^{CK}_1$, known  as the
Church-Kleene ordinal, is
 important in computability theory.
The first
non-countable ordinal $\omega_1$ is  important for the hierarchy theory. From this ordinal one can construct many other
interesting ordinals, in particular 
$\omega_1^{\omega_1}$, $\omega_1^{(\omega_1^{\omega_1})},\ldots$.
Even much bigger ordinals (like the Wadge ordinal discussed below) are of interest
for the hierarchy theory.

\subsection{Partial orders and quasiorders}

We use some standard notation and terminology on partially ordered
sets (posets), which may be found  e.g. in \cite{dp94}.  Recall that a
{\em quasiorder} (QO) is a structure $(P;\leq)$ satisfying the axioms of
reflexivity $\forall x(x\leq x)$ and transitivity $\forall x\forall
y\forall z(x\leq y\wedge y\leq z\to x\leq z)$. {\em Poset} 
is a QO satisfying the antisymmetry axiom $\forall
x\forall y(x\leq y\wedge y\leq x\to x=y)$.   {\em Linear order}  is a partial order satisfying the connectivity axiom $\forall
x\forall y(x\leq y\vee y\leq x)$. A linearly ordered subset of a poset is sometimes called a {\em chain}.

Any partial order $\leq$ on $P$ induces the relation of strict order
$<$ on $P$ defined by $a<b\leftrightarrow a\leq b\wedge a\neq b$ and
called the strict order related to $\leq$. The relation $\leq$ can
be restored from $<$, so we may  safely apply the terminology on
partial orders also to the strict orders. A poset $(P;\leq)$ will be
often shorter denoted just by $P$. Any subset of a poset $P$ may be
considered as a poset with the induced partial order. 

It is well known that any QO $(P;\leq)$ induces the partial
order $(P^*;\leq^*)$ called  {\em the
quotient} of $P$. The set $P^*$ is the quotient set of $P$ under the
equivalence relation defined by $a\equiv b\leftrightarrow a\leq
b\wedge b\leq a;$ the set $P$ consists of all equivalence classes
$[a]=\{x\mid x\equiv a\}$, $a\in P$. The partial order $\leq^*$ is defined
by $[a]\leq^*[b]\leftrightarrow a\equiv b$. We will  not be 
cautious when applying notions about  posets also to QOs; in
such cases we mean the corresponding quotient-poset of the QO.

A partial order $(P;\leq)$ is  {\em well-founded} if it has no
infinite descending chains. In this case there are a unique ordinal
$rk(P)$ and a unique rank function $rk_P$ from $P$ onto $rk(P)$
satisfying $a<b\to rk_P(a)<rk_P(b)$. It is defined by induction
$rk_P(x)=sup\{rk_P(y)+1\mid y<x\}. $ The ordinal $rk(P)$ is called
the {\em rank} (or {\em height}) of $P$, and the ordinal $rk_P(x)$ is called the {\em rank of the
element $x\in P$ in $P$}.

In the sequel we will often deal with semilattices expanded by some additional operations. In particular the following notions introduced in \cite{s82,s07} will often be mentioned. The abbreviation ``dc-semilattice'' refers to ``semilattice with discrete closures''.

\begin{definition}\label{dc1}
By {\em $dc$-semilattice} we mean a structure
$(S;\leq,\cup,p_0,\ldots,p_{k-1})$ such that:
 \begin{enumerate}\itemsep-1mm
 \item $(S;\cup)$ is an upper semilattice, i.e. it satisfies $(x\cup
y)\cup z=x\cup (y\cup z)$, $x\cup y= y\cup x$  and $x\cup x=x$, for
all $x,y,z\in S$.

 \item $\leq$ is the  partial order on $S$ induced by $\cup$, i.e.
$x\leq y$ iff $x\cup y=y$, for all $x,y\in S$.

 \item Every $p_i$, $i<k$, is a closure operation on $(S;\leq)$, i.e. it
satisfies $x\leq p_i(x)$, $x\leq y\rightarrow p_i(x)\leq p_i(y)$ and
$p_i(p_i(x))\leq p_i(x)$, for all $x,y\in S$.

 \item The operations $p_i$ have the following discreteness property:
for all distinct $i,j<k$, $p_i(x)\leq p_j(y)\rightarrow p_i(x)\leq
y$, for all $x,y\in S$.

 \item Every $p_i(x)$ is join-irreducible, i.e. $p_i(x)\leq y\cup
z\rightarrow (p_i(x)\leq y\vee p_i(x)\leq z)$, for all $x,y,z\in S$.
 \end{enumerate} 
\end{definition}

By {\em  $dc\sigma$-semilattice} we mean a $dc$-semilattice which is also a $\sigma$-semilattice (i.e.,  the supremums of countably many elements exist),
 and the axiom (5)  holds
also for the supremums of countable subsets of $S$, (i.e.,
$p_i(x)\leq\bigcup_{j<\omega}y_j$ implies that $p_i(x)\leq y_j$ for
some $j<\omega$; we express this by saying that $p_i(x)$ is
$\sigma$-join-irreducible).

\subsection{Topological spaces}\label{tspaces}

Here we recall some topological notions and facts relevant to this
paper.
We assume the reader to be familiar with the basic notions of
topology \cite{en89}. For the underlying set  of a topological space $X$ we will write $X$, in abuse
of notation. We will often abbreviate ``topological space'' to
``space''. A space is \emph{zero-dimensional} if it has a basis of
clopen sets. Recall that a \emph{basis} for the topology on $X$ is a
set $\cal B$ of open subsets of $X$ such that for every $x\in X$ and
open $U$ containing $x$ there is $B\in \cal B$ satisfying $x\in
B\subseteq U$.

Let $\omega$ be the space of non-negative integers with the
discrete topology. Of course, the spaces
$\omega\times\omega=\omega^2$, and $\omega\sqcup\omega$ are
homeomorphic to $\omega$, the first homeomorphism is realized by
the Cantor pairing function $\langle \cdot,\cdot\rangle$.
Let $\calN=\omega^\omega$ be the set of all infinite
sequences of natural numbers (i.e., of all functions $\xi \colon
\omega \to \omega$). Let $\omega^*$ be the set of finite sequences
of elements of $\omega$, including the empty sequence. For
$\sigma\in\omega^*$ and $\xi\in\calN$, we write
$\sigma\sqsubseteq \xi$ to denote that $\sigma$ is an initial
segment of the sequence $\xi$. By $\sigma\xi=\sigma\cdot\xi$ we
denote the concatenation of $\sigma$ and $\xi$, and by
$\sigma\cdot\calN$ the set of all extensions of $\sigma$ in
$\calN$. For $x\in\calN$, we can write
$x=x(0)x(1)\dotsc$ where $x(i)\in\omega$ for each $i<\omega$. For
$x\in\calN$ and $n<\omega$, let $x\upharpoonright n=x(0)\dotsc x(n-1)$
denote the initial segment of $x$ of length $n$. Notations in the
style of regular expressions like $0^\omega$, $0^\ast 1$ or
$0^m1^n$ have the obvious standard meaning.

By endowing $\calN$ with the product of the discrete
topologies on $\omega$, we obtain the so-called \emph{Baire space}.
The product topology coincides with the topology
generated by the collection of sets of the form
$\sigma\cdot\calN$ for $\sigma\in\omega^*$. The Baire space
is of primary importance for Descriptive Set Theory and Computable Analysis.
The importance stems from
the fact that many countable objects are coded straightforwardly
by elements of $\calN$, and it has very specific topological
properties. In particular, it is a perfect zero-dimensional space such that any countably based zero-dimensional $T_0$-space topologically embeds into it. 
The subspace $\mathcal{C}:=2^\omega$ of $\calN$ formed by
the infinite binary strings (endowed with the relative topology
inherited from $\calN$) is known as the \emph{Cantor space}.

We recall the well-known (see e.g. \cite{ke95}) relation of closed subsets of $\calN$ to trees. A {\em tree} is a non-empty set $T\subseteq\omega^*$ which is closed downwards under $\sqsubseteq$. A {\em leaf} of $T$ is a maximal element of $(T;\sqsubseteq)$.  A {\em pruned tree} is a tree without leafs. A {\em path through} a tree $T$ is an element $x\in\calN$ such that $x\upharpoonright n\in T$ for each $n\in\omega$. For any tree $T$, the set $[T]$ of paths through $T$ is closed in $\calN$. For any non-empty closed set $A\subseteq\calN$ there is a unique pruned tree $T$ with $A=[T]$ and, moreover, there is a continuous  surjection $t:\calN\to A$ which is constant on $A$ (such a surjection is called a retraction onto $A$). Therefore, there is a bijection between the pruned trees and the non-empty closed sets. Note that the well founded trees $T$ (i.e., trees with $[T]=\emptyset$) and non-empty well founded forests of the form $F:=T\setminus\{\varepsilon\}$ will be used below, in particular in defining the $h$-quasiorders in Sections \ref{wqo} and \ref{bqo}.

The  \emph{Sierpinski space} $\mathbb{S}$ is the two-point set
$\{\bot,\top\}$ where the set $\{\top\}$ is open but not closed. The
space $\Pomega$ is formed by the set of subsets of $\omega$ equipped
with the Scott topology \cite{aj}. A countable base of the Scott topology is
formed by the sets $\{A\subseteq\omega\mid F\subseteq A\}$, where
$F$ ranges over the finite subsets of $\omega$. 

Recall  that a space $X$ is \emph{Polish} if it is countably based
and metrizable with a metric $d$ such that $(X,d)$ is a complete
metric space. Important examples of Polish spaces are $\omega$,
$\calN$, $\mathcal{C}$, the space of reals $\IR$ and its Cartesian
powers $\IR^n $ ($ n < \omega $), the closed unit interval $ [0,1]
$, the Hilbert cube $ [0,1]^\omega $ and the space
$\IR^\omega $. Simple examples of non-Polish spaces are $\IS$,
$\Pomega$ and the space $\IQ$ of rationals.

 A {\em
quasi-metric} on $X$ is a function from $X\times X$ to the
nonnegative reals such that $d(x,y)=d(y,x)=0$ iff $x=y$, and
$d(x,y)\leq d(x,z)+d(z,y)$. Every quasi-metric on \( X \)  induces the topology \( \tau_d \) on \( X \) generated by the open balls $\{ y \in X \mid d(x,y) < \varepsilon \}$ for \( x \in X \) and \( 0 < \varepsilon \). A  space \( X \) is  \emph{quasi-metrizable} if there is a quasi-metric on \( X \) which generates its topology.
If \( d \) is a quasi-metric on \( X \), let  \( \hat{d} \) be the metric on \( X \) defined by 
\( \hat{d}(x,y)= max\{d(x,y),d(y,x)\} \). A sequence $\{x_n  \}$ is  \emph{\( d \)-Cauchy} if for every \( \varepsilon>0\) there is \( p \in \omega \) such that \( d(x_n,x_m) < \varepsilon \) for all \( p \leq n \leq m \). We say that the quasi-metric \( d \) on \( X \) is \emph{complete} if every \( d \)-Cauchy sequence converges with respect to \( \hat{d} \).
A \( T_0 \) space \( X \) is called \emph{quasi-Polish} \cite{br} if it is countably based and there is a complete quasi-metric which generates its topology. 

Note that the spaces $\IS$, $\Pomega$ are quasi-Polish while the space $\IQ$ is not. A complete quasi-metric which is compatible with the topology of  \( P \omega \) is given by \( d(x,y) = 0 \) if \( x \subseteq y \) and \( d(x,y) = 2^{-(n+1)} \) otherwise, where \( n \) is the smallest element in \( x \setminus y \) (for every, \( x,y \subseteq \omega \)). As shown in \cite{br}, a space is quasi-Polish iff it is homeomorphic to a $\bfPi^0_2$-subset of  \cite{br} (the well known definition of $\bfPi^0_2$-subset is recalled in the next subsection). There are some other interesting characterizations of quasi-Polish spaces. For this paper the following characterization in terms of total admissible representations is relevant.

A \emph{representation} of a space $X$ is a surjection of a subspace
of the Baire space $\calN$ onto $X$. A basic notion of Computable
Analysis \cite{wei00} is the notion of admissible representation. A
representation $\delta$ of $X$ is \emph{admissible}, if it is
continuous and any continuous function $\nu:Z \to X$ from a subset
$Z\subseteq\calN$ to $X$ is continuously reducible to $\delta$, i.e.
$\nu=\delta\circ g$ for some continuous function $g:Z \to \calN$. 
In \cite{br} the following  characterization of
quasi-Polish spaces  was obtained:
A space $X$ is quasi-Polish iff it has a total admissible representation $\delta:\calN\to X$.

%%%%%%%%%%%%%%%%%%%%%%%%%%%%%%%%%%%%%%%%%%%%%%%%%%%%%%%%%%%%%%%%%
%

\subsection{Classical hierarchies in quasi-Polish spaces}\label{hquasi}

Here we recall some notions and facts on the classical hierarchies in quasi-Polish spaces \cite{ke95,br}. Note that the definitions of Borel and Luzin hierarchies look slightly different  from the well-known definitions for Polish spaces \cite{ke95}, in order to behave correctly also on non-Hausdorff spaces (for Polish spaces the definitions are equivalent to the usual ones).

A \emph{pointclass} on $ X $ is simply a collection $\bfGamma(X) $ of subsets of $ X $. 
A \emph{family of pointclasses} \cite{s13} is a family $ \bfGamma=\{\bfGamma(X)\} $ 
indexed by arbitrary topological spaces $X$ (or by spaces in a reasonable class) such that each $ \bfGamma(X) $ is
a pointclass on $ X $ and $ \bfGamma $ is closed under
continuous preimages, i.e. $ f^{-1}(A)\in\bfGamma(X) $
for every $ A\in\bfGamma(Y) $ and every continuous function $ f \colon X\to Y $. 
A basic example of a family of pointclasses is given by the family
$\mathcal{O}=\{\tau_X\}$ of the topologies of all the spaces $X$.

We will use some operations on families of pointclasses. First, the usual
set-theoretic operations will be applied to the families of
pointclasses pointwise: for example, the union $\bigcup_i
\bfGamma_i$ of the families of pointclasses
$\bfGamma_0,\bfGamma_1,\ldots$ is defined by
$(\bigcup_i\bfGamma_i)(X)=\bigcup_i\bfGamma_i(X)$.

Second, a large class of such operations is induced by the
set-theoretic operations of L.V. Kantorovich and E.M. Livenson
(see e.g. \cite{s13} for the general definition). 
Among them are the operation $\bfGamma\mapsto\bfGamma_\sigma$,
where $\bfGamma(X)_\sigma$ is the set of all countable unions of sets in $\bfGamma(X)$, 
the operation $\bfGamma\mapsto\bfGamma_\delta$,
where $\bfGamma(X)_\delta$ is the set of all countable intersections of sets in $\bfGamma(X)$, 
the operation $\bfGamma\mapsto\bfGamma_c$,
where $\bfGamma(X)_c$ is the set of all complements of sets in $\bfGamma(X)$, 
the operation $\bfGamma\mapsto\bfGamma_d$,
where $\bfGamma(X)_d$ is the set of all differences of sets in $\bfGamma(X)$, 
the operation $\bfGamma\mapsto\bfGamma_\exists$ defined by 
$\bfGamma_\exists(X):=\{\exists^\mathcal{N}(A)\mid A\in\bfGamma(\calN\times X)\}$,
where $\exists^\mathcal{N}(A):=\{x\in X\mid \exists p\in\calN.(p,x)\in A\}$ is the projection of $A\subseteq\calN\times X$ 
along the axis $\calN$,
and finally the operation $\bfGamma\mapsto\bfGamma_\forall$ defined by 
$\bfGamma_\forall(X):=\{ \forall^\mathcal{N}(A) \mid A\in\bfGamma(\calN\times X)\}$,
where $\forall^\mathcal{N}(A):=\{x\in X\mid \forall p\in\calN.(p,x)\in A\}$.

The  operations on families of pointclasses enable to provide short uniform descriptions 
of the classical hierarchies in quasi-Polish spaces. 
E.g., the {\em Borel  hierarchy} is the sequence of families of pointclasses
$\{\bfSig^0_\alpha\}_{ \alpha<\omega_1}$ defined by induction on $\alpha$ as follows \cite{s06,br}:
$\bfSig^0_0(X):=\{\emptyset\}$,
$\bfSig^0_1 := \mathcal{O}$ (the family of open sets), $\bfSig^0_2 := (\bfSig^0_1)_{d\sigma}$,
and  $\bfSig^0_\alpha(X)
:=(\bigcup_{\beta<\alpha}\bfSig^0_\beta(X))_{c\sigma}$ for
$\alpha>2$.
The sequence $\{\bfSig^0_\alpha(X)\}_{ \alpha<\omega_1}$ is called \emph{the Borel hierarchy} in $X$.
We also let $\bfPi^0_\beta(X) := (\bfSig^0_\beta(X))_c $
and $\bfDelta^0_\alpha(X) := \bfSig^0_\alpha(X) \cap \bfPi^0_\alpha (X)$.
The classes $\bfSig^0_\alpha(X),\bfPi^0_\alpha(X),{\bfDelta}^0_\alpha(X)$
are called the \emph{levels} of the Borel hierarchy in $X$. 
 The class $\mathbf{B}(X)$ of \emph{Borel sets} in $X$ is defined as the union of all levels 
 of the Borel hierarchy in $X$; it coincides with the smallest $\sigma$-algebra of subsets of $X$ containing the open sets.
We have $\bfSig^0_\alpha(X)\cup\bfPi^0_\alpha(X)\subseteq\bfDelta^0_\beta(X)$ for all $\alpha<\beta<\omega_1$. For any uncountable quasi-Polish space $X$ and any $\alpha<\omega_1$, $\bfSig^0_\alpha(X)\not\subseteq\bfPi^0_\alpha(X)$.

 The {\em hyperprojective hierarchy} is the sequence of families of pointclasses
 $\{\bfSig^1_\alpha\}_{ \alpha<\omega_1}$ defined by induction on $\alpha$ as follows:  
 $\bfSig^1_0=\bfSig^0_2$, 
 $\bfSig^1_{\alpha+1}=(\bfSig^1_\alpha)_{c\exists}$, 
 $\bfSig^1_\lambda=(\bfSig^1_{<\lambda})_{\delta \exists}$, 
 where $\alpha,\lambda<\omega_1$, $\lambda$ is a limit ordinal, 
 and $\bfSig^1_{<\lambda}(X):=\bigcup_{\alpha<\lambda}\bfSig^1_\alpha(X)$.
 In this way, we obtain for any quasi-Polish space $X$ the sequence 
 $\{\bfSig^1_\alpha(X)\}_{\alpha<\omega_1}$, which we call here \emph{the hyperprojective hierarchy in $X$}. 
 The pointclasses $\bfSig^1_\alpha(X)$, $\bfPi^1_\alpha(X):=(\bfSig^1_\alpha(X))_c$ and 
 $\bfDelta^1_\alpha(X):=\bfSig^1_\alpha(X)\cap\bfPi^1_\alpha(X)$ are called 
 \emph{levels of the hyperprojective hierarchy in $X$}. 
 The finite non-zero levels of the hyperprojective hierarchy coincide with the corresponding levels 
 of the Luzin projective hierarchy. 
 The class of \emph{hyperprojective sets} in $X$ is defined as the union of all levels 
 of the hyperprojective hierarchy in $X$. We have $\bfSig^1_\alpha(X)\cup\bfPi^1_\alpha(X)\subseteq\bfDelta^1_\beta(X)$ for all $\alpha<\beta<\omega_1$. For any uncountable Polish space $X$ and any $\alpha<\omega_1$, $\bfSig^1_\alpha(X)\not\subseteq\bfPi^1_\alpha(X)$. For any quasi-Polish space $X$, $\mathbf{B}(X)=\bfDelta^1_1(X)$ (the Suslin theorem \cite{ke95,br}).
As mentioned in the Introduction, in this paper we mostly stay within the Borel sets, hence the very important Luzin hierarchy will not be considered. We recalled its definition mainly to illustrate the general notions of hierarchy theory.

For any non-zero ordinal $\theta<\omega_1$, let
$\{\bfSig^{-1,\theta}_\alpha\}_{\alpha<\omega_1}$ be the
Hausdorff difference hierarchy over $\bfSig^0_\theta$. We
recall the definition. An ordinal $\alpha$ is {\em even}  (resp.
{\em odd}) if $\alpha=\lambda+n$ where $\lambda$ is either zero or
a limit ordinal and $n<\omega$, and the number $n$ is even (resp.,
odd). For an ordinal $\alpha$, let $r(\alpha)=0$ if $\alpha$ is
even and $r(\alpha)=1$, otherwise. For any ordinal $\alpha$,
define the operation $D_\alpha$ sending sequences of sets
$\{A_\beta\}_{\beta<\alpha}$ to sets by
 $$D_\alpha(\{A_\beta\}_{\beta<\alpha})=
\bigcup\{A_\beta\setminus\bigcup_{\gamma<\beta}A_\gamma\mid
\beta<\alpha,\,r(\beta)\not=r(\alpha)\}.$$
 For any ordinal $\alpha<\omega_1$ and  any pointclass ${\cal E}$ in $X$,
let $D_\alpha({\cal E})$ be the class of all sets
$D_\alpha(\{A_\beta\}_{\beta<\alpha})$, where $A_\beta\in{\cal E}$
for all $\beta<\alpha$. Finally, let
$\bfSig^{-1,\theta}_\alpha(X)=D_\alpha(\bfSig^0_\theta(X))$
for any space $X$ and for all $\alpha,\theta<\omega$, $\theta>0$.
It is well
known  that  $\bfSig^{-1,\theta}_\alpha(X)\cup\bfPi^{-1,\theta}_\alpha(X)\subseteq\bfDelta^{-1,\theta}_{\beta}(X)$ and $\bigcup_{\alpha<\omega_1}\bfSig^{-1,\theta}_\alpha(X)\subseteq
\bfDelta^0_{\theta+1}(X)$ for all $\alpha<\beta<\omega_1$.  For any quasi-Polish space $X$ and any $0<\theta<\omega_1$, $\bigcup_{\alpha<\omega_1}\bfSig^{-1,\theta}_\alpha(X)=
\bfDelta^0_{\theta+1}(X)$ (the Hausdorff-Kuratowski theorem  \cite{ke95,br}).

%%%%%%%%%%%%%%%%%%%%%%%%%%%%%%%%%%%%%%%%%%%%%%%%%%%%%%%

\subsection{Wadge hierarchy}\label{wadgedst}

Here we briefly discuss the Wadge reducibility in the Baire space. For subsets $A,B$ of the Baire space $\mathcal{N}$, $A$ is {\em Wadge reducible} to $B$ ($A\leq_WB$), if $A=f^{-1}(B)$ for some continuous function $f$ on $\mathcal{N}$. The quotient-poset of the QO $(P(\mathcal{N});\leq_W)$ under the induced equivalence relation $\equiv_W$ on the power-set of $\mathcal{N}$ is called {\em the structure of Wadge degrees} in $\mathcal{N}$.

In \cite{wad84} W. Wadge  (using the Martin
determinacy theorem) proved the following result:
The structure $({\mathbf B}(\calN);\leq_W)$ of the Borel sets in the Baire
space is semi-well-ordered (i.e., it is well-founded and for all
$A,B\in{\mathbf B}(\mathcal{N})$ we have $A\leq_WB$ or $\overline{B}\leq_WA$). In particular, there is no antichain of size 3 in $({\mathbf B}(\calN);\leq_W)$. He has also computed the rank $\nu$ of $({\mathbf B}(\calN);\leq_W)$ which we call the Wadge ordinal. Recall that a set $A$ is {\em self-dual} if $A\leq_W\overline{A}$. W. Wadge has shown that
if a Borel set  is self-dual (resp. non-self-dual) then any Borel set of the next Wadge rank is non-self-dual (resp. self-dual), a Borel set of Wadge rank of countable cofinality is self-dual, and a Borel set of Wadge rank of uncountable cofinality is non-self-dual. This characterizes the  structure of Wadge degrees of Borel sets up to isomorphism.

Recall that a pointclass $\bfGamma\subseteq P(\calN)$ has the {\em separation property} if for all disjoint sets $A,B\in\bfGamma$ there is $S\in\bfGamma\cap\bfGamma_c$ with $A\subseteq S\subseteq\overline{B}$. In
\cite{vw76,ste80} the following deep relation of the Wadge
reducibility to the separation property 
was established:
For any Borel set $A$ which is non-self-dual  exactly one of the principal ideals
$\{X\mid X\leq_WA\}$, $\{X\mid X\leq_W\overline{A}\}$ has the
separation property. 
The mentioned results give rise to the {\em Wadge hierarchy} which
is, by definition, the sequence
$\{\bfSig_\alpha(\calN)\}_{\alpha<\nu}$ (where $\nu$ is the Wadge ordinal) of all non-self-dual principal
ideals of $({\bf B}(\calN);\leq_W)$ that do not have the separation
property  and satisfy for all $\alpha<\beta<\nu$ the strict
inclusion $\bfSig_\alpha(\calN)\subset\bfDelta_\beta(\calN)$ where, as usual, $\bfDelta_\beta(\calN)=\bfSig_\alpha(\calN)
\cap\bfPi_\alpha(\calN)$. 

The Wadge hierarchy subsumes the classical hierarchies in the Baire space, in particular
${\bfSig}_\alpha(\calN)={\bfSig}^{-1}_\alpha(\calN)$ for each
$\alpha<\omega_1$, 
${\bfSig}_1(\calN)={\bfSig}^0_1(\calN)$,
${\bfSig}_{\omega_1}(\calN)={\bfSig}^0_2(\calN)$,
${\bfSig}_{\omega_1^{\omega_1}}(\calN)={\bfSig}^0_3(\calN)$ and so on.
Thus, the sets  of finite Borel rank coincide with the sets of Wadge
rank less than
$\lambda=sup\{\omega_1,\omega_1^{\omega_1},\omega_1^{(\omega_1^{\omega_1})},\ldots\}$.
Note that $\lambda$ is the smallest solution of the ordinal equation
$\omega_1^\varkappa=\varkappa$. Hence, we warn the reader not to
mistake $\bfSig_\alpha(\calN)$ with $\bfSig^0_\alpha(\calN)$. To give
the reader a first impression about the Wadge ordinal we note that
the rank of the QO $(\bfDelta^0_\omega(\calN);\leq_W)$ is the
$\omega_1$-st  solution of the ordinal equation
$\omega_1^\varkappa=\varkappa$ \cite{wad84}. As mentioned in the Introduction, Wadge hierarchy is non-standard in the sense that it is based on a highly original tool of infinite games, in contrast to the set-theoretic and topological methods used to investigate the classical hierarchies. As a result, Wadge hierarchy was originally defined only for the Baire space and its close relatives, and it is not straightforward to extend it to non-zero-dimensional spaces. 

The Wadge hierarchy was an important development in classical DST not only as a unifying concept but also as a useful tool to investigate countably based zero-dimension spaces. We illustrate this with two examples. In \cite{engelen}  a complete classification (up to homeomorphism) of homogeneous zero-dimensional absolute Borel sets was achieved, completing a series of earlier results in this direction. In \cite{ems87} it was shown that any Borel subspace of the Baire space with more than one point has a non-trivial auto-homeomorphism.

%%%%%%%%%%%%%%%%%%%%%%%%%%%%%%%%%%%%%%%%%%%%%%%%%%%%%%%%%%%%
%
%
%%%%%%%%%%%%%%%%%%%%%%%%%%%%%%%%%%%%%%%%%%%%%%%%%%%%%%%%%%%%

%%%%%%%%%%%%%%%%%%%%%%%%%%%%%%%%%%%%%%%%%%%%%%%%%%%%%%%%%%%%
%
%
%%%%%%%%%%%%%%%%%%%%%%%%%%%%%%%%%%%%%%%%%%%%%%%%%%%%%%%%%%%%

%\subsection{Gale-Stuwart games}

\section{Well and better quasiorders}\label{wbqo}

In this section we briefly discuss some notions and facts of WQO-theory relevant to our main theme. We do not try to be comprehensive in this survey or with references and from the numerous important concrete WQOs in the literature we choose mainly those directly relevant.

\subsection{Well quasiorders}\label{wqo}

A {\em well quasiorder} (WQO) is a QO $Q=(Q;\leq)$ that has neither infinite
descending chains nor infinite antichains. Note that for this paper WQOs are equivalent to the associated well partial orders (WPOs) and are only used to simplify some notation. 

 With any WQO
$Q$ we associate its rank and also its {\em width} $w(P)$ defined as
follows: if $P$ has antichains with any finite number of elements,
then $w(Q)=\omega$, otherwise $w(Q)$ is the greatest natural number
$n$ for which $Q$ has an antichain with $n$ elements. For instance, the structure of Wadge degrees of Borel sets is of width 2. Note that the notion of width maybe naturally refined in order to stratify the WQOs of infinite width (in the above ``naive'' sense) using ordinals. 

There are several useful characterizations of WQOs. Some of them are collected in the following proposition. An infinite  sequence $\{x_n\}$ in $Q$ is {\em good} if $x_i\leq x_j$ for some $i<j<\omega$, and {\em bad} otherwise. Let $\mathcal{F}(Q)$ be the class of all upward closed subsets of a QO $Q$.

\begin{proposition}\label{wqo1}
For a quasiorder $Q=(Q;\leq)$ the following are equivalent:
\begin{enumerate}\itemsep-1mm
\item $Q$ is WQO;
\item every infinite sequence in $Q$ is good;
\item every infinite sequence in $Q$ contains an increasing subsequence;
\item any non-empty upward closed set in $Q$ has a finite number of minimal elements;
\item the poset $(\mathcal{F}(Q);\supseteq)$ is well-founded;
\item every linear order on $Q$ which extends $\leq$ is a well-order.
\end{enumerate}
\end{proposition}

It is easy to see that if $Q$ is WQO then any QO on $Q$ that extends $\leq$ is WQO, as well as any subset of $Q$ with the induced QO. Also, the cartesian product of two WQOs is WQO, and if $P,Q$ are WQOs which are substructures of some QO, then $P\cup Q$ is WQO. There are also many other useful closure properties of WQOs including the following two examples:

\begin{enumerate}\itemsep-1mm
\item If $Q$ is WQO then  $(Q^*;\leq^*)$ is WQO where $Q^*$ is the set of finite sequences in $Q$ and  $(x_1,\ldots,x_m)\leq^*(y_1,\dots,y_n)$ means that for some strictly increasing $\varphi:\{1,\ldots,m\}\to\{1,\ldots,n\}$ we have $x_i\leq y_{\varphi(i)}$ for all $i$ (Higman's lemma \cite{hi52}). 

\item  If $Q$ is WQO then  $({\mathcal T}_Q;\leq_h)$ is WQO where ${\mathcal T}_Q$ is the set of finite $Q$-labeled trees $(T,c)$, $c:T\to Q$, and $\leq_h$ is the homomorphism QO ($h$-QO for short) defined as follows: $(T,c)\leq_h(S,d)$ if there is a monotone function $\varphi:(T,\sqsubseteq)\to (S,\sqsubseteq)$ such that $c(t)\leq d(\varphi(t))$ for all $t\in T$ (a consequence of Kruskal's theorem \cite{kr60}). Recall that our trees are initial segments of $(\omega^*;\sqsubseteq)$.
\end{enumerate}

We proceed with some concrete examples of WQOs. In Section \ref{wqoreg} we will consider  the important particular case of Higman's lemma $Q=\overline{k}=(k;=)$ for $2\leq k<\omega$; in this case $(Q^*;\leq^*)$ is  the subword relation on the set $k^*$ of finite words over $k$-letter alphabet $\{0,\ldots,k-1\}$. The Kruskal's theorem and Higman's lemma  are close to optimal in the sense that the sets of finite structures in many natural classes (for instance, the set of finite distributive lattices of width 2 \cite{s88}) are not WQOs under the embeddability relation. 

Similarly, we will be interested  in the $h$-QO  on the set ${\mathcal F}_k={\mathcal F}_Q$ (where $Q=\overline{k}=(k;=)$) of finite $k$-labeled forests defined in the same manner as for trees; this QO first appeared in \cite{he93}. Also some weaker QOs $\leq_0,\leq_1,\leq_2$ on ${\mathcal F}_k$  are of interest \cite{he93,s13}. They are defined as follows: $(T,c)\leq_0(S,d)$ (resp. $(T,c)\leq_1(S,d)$, $(T,c)\leq_2(S,d)$) if there is a monotone function $\varphi:(T,\sqsubseteq)\to (S,\sqsubseteq)$ such that $c=g\circ d\circ \varphi$ for some permutation $g:k\to k$ (resp. $c=g\circ d\circ \varphi$ for some $g:k\to k$, $\forall x,y\in T ((x\sqsubseteq y\wedge c(x)\not=c(y))\rightarrow d(\varphi(x))\not=d(\varphi(y)))$. Obviously, $\leq_h\subseteq\leq_0\subseteq\leq_1\subseteq\leq_2$.

To obtain further interesting examples, we   iterate the construction
$Q\mapsto\mathcal{T}_Q$ starting with the antichain
$\overline{k}$ of $k$ elements $\{0,\ldots,k-1\}$ (or with any other WQO $P$ in place of $\bar{k}$). Define the
sequence $\{\mathcal{T}_k(n)\}_{n<\omega}$ of QOs
by induction on $n$ as follows:
$\mathcal{T}_k(0)=\overline{k}$  and
$\mathcal{T}_k(n+1)=\mathcal{T}_{\mathcal{T}_k(n)}$. Note that
$\mathcal{T}_k(1)=\mathcal{T}_k$.
Identifying the elements $i<k$ of $\overline{k}$ with the
corresponding minimal elements $s(i)$ of
$\mathcal{T}_k(1)$, we may think that
$\mathcal{T}_k(0)$ is an initial segment of $\mathcal{T}_k(1)$.
Then
$\mathcal{T}_k(n)$ is an initial segment of $\mathcal{T}_k(n+1)$
for each $n<\omega$, and hence
$\mathcal{T}_k(\omega)=\bigcup_{n<\omega}\mathcal{T}_k(n)$
is WQO. Note that $\mathcal{T}_{\mathcal{T}_k(\omega)}=\mathcal{T}_k(\omega)$. The set $\mathcal{T}^\sqcup_k(\omega)$ of forests generated by the trees in $\mathcal{T}_k(\omega)$ is also WQO. The iterated $h$-QOs were first defined and studied in \cite{s12}.

By a result in \cite{s12}, in the case $k=2$ the QO $(\mathcal{T}_k(\omega);\leq)$ is semi-well-ordered with the order type $\varepsilon_0$. This indicates a possible relation to the hierarchy theory. We will see its close relation to the so called fine hierarchy of sets in Section \ref{hset}

Finally, we mention the remarkable example  of the QO of finite graphs with the graph-minor relation (we do not define this relation because do not discuss it in the sequel). Robertson-Seymour theorem \cite{rs04} stating that this structure is WQO is one of the deepest known facts about finite graphs. The above-mentioned Higman's lemma and Kruskal's theorem are certainly much easier to prove than Robertson-Seymour's but their proofs are also  non-trivial. Robertson-Seymour theorem is important for computer science because it implies that many graph problems are solvable in polynomial time, although such algorithms are hard to discover because it is hard to compute the minimal (under the minor relation) elements of a given upward closed sets of graphs (see e.g. \cite{df98} for details).

Along with the rank and width, there are some other important invariants of a WPO $(P;\leq)$. The most important is probably the {\em maximal order type} $o(P)$ which is the supremum of the order types of linearizations of $\leq$ (i.e., linear orders on $P$ which extend $\leq$). By a nice result of D. De Jongh and  R. Parikh \cite{jp77}, every WPO $P$ has a linearization of order type $o(P)$.
The computation of $o(P)$ for natural WPOs turned out an interesting and challenging task. D. Scmidt \cite{sc79} computed the maximal order type of the Higman's WPO $(k^*;\leq^*)$ and gave upper bounds on the maximal order types of some other important WPOs including that of Kruskal's.

To our knowledge, the maximal order types of the other above-mentioned concrete WPOs are still unknown. Also the problem of relating $rk(P)$ and $o(P)$ discussed in  \cite{sc79} seems still to be  open. In particular, there is no known characterization of pairs of ordinals $(\alpha,\beta)$ such that $\alpha=rk(P)$ and $\beta=o(P)$ for some WPO $P$ \cite{mon07}.

The structure $(\mathcal{T}^\sqcup_k(\omega);\leq_h)$ may be expanded by natural operations inducing a rich algebraic structure on the quotient-poset. These operations, introduced and studied in  \cite{s11,s12,s16}, are important for relating the $h$-QOs to hierarchy theory.

 The binary operation $\oplus$ of disjoint union on  
$\mathcal{T}^\sqcup_k(\omega)$ is defined in the obvious way. 
For any $i<k$ and
$F\in\mathcal{T}^\sqcup_k(\omega)$, let $p_i(F)$ be the tree in
$\mathcal{T}_k(\omega)$ obtained from $F$ by adjoining the empty string labeled by $i$. Let $\bf{i}$ be the singleton tree $\{\varepsilon\}$ labeled by $i$.
Define the binary operation  $+$ on $\mathcal{T}^\sqcup_k(\omega)$ as
follows: $F+G$ is obtained by adjoining a copy of $G$ below any
leaf of $F$. One easily checks that ${\bf i}+F\equiv_hp_i(F)$, $F\leq_hF+G$,
$G\leq_hF+G$, $F\leq_hF_1\to F+G\leq_hF_1+G$, $G\leq_hG_1\to
F+G\leq_hF+G_1$, $(F+G)+H\equiv_hF+(G+H)$. Note that the set $\mathcal{T}^\sqcup_k(n)$ is closed under the operation $+$ for each $1\leq n\leq\omega$.
Define the function $s$ on $\mathcal{T}_k(\omega)$ as
follows: $s(F)$ is the singleton tree carrying the  label $F$. Note that $s(\bf i)=\bf{i}$ for each $i<k$, and $T\leq_hS$ iff $s(T)\leq_hs(S)$, for all $S,T\in\widetilde{\mathcal{T}}_k$.  One easily
checks the following properties:

\begin{proposition}\label{dcq}
 \begin{enumerate}\itemsep-1mm
 \item For each $1\leq n\leq\omega$, $(\mathcal{T}^\sqcup_k(n);\oplus,\leq_h,p_0,\ldots,p_{k-1})$ is a $dc$-semilattice.
 \item For any $T\in\mathcal{T}_k(\omega)$,  $F\mapsto s(T)+F$
is a closure operator on $(\mathcal{T}^\sqcup_k(\omega);\leq_h)$.
 \item For all $T,T_1\in\mathcal{T}_k(\omega)$ and $F,F_1\in\mathcal{T}^\sqcup_k(\omega)$, if
$s(T)+F\leq_h s(T_1)+F_1$ and $T\not\leq_h T_1$ then $s(T)+F\leq_h F_1$.
\item The QO $(\mathcal{T}^\sqcup_2(\omega);\leq_h)$ is semi-well-ordered with order type $\varepsilon_0$. 
 \end{enumerate}
  \end{proposition}

\subsection{Better quasiorders}\label{bqo}

As we know, the closure properties of WQOs suffice to obtain nice WQOs using finitary constructions like forming finite labeled words or trees. But they do not suffice to establish that similar structures on, say, infinite words and trees are WQOs. A typical example is the attempt to extend the example (1) from the previous subsection to the set $Q^\omega$ of infinite $Q$-labeled sequences. As shown by R. Rado (see e.g. \cite{la71}), there is WQO $Q$ such that $Q^\omega$ is not WQO.

Nevertheless, it turns out possible to find a natural subclass of WQOs, called better quasiorders (BQOs) which contains most of the ``natural'' WQOs and has strong closure properties also for many infinitary constructions. In particular, if  $Q$ is BQO then $Q^\omega$ is BQO. In this way it is possible to show that many important QOs are BQOs and hence also WQOs. The notion of BQO is due to C. Nash-Williams \cite{nw65}, we recall an alternative equivalent definition due to S. Simpson \cite{si85}, see also \cite{km16}.

Let $[\omega]^\omega$ be the subspace of the Baire space formed by all strictly
increasing functions $p$ on $\omega$. Given $p\in[\omega]^\omega$, by $p^-$ we denote
the result of dropping the first entry from $p$ (or equivalently, $p^- = p\setminus \{min X\}$, if we
think of $p$ as an infinite subset $X=rng(p)$ of $\omega$).
A QO Q is called  BQO if, for any continuous function $f : [\omega]^\omega\to Q$ ($Q$ is assumed to carry the discrete topology), there is $p\in[\omega]^\omega$ with $f(p) \leq f(p^{-})$.

It is easy to see that: any BQO is WQO; if $Q$ is BQO then any QO on $Q$ that extends $\leq$ is BQO, as well as any subset of $Q$ with the induced QO. Also, the cartesian product of two BQOs is BQO, and if $P,Q$ are BQO which are substructures of some QO, then $P\cup Q$ is BQO. There are also many other useful closure properties of BQOs including the following:

\begin{enumerate}\itemsep-1mm

\item If $Q$ is BQO then  $(Q^\omega;\leq^\omega)$ is BQO  (in fact, this holds for sequences of arbitrary transfinite length \cite{nw65,nw68}). 

\item  If $Q$ is BQO then  $(\widetilde{{\mathcal T}}_Q;\leq_h)$ is BQO where $\widetilde{{\mathcal T}}_Q$ is the set of well-founded $Q$-labeled trees $(T,c)$, $c:T\to Q$, and $\leq_h$ is the homomorphism relation defined as follows: $(T,c)\leq_h(S,d)$ if there is a monotone function $\varphi:(T,\sqsubseteq)\to (S,\sqsubseteq)$ such that $c(t)\leq d(\varphi(t))$ for all $t\in T$ (a consequence of the extension of Kruskal's theorem to infinite trees \cite{nw65,nw68}).

\end{enumerate}

We proceed with some concrete examples of BQOs. In Section \ref{wqoreg} we will consider  the  particular case of (1) for $Q=\overline{k}=(k;=)$, $2\leq k<\omega$; in this case $(Q^\omega;\leq^\omega)$ is the the subword relation on the set $k^\omega$ of infinite words over $k$-letter alphabet $\{1,\ldots,k-1\}$. 

Similarly, we will be interested in the $h$-QO on the set $\widetilde{{\mathcal F}}_k=\widetilde{{\mathcal F}}_Q$, $Q=\overline{k}=(k;=)$, of well-founded $k$-labeled forests. 
We can also iterate the construction
$Q\mapsto\widetilde{\mathcal{T}}_Q$ starting with the antichain
$\overline{k}$ of $k$ elements $\{0,\ldots,k-1\}$. Define the
sequence $\{\widetilde{\mathcal{T}}_k(\alpha)\}_{\alpha<\omega_1}$ of QOs
by induction on $\alpha$ as follows:
$\widetilde{\mathcal{T}}_k(0)=\overline{k}$,  
$\widetilde{\mathcal{T}}_k(\alpha+1)=\widetilde{\mathcal{T}}_{\widetilde{\mathcal{T}}_k(\alpha)}$, and $\widetilde{\mathcal{T}}_k(\lambda)=\bigcup_{\alpha<\lambda}\widetilde{\mathcal{T}}_{\widetilde{\mathcal{T}}_k(\alpha)}$ for limit $\lambda<\omega_1$. 
Note that
$\widetilde{\mathcal{T}}_k(1)=\widetilde{\mathcal{T}}_k$.
Identifying the elements $i<k$ of $\overline{k}$ with the
corresponding minimal elements $s(i)$ of
$\widetilde{\mathcal{T}}_k(1)$, we may think that
$\widetilde{\mathcal{T}}_k(0)$ is an initial segment of $\mathcal{T}_k(1)$.
Then
$\widetilde{\mathcal{T}}_k(\alpha)$ is an initial segment of $\widetilde{\mathcal{T}}_k(\beta)$
for all $\alpha<\beta<\omega_1$,  hence
$\widetilde{\mathcal{T}}_k(\omega_1)=\bigcup_{\alpha<\omega_1}\widetilde{\mathcal{T}}_k(\alpha)$
is BQO. Note that $\widetilde{\mathcal{T}}_{\widetilde{\mathcal{T}}_k(\omega_1)}=\widetilde{\mathcal{T}}_k(\omega_1)$. The set $\widetilde{\mathcal{T}}^\sqcup_k(\omega_1)$ of countable disjoint unions of trees in $\widetilde{\mathcal{T}}_k(\omega_1)$ is also BQO. Similar iterated h-QOs were first studied in \cite{s16}.

By a result in \cite{s16}, in the case $k=2$  the QO $\widetilde{\mathcal{T}}_k(\omega)$  is semi-well-ordered (in fact, $\widetilde{\mathcal{T}}_2(\omega_1)$ is also semi-well-ordered). This indicates a possible relation to the Wadge  hierarchy from Section \ref{wadgedst}.

We also mention the remarkable example  of the QO of countable linear orders with the embeddability relation. Laver's theorem \cite{la71} (see also \cite{si85}) stating that this structure is WQO (and thus resolving the Fra\"iss\'e conjecture) is one of the deepest applications of BQO-theory.

The  maximal order types of the concrete BPOs introduced above seem to be unknown. 

We conclude the list of examples of BQOs by a deep fact related to Wadge reducibility. For any QO $Q$ (equipped with the discrete topology), let $(Q^*;\leq^*)$ be the QO of Borel functions $A:\calN\to Q$ with countable range, where $A\leq^*B$ means that for some continuous function $f$ on $\calN$ we have $A(x)\leq B(f(x))$ for all $x\in\calN$. Note that for $Q=(2,=)$ the QO $(Q^*;\leq^*)$ coincides with $(\mathbf{B}(\calN);\leq_W)$.  Theorem 3.2 in \cite{ems87} states the following:

\begin{theorem}\label{emsbqo}
If $(Q;\leq)$ is BQO then $(Q^*;\leq^*)$ is BQO. 
\end{theorem}

Let us briefly recall from \cite{s16} some operations on the iterated labeled forests and collect some of their properties used in the sequel.  The $\omega$-ary operation $\bigoplus$ of disjoint union on $\widetilde{\mathcal{T}}^\sqcup_k(\omega_1)$ is defined in the obvious way. 
For any $i<k$ and
$F\in\widetilde{\mathcal{T}}^\sqcup_k(\omega_1)$, let $p_i(F)$ be the tree in
$\widetilde{\mathcal{T}}_k(\omega_1)$ obtained from $F$ by adjoining the empty string labeled by $i$. 
Define the binary operation  $+$ on $\widetilde{\mathcal{T}}^\sqcup_k(\omega_1)$ as
follows: $F+G$ is obtained by adjoining a copy of $G$ below any
leaf of $F$. One easily checks that ${\bf i}+F\equiv_hp_i(F)$, $F\leq_hF+G$,
$G\leq_hF+G$, $F\leq_hF_1\to F+G\leq_hF_1+G$, $G\leq_hG_1\to
F+G\leq_hF+G_1$, $(F+G)+H\equiv_hF+(G+H)$. Note that the set $\widetilde{\mathcal{T}}^\sqcup_k(\alpha)$ is closed under the operation $+$ for each $1\leq\alpha\leq\omega_1$.
Define the function $s$ on $\widetilde{\mathcal{T}}_k(\omega_1)$ as
follows: $s(F)$ is the singleton tree carrying the  label $F$. Note that $s(\bf i)=\bf{i}$ for each $i<k$, and $T\leq_hS$ iff $s(T)\leq_hs(S)$, for all $S,T\in\widetilde{\mathcal{T}}_k$.  One easily
checks the following properties:

\begin{proposition}\label{dcq1}
 \begin{enumerate}\itemsep-1mm
 \item For any $1\leq\alpha\leq\omega_1$, $(\widetilde{\mathcal{T}}^\sqcup_k(\alpha);\bigoplus,\leq_h,p_0,\ldots,p_{k-1})$ is a $dc\sigma$-semilattice.
 \item For any $T\in\widetilde{\mathcal{T}}_k(\omega_1)$,  $F\mapsto s(T)+F$
is a closure operator on $(\widetilde{\mathcal{T}}^\sqcup_k(\omega_1);\leq_h)$.
 \item For all $T,T_1\in\widetilde{\mathcal{T}}_k(\omega_1)$ and $F,F_1\in\widetilde{\mathcal{T}}^\sqcup_k(\omega_1)$, if
$s(T)+F\leq_h s(T_1)+F_1$ and $T\not\leq_h T_1$ then $s(T)+F\leq_h F_1$.
 \item  The QO $(\widetilde{\mathcal{T}}^\sqcup_2(\omega);\leq_h)$ is semi-well-ordered with order type $sup\{\omega_1,\omega_1^{\omega_1},\omega_1^{(\omega_1^{\omega_1})},\ldots\}$.
 \end{enumerate}
  \end{proposition}

\subsection{Computable well partial orders}

Here we briefly discuss  computability properties of WQOs. The investigation of computable structures (see e.g. \cite{eg99,ak00}) is an important direction of computability theory. 
Recall that an algebraic structure  $\mathbb{A}=(A;\sigma)$  of a finite signature $\sigma$
is {\em computable} if $A$ is a computable subset of $\omega$ and  all signature relations and functions  are computable. A structure is {\em computably presentable} if it is isomorphic to a computable structure. The notions of polynomial-time computable and polynomial-time computably presentable structure are defined in a similar manner using, say, the set $2^*$ of binary words instead of $\omega$.

It is easy to see that all concrete WQOs from Section \ref{wqo}, as well as the expansions of the $h$-QOs by functions, are computably presentable (as well as many other natural countable WQOs). As observed in \cite{mon07}, from results in \cite{jp77} it follows that if a WPO is computable then its maximal order type is a computable ordinal. Moreover, the following result was established in \cite{mon07}: 

\begin{theorem}\label{mont}
\begin{enumerate}\itemsep-1mm
\item Every computable WPO has a computable linearization of maximal
order type.
\item There is no computable (even hyperatithmetical) function  which, given an index for a computable WPO, returns
an index for a computable maximal linearization of this WPO.
\end{enumerate}
\end{theorem}

Which of the concrete WQOs in Section \ref{wqo} (and of their functional expansions) are polynomial-time presentable? By a well-known general fact \cite{ce91}, any computably presentable structure of a relational signature is in fact polynomial-time presentable. Therefore, any of the structures $(\mathcal{F}_k;\leq_h)$ and $(\mathcal{F}_k;\leq_i)$ for $k\leq\omega, i\leq2$ (and other WQOs in Section \ref{wqo}) is polynomial-time presentable. Since the  presentations given by the proof in \cite{ce91} are often artificial, one can ask further natural questions related to feasibility of our structures. We give some examples.

The sets $\mathcal{F}_k$, $k\leq\omega$, may be encoded in a natural way by words over a finite alphabet \cite{hs14}. Will the relations $\leq_i$ be polynomial-time computable w.r.t. this coding? In \cite{hs14} the following results  were obtained:

\begin{theorem}\label{polyn}
\begin{enumerate}\itemsep-1mm
\item The relation $\leq_h$ on $\mathcal{F}_\omega$ is computable in polynomial time.
\item The relations $\leq_1,\leq_2$ on $\mathcal{F}_k$ are computable in polynomial time for $k<\omega$ and are NP-complete for $k=\omega$.
\end{enumerate}
\end{theorem}

Many natural questions concerning computability properties of WPO remain open, e.g.:

\begin{enumerate}\itemsep-1mm
\item Characterize the maximal order types of computable WPOs of rank $\omega$.

\item Recall that the {\em degree spectrum} of a countable structure $\mathbb{A}$ is the set of Turing degrees $\bf{a}$ such that $\mathbb{A}$ is computably presentable relative to $\bf{a}$. What are the degree spectra of countable WPOs? In particular, is it true that for any given countable graph there is a countable WPO with the same degree spectrum?

\item Associate with any WPO $P$ the function $f_P:rk(P)\to\omega$ by: $f_P(\alpha)$ is the cardinality of $\{x\in P\mid rk_P(x)=\alpha\}$. Is $f_P$ computable provided that $P$ is computably presentable? It maybe shown that for several concrete examples of WPOs in from Section \ref{wqo} the answer is positive, though in general we expect the negative answer.   

\end{enumerate}

\subsection{Definability and decidability issues}\label{defwqo}

The study of  definability and
(un)decidability of first order theories is a central 
topic in logic and model theory. Here we briefly discuss such questions for some WQOs from Sections \ref{wqo} and \ref{bqo}. Along with WQOs, we also mention the infix order $\leq$ on the set of words $k^*$ where $u\leq v$ means that $v=xuy$ for some $x,y\in k^*$ (this relation is well founded but has infinite antichains).  

Let
$\mathbb{A}=(A;\sigma)$ be a structure of a given finite signature $\sigma$. As the understanding of definability in $\mathbb{A}$ assumes understanding of its automorphism group $Aut(\mathbb{A})$, we start with citing some facts following from the results in \cite{ks09,ks09a,ksz09,ksy10}.

\begin{theorem} \label{aut}
\begin{enumerate}\itemsep-1mm
\item For any $k\geq3$,  the automorphism groups of the quotient-posets of $({\mathcal F}_k;\leq_h)$ and $(\widetilde{{\mathcal F}}_k;\leq_h)$ are isomorphic to the symmetric group $\mathbf{S}_k$ of permutation of $k$ elements.
\item For any $k\geq2$, $Aut(k^\ast;\leq^*)\simeq Aut(k^\ast;\leq)\simeq{\bf S}_k\times{\bf
S}_2$.
\end{enumerate}
\end{theorem}

Recall that a relation $R\subseteq
A^k$ is {\em definable in $\mathbb{A}$} if there is a first-order $\sigma$-formula
$\phi(x_1,\ldots,x_k)$ with
 $R=\{(x_1,\ldots,x_k)\in
A^k\mid\mathbb{A}\models \phi(x_1,\ldots,x_k)\}.$ A function on $A$ is definable  if its graph is definable. An
element  $a\in A$ is definable if the  set
$\{a\}$ is definable. A structure is definable if its universe and
all signature predicates and functions are definable. 

The characterization of definable relations in a structure is quite important for understanding the structure. In a series of papers by  O. Kudinov and V. Selivanov \cite{ks06,ks07,ks07a,ks09,ks09a,ksz09,ksy10} a method of characterizing the  definable relations was developed  which might be of use for many similar structures on words, trees, graphs and so on (currently, the method mainly applies to well founded partial orders of rank $\omega$). We cite some facts following from  \cite{ks09,ks09a,ksz09,ksy10} which  characterize the definable relations in some of the mentioned structures.

Recall that a structure ${\mathbb A}$ equipped with a numbering $\alpha$
(i.e., a surjection from $\omega$ onto $A$) is {\em arithmetical},
if the equality predicate and all signature predicates and functions are
arithmetical modulo $\alpha$. Obviously, any  definable predicate on
an arithmetical structure $({\mathbb A};\alpha)$ is arithmetical (w.r.t.
$\alpha$) and invariant under the automorphisms of $\mathbb{A}$; we say
that $({\mathbb A};\alpha)$ has the {\em maximal definability property}
if the converse is also true, i.e., any arithmetical predicate
invariant under the automorphisms of ${\mathbb A}$ is definable. The natural numberings of ${\mathcal F}_k$ and $k^*$ (which are not mentioned explicitly in the next theorem) make the structures $({\mathcal F}_k;\leq_h)$, $(k^\ast;\leq^*)$ and $(k^\ast;\leq)$ arithmetical (even computable).

\begin{theorem}\label{main}
Let $\mathbb{A}$ be one of the structures $(k^\ast;\leq^*),(k^\ast;\leq)$ for $k\geq2$ or the quotient-posets of $({\mathcal F}_k;\leq_h)$ for $k\geq3$. Then $\mathbb{A}$ has the maximal definability property.
\end{theorem}

Recall (cf. \cite{ho93,ni00}) that a structure $\mathbb{B}$ of a finite
relational signature $\tau$ is {\em biinterpretable} with a
structure $\mathbb{C}$ of a finite relational signature $\rho$ if $\mathbb{B}$ is definable in $\mathbb{C}$ (in particular, there is a bijection
$f:B\rightarrow B_1$ on a definable set $B_1\subseteq C^m$ for some
$m\geq1$ which induces an isomorphism on the $\tau$-structure $\mathbb{B}_1$ definable in $\mathbb{C}$), $\mathbb{C}$ is definable in $\mathbb{B}$
(in particular, there is a similar bijection $g:C\rightarrow C_1$ on
a definable set $C_1\subseteq B^n$ for some $n\geq1$),  the function
$g^m\circ f:B\rightarrow B^{nm}$ is definable in $\mathbb{B}$ and  the
function $f^n\circ g:C\rightarrow C^{mn}$ is definable in $\mathbb{C}$.
Though the notion of biinterpretability looks sophisticated, its role in model theory is increasing because it gives a natural and strong equivalence relations on structures.

\begin{theorem}\label{main1}
The expansions of the structures $(k^\ast;\leq^*)$ and $(k^\ast;\leq)$, $k\geq2$, by the constants for words of lengths 1 and 2, and the expansion of the quotient-posets of $({\mathcal F}_k;\leq_h)$,  $k\geq3$, by the constants $\bf{0},\ldots,\bf{k-1}$ for  singleton trees, are biinterpretable with
$\mathbb{N}=(\omega;+,\cdot)$.
\end{theorem}

The closely related definability issues for embeddability relations on graphs and different classes of finite structures are now actively studied (note that most of these QOs have infinite antichains), see e.g. \cite{jm09,jm09a,jm09b,jm10,ku15,rt16,wi12,wi16}. Nevertheless, there are still many interesting open questions, including those for some structures mentioned in Sections \ref{wqo}, \ref{bqo}.

Recall that the {\em first-order theory} $FO(\mathbb{A})$ of a structure
$\mathbb{A}$ of signature $\sigma$ is the set of $\sigma$-sentences true in $\mathbb{A}$. The investigation of algorithmic complexity of first-order theories of natural structures is a big chapter of logic, model theory and computability theory (see e.g. \cite{tmr53,eltt,eg99}). The proof of Theorem \ref{main}  implies that  $FO(\mathbb{A})$ is $m$-equivalent  to first order arithmetic $FO(\mathbb{N})$ for any structure mentioned in that theorem. 

\begin{theorem} \label{main3}
\begin{enumerate}\itemsep-1mm
\item Let $k\geq3$ and $\mathbb{A}$ be the quotient-poset of some of  $({\mathcal F}_k;\leq_h)$,  $({\mathcal T}^\sqcup_k(\omega);\leq_h)$, or $({\mathcal T}^\sqcup_k(n);\leq_h)$, $2\leq n<\omega$. Then $FO(\mathbb{A})\equiv_mFO(\mathbb{N})$.
\item Let $\mathbb{A}$ be the quotient-poset of some of  $(\widetilde{{\mathcal F}}_k;\leq_h)$,  $(\widetilde{{\mathcal T}}^\sqcup_k(\omega_1);\leq_h)$, or of $(\widetilde{{\mathcal T}}^\sqcup_k(\alpha);\leq_h)$ for some $2\leq \alpha<\omega_1$. Then $FO(\mathbb{N})\leq_mFO(\mathbb{A})$

\end{enumerate}
\end{theorem}

Note that in item (2) we have only the lower bound. The natural upper bound is second-order arithmetic, the precise estimation is an interesting open problem. For many other interesting WPOs we do not know so comprehensive results as above, but for many first-order theories undecidability is known. By interpreting the finite structures of two equivalence relations \cite{eltt} the following result about some other structures from Section \ref{wqo} was established in \cite{ksz09,s13}. Recall that a structure of a finite signature is {\em hereditarily undecidable} if any  its subtheory of the same signature is undecidable.

\begin{theorem} \label{main2}
 For any $k\geq3$, the first-order theories of the quotient-posets of
$(\mathcal{F}_k; \leq_0)$, $(\mathcal{F}_k; \leq_1)$, and $(\mathcal{F}_k; \leq_2)$ are hereditarily undecidable.
\end{theorem}

It is easy to see that the first-order theories of $(k^\ast;\leq^*),(k^\ast;\leq)$ for $k=1$ and of the quotient-poset of $({\mathcal F}_2;\leq_h)$ are decidable. For most of the non-countable structures in Section \ref{bqo} the complexity of first-order theories seem to be  open.

Since the first-order theories of most of the mentioned structures are  undecidable, it is  natural to look for their decidable fragments. Such questions are interesting because they often originate from the computer science community, have many applications and, unlike most structures originated in mathematics, were considered relatively recently and many natural questions  remain open (see e.g. \cite{ku06} and references therein). 

We illustrate this by the subword order on words.
The study of subword order is important in many areas of computer science,
e.g., in pattern matching,  coding theory,
 theorem proving,  algorithmics,  automatic
verification of unreliable channel systems \cite{ab04, ksc15a}. The
reasoning about subwords involves ad
hoc techniques quite unlike the standard tools that work well with prefixes and
suffixes \cite{kns17}.

The study of $FO(A^*;\leq^*)$ was started by Comon and Treinen who showed undecidability for
an expanded signature where $A$ has at least three letters. Kuske \cite{ku06} showed that  the 3-quantifier fragment of  $FO(A^*;\leq^*)$ is undecidable. 
Karandikar and Schnoebelen showed that already the 2-quantifier theory is undecidable \cite{ksc15a} and this is tight since the 1-quantifier fragment is decidable, in fact NP-complete \cite{ku06, ksc15}. Karandikar and Schnoebelen also showed that the two-variable
fragment  is decidable \cite{ksc15} and that it has an elementary complexity
upper bound \cite{ksc16}. 
Recently, it was shown \cite{kns17} that, when constants are allowed, the
1-quantifier fragment   is actually undecidable. This holds as soon as $A$
contains two distinct letters and exhibits a strong dependence on the presence of constants in the signature. To our knowledge, a similar detailed study for most of the above-mentioned structures is still to be done.

%%%%%%%%%%%%%%%%%%%%%%%%%%%%%%%%%%%%%%%%%%%%%%%%%%%%%%%%%%%%%%%%%

%%%%%%%%%%%%%%%%%%%%%%%%%%%%%%%%%%%%%%%%%%%%%%%%%%%%%%%%%%%%%%%%%

\section{Wadge-like reducibilities in quasi-Polish spaces}\label{wlike}

As we know from Section \ref{wadgedst}, the structure of Wadge degrees in the Baire space refines the structure of levels of several popular
hierarchies and serves as a  tool to measure the topological complexity
of some problems of interest in  set-theoretic topology.
There are several reasons and several ways to generalize the
Wadge reducibility on the Baire space. For example, one can consider

\begin{enumerate}\itemsep-1mm

\item more complicated topological spaces instead
of \( \mathcal{N} \) (the notion of Wadge reducibility makes
sense for arbitrary topological spaces);

\item
other natural classes of reducing functions
in place of the continuous functions;

\item reducibilities between functions rather than reducibilities between sets (the sets may be
identified with their characteristic functions).

\end{enumerate}

In any of the mentioned directions a certain progress has been
achieved, although in many cases the situation typically becomes more
complicated than in the classical case. In this section we mention some results in this direction.

\subsection{Wadge-like reducibilities in the Baire space}\label{cf}

For any family of pointclasses $\bfGamma$ and for any spaces $X,Y$, let  $\bfGamma(X,Y)$ be the class of functions $f:X\to Y$ such that $f^{-1}(A)\in\bfGamma(X)$ for each $A\in\bfGamma(Y)$, and let $\bfGamma(X)=\bfGamma(X,X)$. Clearly, $\bfGamma(X)$ is closed under composition and contains the identity function, hence it induces a reducibility $\leq_\bfGamma$ on subsets of $X$ similar to the Wadge reducibility. For any $1\leq\alpha<\omega_1$, let $D_\alpha(X,Y)$ denote $\bfSig^0_\alpha(X,Y)$ and let $\leq_\alpha$ abbreviate $\leq_{\bfSig^0_\alpha}$. Then $\leq_\alpha$ is a QO on $P(X)$. In particular $\leq_1$ coincides with the Wadge reducibility.

For any $\alpha<\omega_1$ and any spaces $X,Y$, let $D_\alpha^W(X,Y)$ be the  class of functions $f:X\to Y$ such that there is a partition $\{D_n\}$ of $X$ to $\bfSig^0_\alpha$-sets and a sequence $f_n:D_n\to Y$ of continuous functions with  $f=\bigcup_{n<\omega}f_n$. Note that $D_\alpha^W(X,Y)\subseteq D_\alpha(X,Y)$.  We again set $D_\alpha^W(X)=D_\alpha^W(X,X)$. Clearly, $D_\alpha^W(X)$ is closed under composition and contains the identity function, hence it induces a reducibility $\leq_\alpha^W$ on subsets of $X$.

The study of the reducibility by Borel functions on the Baire space was initiated by A. Andretta and D. Martin in \cite{am03}, the reducibility $\leq_2$  was studied by A. Andretta \cite{an06}, the other of just defined reducibilities (as well as many other so called amenable reducibilities) were comprehensively investigated by L. Motto-Ross \cite{ros09}. The next result, which is a very particular case of the results in \cite{ros09}, shows that these reducibilities behave similarly to the Wadge reducibility:

\begin{theorem}\label{luca}
For any $1\leq\alpha<\omega_1$, the quotient-posets of $({\mathbf B}(\calN);\leq_\alpha)$ and $({\mathbf B}(\calN);\leq_\alpha^W)$ are isomorphic to that of $({\mathbf B}(\calN);\leq_W)$.
\end{theorem}

\subsection{Wadge reducibility of $k$-partitions in the Baire space}\label{wpartbaire}

Let  $2\leq k<\omega$. By a {\em $k$-partition of $\mathcal{N}$} we
mean a function $A:\mathcal{N}\to k=\{0,\ldots,k-1\}$ often identified with
the sequence $(A_0,\ldots,A_{k-1})$ where $A_i=A^{-1}(i)$ are the components of $A$.
Obviously, 2-partitions of $\mathcal{N}$ can be identified with the subsets of
$\mathcal{N}$ using the characteristic functions. The set of all
$k$-partitions of $\mathcal{N}$ is denoted $k^\mathcal{N}$, thus $2^\mathcal{N}=P(\mathcal{N})$. 
The Wadge  reducibility on subsets of $\mathcal{N}$ is naturally extended to
$k$-partitions: for $A,B\in k^\mathcal{N}$, $A\leq_W B$ means that $A=B\circ
f$ for some continuous function $f$ on $\mathcal{N}$. In this way, we obtain
the QO $(k^\mathcal{N};\leq_W)$. For any pointclass  $\bfGamma\subseteq P(\mathcal{N})$, let $\bfGamma(k^\mathcal{N})$ be the set of $k$-partitions of $\calN$ with components in $\bfGamma$. 

In contrast with the Wadge degrees of sets, the structure $(\mathbf{B}(k^\calN);\leq_W)$  for  $k>2$ has antichains of any finite size. Nevertheless, a basic property of the Wadge degrees of sets may be lifted to $k$-partitions, as the following very particular case of Theorem 3.2 in \cite{ems87} (see Theorem \ref{emsbqo} in Section \ref{bqo}) shows:

\begin{theorem}\label{ems}
For any $2\leq k<\omega$, the structure
$(\mathbf{B}(k^\mathcal{N});\leq_W)$ is WQO.
\end{theorem}

Although this result gives an important information about the Wadge degrees of Borel $k$-partitions, it is  far from a  characterization.
Here we briefly discuss some steps to such a characterization made in \cite{he93,s07a,s16,s17}). The approach of \cite{s07a,s17} is to characterize the initial segments $(\bfDelta^0_\alpha(k^\mathcal{N});\leq_W)$ for bigger and bigger ordinals $2\leq\alpha<\omega_1$. In \cite{s07a}  this was done for $\alpha=2$, in \cite{s17} for $\alpha=3$ where also a way to the general characterization was sketched. This characterization uses the iterated $h$-QO from Section \ref{bqo}. The main idea is to expand the structure by suitable operations whose properties are similar to those on the labeled forests and use the similarity to prove isomorphism. Some of these operations extend the corresponding operations on sets from \cite{wad84} to $k$-partitions.

Let $\bigoplus_iA_i$ be the disjoint union of a sequence of elements $A_0,A_1,\ldots$ of $k^\mathcal{N}$. Let $\mathcal{N}^+:=\{1,2,\ldots\}^\omega$ and for $x\in\mathcal{N}^+$ let $x^-:=\lambda i.x(i)-1$, so $x^-\in\mathcal{N}$. Define the binary operation $+$ on $k^\mathcal{N}$ as follows: $(A+B)(x):=A(x^-)$ if $x\in\mathcal{N}^+$, otherwise $(A+B)(x):=B(y)$ where $y$ is the unique element of $\mathcal{N}$ such that $x=\sigma 0y$ for a unique finite sequence $\sigma$ of positive integers. For any $i<k$, define a unary operation $p_i$ on $k^\mathcal{N}$ by $p_i(A):=\mathbf{i}+A$ where $\mathbf{i}:=\lambda x.i$ are the constant $k$-partitions (which are precisely the distinct minimal elements of $(k^\mathcal{N};\leq_W)$). For any $i<k$, define a unary operation $q_i$ on $k^\mathcal{N}$ (for $k=2$, $q_0$ and $q_1$ coincide with the Wadge's operations $\sharp$ and $\flat$ from Section III.E of \cite{wad84}) as follows: $q_i(A)(x):=i$ if $x$ has infinitely many zeroes, $q_i(A)(x):=A(x^-)$ if $x$ has no zeroes, and $q_i(A)(x):=A(y^-)$ otherwise where $y$ is the unique element of $\mathcal{N}^+$ such that $x=\sigma 0y$ for a string $\sigma$ of non-negative integers. The introduced operations are correctly defined on Wadge degrees.

The result next from \cite{s17} characterizes some subalgebras of the Wadge degrees generated from the minimal degrees $\{\mathbf{0}\},\ldots,\{\mathbf{k-1}\}$. The proof uses Theorem \ref{ems}. 

\begin{theorem}\label{generate}
 The quotient-poset of $(\bfDelta^0_2(k^\mathcal{N});\leq_W)$ is generated from the degrees $\{\bf{0}\},\ldots,\{\bf{k-1}\}$ by the operations  $\bigoplus,  p_0,\ldots,p_{k-1}$.
 The quotient-poset of $(\bfDelta^0_3(k^\mathcal{N});\leq_W)$ is generated from $\{\bf{0}\},\ldots,\{\bf{k-1}\}$ by the operations $\bigoplus,+,q_0,\ldots,q_{k-1}$.

 \end{theorem}
 
The next result from \cite{s17} characterizes the structures above using Proposition \ref{dcq1}.

\begin{theorem}\label{char1}
\begin{enumerate}\itemsep-1mm
 \item The quotient-posets of $(\bfDelta^0_2(k^\mathcal{N});\leq_W)$ and of $(\widetilde{{\mathcal F}}_k;\leq_h)$ are isomorphic.
 \item The quotient-posets of $(\bfDelta^0_3(k^\mathcal{N});\leq_2)$ and of $(\widetilde{{\mathcal F}}_k;\leq_h)$ are isomorphic.
 \item The quotient-posets of $(\bfDelta^0_3(k^\mathcal{N});\leq_W)$ and of $(\widetilde{{\mathcal T}}^\sqcup_k(2);\leq_h)$ are isomorphic.
\end{enumerate}
 \end{theorem}

We describe functions that induce the isomorphisms of the quotient-posets. For (1), let $(T;c)\in\widetilde{\mathcal{T}}_k$. Associate with any node $\sigma\in T$ the $k$-partition $\mu_T(\sigma)$ by induction on the rank $rk(\sigma)$ of $\sigma$ in $(T;\sqsupseteq)$ as follows: if $rk(\sigma)=0$, i.e. $\sigma$ is a leaf of $T$ then $\mu_T(\sigma):=\bf{i}$ where $i=c(\sigma)$; otherwise, $\mu_T(\sigma):=p_i(\bigoplus\{\mu_T(\sigma n)\mid n<\omega,\sigma n\in T\})$. 
Now, define a function $\mu:\widetilde{\mathcal{T}}_k\to k^\calN$ by $\mu(T):=\mu_T(\varepsilon)$. 
Next  extend $\mu$ to $\widetilde{\mathcal{F}}_k$ by $\mu(F):=\bigoplus\{\mu_T(n)\mid n<\omega, (n)\in T\}$. Then $\mu\to \widetilde{\mathcal{F}}_k$ induces the isomorphism in (1).

The isomorphism $\nu$ in (2) is constructed just as $\mu$ but with  $q_i$ instead of $p_i$.

Towards the isomorphism in (3), let $(T;c)\in\widetilde{\mathcal{T}}_k(2)$. Relate to any node $\sigma\in T$ the $k$-partition $\rho_T(\sigma)$ by induction on the rank $rk(\sigma)$ of $\sigma$ in $(T;\sqsupseteq)$ as follows: if $rk(\sigma)=0$ then $\rho_T(\sigma):=\nu(Q)$ where $Q=c(\sigma)\in\widetilde{\mathcal{T}}_k$; otherwise, $\rho_T(\sigma):=\nu(Q)+(\bigoplus\{\rho_T(\sigma n)\mid n<\omega,\sigma n\in T\})$. 
Now  define a function $\rho:\widetilde{\mathcal{T}}_{\widetilde{\mathcal{T}}_k}\to k^\calN$ by $\rho(T):=\rho_T(\varepsilon)$. 
Finally, extend $\rho$ to $\widetilde{\mathcal{T}}^\sqcup_k(2)$ by $\rho(F):=\bigoplus\{\rho_T(n)\mid n<\omega, (n)\in T\}$ where $T:=\{\varepsilon\}\cup F$. Then $\rho$ induces the isomorphism in (3).

As conjectured in \cite{s17},
 the results above may be extended to larger segments $(\bfDelta^0_\alpha(k^\mathcal{N});\leq_W)$, $4\leq\alpha<\omega_1$. Using the Kuratowski relativization technique \cite{wad84,am03,ros09}, we can define for any $1\leq\beta<\omega_1$ the binary operation $+_\beta$ on $k^\mathcal{N}$ such that $+_1$ coincides with $+$ and, for any  $2\leq\alpha<\omega_1$, the quotient-poset of $(\bfDelta^0_\alpha(k^\mathcal{N});\leq_W)$ is generated from $\{\bf{0}\},\ldots,\{\bf{k-1}\}$ by the operations $\bigoplus$ and $+_\beta$ for all $1\leq\beta<\alpha$. The extension of Theorem \ref{char1} could probably be obtained by defining  suitable iterated versions of the $h$-quasiorder in the spirit of \cite{s16}. 
Since $\mathbf{B}(k^\mathcal{N})=\bigcup_{\alpha<\omega_1}\bfDelta^0_\alpha(k^\mathcal{N})$, we obtain the characterization of Wadge degrees of Borel $k$-partitions. Note that item (2) suggests that the extension of Theorem \ref{luca} to $k$-partitions holds.

\subsection{Wadge reducibility in quasi-Polish spaces}\label{wadpol}

A straightforward way to extend the Wadge hierarchy to non-zero-dimensional spaces would be to show that Wadge reducibility in such spaces behaves similarly to its behaviour in the Baire space, e.g. it is a semi-well-order. Unfortunately, this is not the case for many natural spaces: Wadge reducibility is often far from being WQO.

 Using the methods of \cite{wad84} it is easy to check that the structure $({\mathbf B}(X);\leq_W)$ of Wadge degrees of Borel sets in any zero-dimensional Polish space $X$ remains semi-well-ordered. 
In contrast, the structure of Wadge degrees in non-zero-dimensional spaces is typically more complicated. P.~Hertling demonstrated this in \cite{he96}  by showing that there are infinite antichains and infinite descending chains in
the structure of Wadge degrees of $\bfDelta^0_2(\RR)$-sets. This result has been strengthened in \cite{sc10} to the result that any poset of cardinality $\omega_1$ embeds into $({\mathbf B}(\RR);\leq_W)$. 

P.~Schlicht  showed  in \cite{sc11} that the structure of Wadge degrees on \emph{any} non zero-dimensional Polish space must contain infinite antichains. Thus, the class of zero-dimensional Polish spaces maybe characterized in terms of Wadge reducibility within the Polish spaces.

V.~Selivanov  showed in \cite{s05} that the structure of  Wadge degrees of finite Boolean combinations of open sets in many \( \omega \)-algebraic domains is semi-well-ordered, but already for $\bfDelta^0_2$-sets the structure contains antichains of size 4. Additional information on the structure of Wadge-degrees in non-zero-dimensional spaces maybe found e.g. in \cite{s05,ik10,sc10}.

The mentioned results show that it is not straightforward to extend Wadge hierarchy to quasi-Polish spaces using the Wadge reducibility in those spaces. We return to this question in Section \ref{opens}.

%%%%%%%%%%%%%%%%%%%%%%%%%%%%%%%%%%%%%%%%%%%%%%%%%%%%%%%
\subsection{Weak homeomorphisms between quasi-Polish spaces}

As the  Wadge reducibility in non-zero-dimensional quasi-Polish spaces is often far from being WQO, one could hope to find natural weaker notions of reducibility that induce semi-well-ordered degree structures. 
Good candidates are $\leq_\alpha$ and $\leq^W_\alpha$, but before looking on them we briefly discuss here some properties of the corresponding classes of functions. All uncredited results in this section are from \cite{mss12}. 

It is a classical result of DST that  every two uncountable Polish
spaces \( X , Y \) are Borel-isomorphic (see e.g. Theorem 15.6 in
\cite{ke95}). The next proposition extends this
result to the context of uncountable quasi-Polish spaces and
computes an upper bound for the complexity of the
Borel-isomorphism. We write \( X \simeq_\alpha Y \) to denote that there is a bijection $f:X\to Y$ such that $f\in D_\alpha(X,Y)$ and $f^{-1}\in D_\alpha(Y,X)$. The relation $\simeq^W_\alpha$ is defined in the same way.

\begin{proposition}\label{propexamples}
\begin{enumerate}
\item Let  \( X,Y \) be two uncountable quasi-Polish spaces. Then   \( X \simeq_\omega Y \).

\item Every quasi-Polish space is \( D^W_4 \)-isomorphic to an \(\omega\)-algebraic domain.
 \item \( \mathcal{N}  \simeq^W_2  \omega \sqcup
\mathcal{N}  \).
 \item If \( X \) is a \(\sigma\)-compact
quasi-Polish space then \( \mathcal{N} \not\simeq^W_2 X
\). In particular, \( \mathcal{N} \not\simeq^W_2
\mathcal{C} \), \( \mathcal{N} \not\simeq^W_2 \RR^n \)
for every \( n < \omega \), and \( \mathcal{N}
\not\simeq^W_2 \omega^{\leq \omega} \) where \( \omega^{\leq \omega} \) is the \(\omega\)-algebraic domain  \( (\omega^*\cup\omega^{\omega}, \sqsubseteq) \) endowed with the Scott topology.
 \item \(
\mathcal{N} \simeq^W_3 \mathcal{C}\simeq^W_3 \omega^{\leq \omega}\simeq^W_3 \RR^n \) for every \(1 \leq  n <
\omega \). 
 \end{enumerate}
\end{proposition}

Our next goal is to extend Proposition \ref{propexamples} (5) to a wider class of quasi-Polish spaces (see Theorem \ref{theordim}). Such generalization will involve the following definition of the (inductive) topological dimension of a space \( X \), denoted in this paper by \( \dim(X) \) --- see e.g.\ \cite[p. 24]{hurwal}.

\begin{definition}
The empty set \( \emptyset \) is the only space  with \emph{dimension \( - 1 \)}, in symbols \( \dim(\emptyset) = -1 \).

Let \( \alpha \) be an ordinal and \( \emptyset \neq X \). We say that \( X \) has \emph{dimension \( \leq \alpha \)}, \( \dim(X) \leq \alpha \) in symbols, if every \( x \in X \) has arbitrarily small neighborhoods whose boundaries have dimension \( < \alpha \), i.e.\ for every \( x \in X \) and every open set \( U \) containing \( x \) there is an open \( x \in V \subseteq U \) such that \( \dim(\partial V) \leq \beta \) (where \( \partial V = cl(V) \setminus V \)) for some \( \beta < \alpha \).

We say that a space \( X \) has \emph{dimension \( \alpha \)}, \( \dim(X) = \alpha \) in symbols, if \( \dim(X) \leq \alpha \) and \( \dim(X) \nleq \beta \) for all \( \beta < \alpha \).

Finally, we say that a space \( X \) has \emph{dimension \( \infty \)}, \( \dim(X) = \infty \) in symbols, if \( \dim(X) \nleq \alpha \) for every ordinal \( \alpha  \).
\end{definition}

It is obvious that the dimension of a space is a topological invariant (i.e.\ \( \dim(X)  = \dim(Y) \) whenever \( X \) and \( Y \) are homeomorphic). Moreover, one can easily check that \( \dim(X) \leq \alpha \) (for \(\alpha\) an ordinal) if and only if there is a base of the topology of \( X \) consisting of open sets whose boundaries have dimension \( < \alpha \). Therefore, if \( X \) is countably based and \( \dim(X) \neq \infty \), then \( \dim(X) = \alpha \) for some \emph{countable} ordinal \(\alpha\).

\begin{example}\label{exdim1}
\emph{Finite dimension}.

\begin{enumerate}\itemsep-1mm
\item \( \dim(\mathcal{N}) = \dim(\mathcal{C}) = 0 \).

\item  \( \dim(\RR^n) = n \) for every \(0 \neq  n \leq \omega \).

\item For \( n < \omega\), let \( L_n \) be the (finite) quasi-Polish space obtained by endowing  \( (n, \leq ) \) with the Scott (equivalently, the Alexandrov)  topology: then \( \dim(L_n) = n-1 \).

\end{enumerate}

\end{example}

\begin{example}\label{exdim2}
\emph{Transfinite dimension}.

\begin{enumerate}\itemsep-1mm
\item The disjoint union \( X =  \bigsqcup_{0 \neq n \in \omega} [0,1]^n \) of the \( n \)-dimensional cubes \( [0,1]^n \) is a Polish space of dimension \( \omega \).

\item \( \dim(\omega^{ \leq \omega})  = \omega \).

\item For \( \alpha < \omega_1 \), let \( L_{\alpha+1} \) be the quasi-Polish space obtained by endowing the poset \( (\alpha+1, \leq ) \) with the Scott  topology. Then  \( \dim(L_{\alpha+1}) = \alpha \).

\end{enumerate}
\end{example}

\begin{example}\label{exdim3}
\emph{Dimension \( \infty \)}.

\begin{enumerate}\itemsep-1mm
\item The Hilbert cube \( [0,1]^\omega \), the space \( \RR^\omega \) (both endowed with the product topology), and the Scott domain \( P \omega \) all have  dimension \( \infty \).

\item Let \( \mathsf{C}_\infty \) be the quasi-Polish space obtained by endowing the poset \( (\omega, \geq ) \) with the Scott (equivalently, the Alexandrov) topology. Then \( \mathsf{C}_\infty \) is a (scattered) countable space with \( \dim(\mathsf{C}_\infty) = \infty \). Hence the space \( \mathsf{UC}_\infty = \mathsf{C}_\infty \times \mathcal{N} \), endowed with the product topology, is an (uncountable) quasi-Polish space of dimension \( \infty \).

\end{enumerate}

\end{example}

We are ready to formulate the extension of Theorem \ref{propexamples}.

\begin{theorem}\label{theordim}
Let \( X \) be an uncountable quasi-Polish space.
\begin{enumerate}\itemsep-1mm
\item
If \( \dim(X) \neq \infty \) then  \( \mathcal{N} \simeq^W_3 X \);
\item
If \( \dim(X) = \infty \) and \( X \) is Polish then \( \mathcal{N} \not\simeq^W_\alpha X \) for every \( \alpha < \omega_1 \) and \( \mathcal{N} \not\simeq_n X \) for every \( n < \omega \);
\item
 \( P \omega \not\simeq^W_\alpha \mathcal{N} \) for every \( \alpha < \omega_1 \) and \( P \omega \not\simeq_n \mathcal{N} \) for every \( n < \omega \).  The same result holds when replacing \( P\omega \) with any other quasi-Polish space which is universal for (compact) Polish spaces;
\item
\( \mathsf{UC}_\infty \simeq^W_2 \mathcal{N} \). Therefore, \( \mathsf{UC}_\infty \not\simeq^W_\alpha X \) (\( \alpha < \omega_1 \)) and \( \mathsf{UC}_\infty \not\simeq_n X \) (\( n \in \omega \)) for \( X \) a Polish space of dimension \( \infty \) (e.g.\ \( X = [0,1]^\omega \) or \( X = \RR^\omega \)) or \( X = P \omega \).
\end{enumerate}
\end{theorem}

\subsection{Weak reducibilities in quasi-Polish spaces}

Here we show that most of the reducibilities $\leq_\alpha$, $\leq_\alpha^W$ are in fact semi-well-orders on the Borel sets in quasi-Polish spaces.
The following result from \cite{mss12} is an immediate corollary of Theorem \ref{theordim}.

\begin{theorem}  \label{theorcoruncountable}
Let \( X \) be an uncountable quasi-Polish space.
\begin{enumerate}\itemsep-1mm
%with $\bigcup_{X,Y\in\mathscr{X}}\mathsf{D}^W_3\subseteq\mathcal{F}$ and the $
\item
If $\dim(X)=0$ and $1\leq\alpha<\omega_1$ then $({\mathbf B}(X);\leq_\alpha)$ and $({\mathbf B}(X);\leq_\alpha^W)$ are semi-well-ordered.

\item
Assume $\dim(X)=0$ and $2\leq\alpha<\omega_1$. If $X$ is $\sigma$-compact then  the quotient-posets of $({\mathbf B}(X);\leq_\alpha)$ and $({\mathbf B}(X);\leq_\alpha^W)$ are isomorphic to the quotient-poset of $({\mathbf B}(\calC);\leq_W)$, otherwise they are isomorphic to  the quotient-poset of $({\mathbf B}(\calN);\leq_W)$.

\item
If $\dim(X)\neq\infty$ and $3\leq\alpha<\omega_1$ then the quotient-posets of $({\mathbf B}(X);\leq_\alpha)$ and $({\mathbf B}(X);\leq_\alpha^W)$ are isomorphic to  the quotient-poset of $({\mathbf B}(\calN);\leq_W)$.

\item
If \( X \) is universal for Polish (respectively, quasi-Polish) spaces and $3\leq\alpha<\omega_1$ then  the quotient-posets of $({\mathbf B}(X);\leq_\alpha)$ and $({\mathbf B}(X);\leq_\alpha^W)$ are isomorphic to  the quotient-posets of $({\mathbf B}([0,1]^\omega);\leq_W)$ and of $({\mathbf B}(P\omega);\leq_W)$.

\item
If $\omega\leq\alpha<\omega_1$ then the quotient-posets of $({\mathbf B}(X);\leq_\alpha)$ and $({\mathbf B}(X);\leq_\alpha^W)$ are isomorphic to  the quotient-poset of $({\mathbf B}(\calN);\leq_W)$.
\end{enumerate}
\end{theorem}

Similar results clearly hold also for $k$-partitions which means that, for most of $\alpha$, the relations $\leq_\alpha$ and $\leq_\alpha^W$ on the Borel $k$-partitions of quasi-Polish spaces are again intimately related to the iterated $h$-QO in Section \ref{bqo}.

Theorem \ref{theorcoruncountable} leaves  open the question about the structure of degrees under $\leq_2$. The next two results from \cite{mss12} give some information on this.

\begin{theorem}\label{propR2}
\begin{enumerate}\itemsep-1mm
\item There are infinite antichains in  $(\mathbf{B}([0,1]);\leq_2)$. 

\item The quasiorder $(P(\omega),\subseteq^*)$ of inclusion modulo finite sets on $P(\omega)$ embeds into $(\bfSig^0_2(\mathbb{R}^2),\leq_2)$.
\end{enumerate} 
\end{theorem}

\section{Other  reducibilities and hierarchies}\label{redfun}

In this section we discuss some reducibilities and hierarchies  on objects more complex than sets and $k$-partitions (in particular, on equivalence relations or functions between spaces).
They provide useful tools for measuring descriptive complexity of such objects but, unfortunately, in most  interesting cases they are far from being WQOs.

\subsection{Borel reducibility of Borel equivalence relations}

In
mathematics one often deals with problems of classifying  objects up to some
notion of equivalence by invariants. Via  suitable encodings, these objects can be viewed as elements
of a standard Borel space $X$ and the equivalence turns out to be a Borel
equivalence relation $E$ on $X$. In this and the next subsection we briefly discuss (following \cite{gao09,ak00a}) some reducibilities on Borel equivalence relations which provide a mathematical framework
for measuring the complexity of such classification problems. The most fundamental is probably Borel reducibility defined as follows.

By {\em standard
Borel space} we mean a Polish (equivalently, quasi-Polish) space equipped with its Borel structure. Let $E,F$ be equivalence relations on standard Borel spaces $X, Y$, respectively. We say that
$E$ is {\em Borel reducible} to $F$ (in symbols,
$E \leq_B F$)
if there is a Borel map $f : X \to Y$ such that
$xEy$ iff $f(x)Ff(y)$. We denote by $(BER:\leq_B)$ the QO of Borel equivalence relations with the Borel reducibility.

For any standard Borel space $X$, denote by $X$ also the
equality relation on this space, and let $n$ be any such space of
finite cardinality $n$. Then we clearly have 
$1 <_B 2 <_B \cdots <_B \omega<\mathbb{R}$.
  Let $E_0$ be  the {\em Vitali equivalence relation} on $\mathbb{R}$ defined by:
$xE_0y$ iff $x - y \in\mathbb{Q}$. Then we have the following deep result which includes the so called Silver Dichotomy and General Glimm-Effros Dichotomy (due to Harrington-Kechris-Louveau):

\begin{theorem} \label{dich}
The chain $1 <_B 2 <_B \cdots <_B \omega<\mathbb{R}<_BE_0$ is an initial segment of $(BER;\leq_B).$
\end{theorem}

The QO $(BER;\leq_B)$ is very rich, in particular it has no maximal elements (by the Friedman-Stanley jump theorem). In contrast, the set $CBER$ of countable Borel equivalence relations (those with countable equivalence classes) has the greatest element $E_\infty$ which is equivalent to many natural equivalence relations. For instance, it was recently shown in \cite{ma16,mass} that the equivalence relations on the Cantor space induced by some natural reducibilities  in computability theory (e.g., polynomial-time many-one, polynomial-time Turing, and the arithmetical reducibilities) are Borel equivalent to $E_\infty$. 

According to Feldman-Moore theorem, the countable Borel equivalence relations coincide with the orbit equivalence relations $E_G$ induced by  Borel actions $G\times X\to X$ of  (discrete) countable groups $G$ where $xE_Gy$ iff $g\cdot x=y$ for some $g\in G$. This suggests that the QO $(CBER;\leq_B)$ is rich. Is it a WQO? Unfortunately, this is far from being the case, as the following deep result from \cite{ak00a} demonstrates.

\begin{theorem} \label{rich}
The poset of Borel subsets of $\calN$ under inclusion can be embedded
in the quasiorder $(CBER;\leq_B)$.
\end{theorem}

The last result shows that WQO-theory is not  directly related to the QO $(CBER;\leq_B)$. To find such a relation, one probably would have to search for natural substructures of $(CBER;\leq_B)$ which form WQOs.

\subsection{Continuous reducibility of Borel equivalence relations}\label{cer}

Although the Borel reducibility from the previous section is quite important due to its deep relation to classical mathematics, it is in a sense too coarse. For instance, it does not distinguish between Polish and  quasi-Polish spaces of the same cardinality. There are  natural finer versions, the most important of which is  continuous reducibility. Though it may be defined for arbitrary quasi-Polish spaces, we briefly discuss it here only for the Baire space.

For equivalence relations $E,F$ on
$\mathcal{N}$, $E$ is {\em continuously reducible} 
 to $F$, in symbols $E\leq_cF$, if
there is a continuous  function $f$ on
$\mathcal{N}$ such that for all  $x,y\in\mathcal{N}$, $xEy$ iff $f(x)Ff(y)$.
The QO $(ER(\mathcal{N});\leq_c)$, where $ER(\mathcal{N})$ is the set of
all equivalence relations on $\mathcal{N}$, and  its substructure $(BER(\mathcal{N});\leq_c)$ formed by Borel equivalence relations, are extremely complicated, as it follows from Theorem \ref{rich}.

Let
$(ER_k;\leq_c)$ (resp. $(BER_k;\leq_c)$) be the initial segment  of $(ER(\mathcal{N});\leq_c)$ (resp. $(BER(\mathcal{N});\leq_c)$)
formed by the set $ER_k$ of equivalence relations which have at
most $k$ equivalence classes. We relate this substructure to the
structure $(k^\mathcal{N};\leq_0)$ where  $\mu\leq_0\nu$ iff
$\mu=\varphi\circ\nu\circ f$ for some continuous function $f$ on
$\mathcal{N}$ and for some permutation $\varphi$ of
$\{0,\ldots,k-1\}$. Since $\leq_W$ implies $\leq_0$, the results of Section \ref{wpartbaire} imply that $(BER_k;\leq_c)$ is WQO. The following is straightforward:

 \begin{proposition}\label{er1}
For any $2\leq k<\omega$, the function $\nu\mapsto E_\nu$, where $pE_\nu q$ iff $\nu(p)=\nu(q)$, induces
an isomorphism between the quotient-structures of
$(k^\mathcal{N};\leq_0)$ and $(ER_k;\leq_c)$.
\end{proposition}

Let $\mathcal{B}_k$ be the subset of $k^\mathcal{N}$ formed by the $k$-patitions whose components are finite Boolean combinations of open sets. By a result in \cite{he93}, the quotient-posets of $(\mathcal{B}_k;\leq_W)$ and $(\mathcal{F}_k;\leq_h)$ are isomorphic. It is not hard to modify the proof in \cite{he93} to show that the quotient-posets of $(\mathcal{B}_k;\leq_0)$ and $(\mathcal{F}_k;\leq_0)$ are isomorphic.

\subsection{Hierarchies and  reducibilities of  functions}\label{weired}

$k$-Partitions are of course very special functions between spaces, and it is natural to search for natural hierarchies and reducibilities between functions. Some such hierarchies induced by the classical hierarchies of sets are known from the beginning of DST. 

For any family $\mathbf{\Gamma}$ of pointclasses and any quasi-Polish spaces $X,Y$, a function $f:X\to Y$ is {\em $\mathbf{\Gamma}$-measurable} if $f^{-1}(U)\in\mathbf{\Gamma}(X)$ for each open set $U\subseteq Y$. For each $1\leq\alpha<\omega_1$, let $B_\alpha(X)$ be the class of $\bfSig^0_\alpha$-measurable functions. The ascending sequence $\{B_\alpha(X)\}$, known as the {\em Baire hierarchy in $X$} exhausts all the Borel  functions and is important for DST.

Equally natural is the hierarchy $\{D_\alpha(X)\}$ formed by the classes from Section \ref{cf}. It is also ascending and exhausts the Borel functions \cite{ros09}. A pleasant property of this hierarchy is that all of its levels are closed under composition. A problem with  both hierarchies is that they are coarse, but it is not clear how to extend the classical hierarchy theory of sets (and $k$-partitions) to these hierarchies of functions (in particular, we are not aware of  natural analogues for the Hausdorff-Kuratowski theorem). Also, there is no clear relation to WQO-theory so far.

The Borel hierarchy also induces a natural QO $\leq_B$ on Borel functions which we define only for the Baire space. Associate with any Borel function $f$ on $\calN$ the monotone function $d_f$ on $(\omega_1;\leq)$ as follows: $d_f(\alpha)$ is the smallest $\beta$ with $\forall A\in\bfSig^0_\alpha(\calN)(f^{-1}(A)\in\bfSig^0_\beta(\calN))$. Now define the QO $\leq_B$ by: $f\leq_Bg$ if $\forall\alpha(d_f(\alpha)\leq d_g(\alpha))$. We do not know whether $\leq_B$ is a WQO but  we will see this at least for some reasonable sets of partial functions (note that the definition works for the Borel partial functions). 

In an attempt to measure the discontinuity of functions, K. Weihrauch \cite{w92,hw94} introduced some notions of
reducibility for functions on topological spaces. In
particular, the following three notions of reducibility between
functions $f,g:X\rightarrow Y$ on topological spaces were
introduced: $f\leq_0g$ (resp. $f\leq_1g$, $f\leq_2g$) iff $f=g\circ
H$ for some continuous function $H:X\rightarrow X$ (resp. $f=F\circ
g\circ H$ for some continuous functions $H:X\rightarrow X$ and
$F:Y\rightarrow Y$, $f(x)=F(x,g(H(x)))$ for some continuous functions
$H:X\rightarrow X$ and $F:X\times Y\rightarrow Y$). In this way we
obtain QOs $(Y^X;\leq_i)$, $i\leq2$, on the set  of all
functions from $X$ to $Y$.

There are many variations of Weihrauch reducibilities (in particular, for multi-valued functions or with computable reducing functions in place of  continuous ones), some of which turned out to be
useful for understanding the non-computability and non-continuity of
interesting decision problems in computable analysis
\cite{hw94,bg09a}  and constructive mathematics \cite{bg09}. This research is closely related to reverse mathematics where similar problems (including intersting problems about WQOs) are treated by proof-theoretic means \cite{hir}. This is now a popular research field, but its relation to WQO-theory is not clear. Nevertheless, some restricted versions of Weihrauch reducibilities are relevant.

The notions of Wehrauch reducibilities are nontrivial even for the case of discrete spaces
$Y=k=\{0,\ldots,k-1\}$ with $k<\omega$ points, which brings us back to the $k$-partitions of $X$. In this case, $\leq_0$ coincides with the Wadge reducibility of $k$-partitions already discussed above (for this reason in the next subsection we reserve the notation $\leq_0$ for the QO from the previous subsection). We again concentrate on the case $X=\calN$ because for the non-zero dimensional spaces the degree structures are often complicated.
Let $\mathcal{B}_k$ be the subset of $k^\mathcal{N}$ formed by the $k$-patitions whose components are finite Boolean combinations of open sets. In \cite{he93} P. Hertling proved the following combinatorial description of small fragments of Weihrauch degrees:

\begin{theorem} \label{her}
The quotient-posets of $(\mathcal{B}_k;\leq_i)$ and of $(\mathcal{F}_k;\leq_i)$ are isomorphic for each  $i=1,2$.
\end{theorem}

For the functions with infinite range, some interesting results about the relation $\leq_1$ on partial functions on the Baire space were obtained in \cite{ca13}. The relation $\leq_2$ has the disadvantage that it does not refine the above-defined relation $\leq_B$. Some natural easy properties of $\leq_1$ are collected in the next assertion:

\begin{proposition} \label{ca1}
\begin{enumerate}\itemsep-1mm
\item  For every $A,B\subseteq\calN$, $A\leq_WB$ implies $c_A\leq_1c_B$;  $c_A\leq_1c_B$ implies that $A\leq_WB$ or $\overline{A}\leq_WB$; and $c_A\equiv_1c_{\overline{A}}$, where $c_A(p)=1^\omega$ for $p\in A$ and $c_A(p)=0^\omega$ otherwise.
\item For every $A,B\subseteq\calN$, $id_A\leq_1id_B$ iff $A$ is a retract of $B$, where $id_A$ is the identity on $A$.
\item For every pair of Borel partial functions $f,g$ on $\calN$, $f\leq_1g$ implies $f\leq_Bg$.
\end{enumerate}
\end{proposition}

Let $C_\infty$ (resp. $C,C^*$) be the set of partial continuous functions on $\calN$ with closed domain (resp. with closed domain and countable range, resp. with compact domain).
In \cite{ca13} the notion of Cantor-Bendixon rank of a function in $C$ was defined in such a way that $CB(id_F)$ coincides with the classical Cantor-Bendixon rank $CB(F)$ of a closed set $F$. Let $C_\alpha$ be the set of functions with Cantor-Bendixon rank $\alpha$.
 Although it is open whether $\leq_1$ is WQO on $C$, the
following theorem (that collects some results in \cite{ca13}) shows that it is WQO on some subsets of $C$.

\begin{theorem} \label{ca2}
\begin{enumerate}\itemsep-1mm
\item The quotient-poset of $(C^*\leq_1)$ is a well order of rank $\omega_1$.
\item Any two functions from $C_\infty$ with uncountable range are equivalent w.r.t. $\equiv_1$.
\item Let $Q$ be a subset of $C_\infty$ such that $(Q\cap C_\alpha;\leq_1)$ is BQO for all $\alpha<\omega_1$. Then $(Q;\leq_1)$ is BQO.
\item The relation of being a retract is WQO on $\bfPi^0_1(\calN)$.
\end{enumerate}
\end{theorem}

\subsection{Definability and decidability issues}\label{defdeg}

One may ask why we are interested in characterizing the degrees of $k$-partitions and other degree structures above in terms of ``combinatorial'' objects like labeled forests. First, any such non-obvious characterization gives new information about the structures under investigation. Second, the combinatorial objects are much easier to handle and explore than the degree structures. In particular, the 
definability and decidability issues are much easier to study for the labeled forests than for the Wadge degrees. Such issues (e.g. characterizing the automorphism groups or the complexity of first-order theories) are important and principal for understanding the natural degree structures in DST, though they remain much unexplored, to our knowledge. In contrast, this topic for the natural degree structures in computability theory is a central theme.

From   the results in Sections \ref{wlike} and \ref{defwqo} we immediately obtain many corollaries on definability and decidability in the initial segments of the Wadge degrees of $k$-partitions. In particular, Theorem \ref{main3} implies the following:

\begin{theorem} \label{main31}
\begin{enumerate}
\item Let  $k\geq3$, $Q$ be any set of Borel $k$-partitions of $\calN$ that contains the set $\mathcal{B}_k$ from the previous subsection, and let $\mathbb{A}$ be the quotient-poset of $(Q;\leq_W)$. Then $FO(\mathbb{N})\leq_mFO(\mathbb{A})$.
\item Let $\mathbb{A}$ be the quotient-poset of $(\mathcal{B}_k;\leq_W)$.  Then $FO(\mathbb{N})\equiv_mFO(\mathbb{A})$ and $(\mathbb{A};{\bf{0},\ldots,\bf{k-1}})$ is biinterpretable with $\mathbb{N}$.

\end{enumerate}
\end{theorem}

In computability theory, people actively discuss several versions of
the so called biinterpretability conjecture stating that some
structures of degrees of unsolvability are biinterpretable (in
parameters) with $\mathbb{N}$ (see e.g. \cite{ni00} and
references therein). The conjecture (which seems still open for the
most important cases) is considered as in a sense the best possible
definability result about degree structures. Item (2) above solves a similar question (even without parameters) for a natural object of DST.

The results at the end of Sections \ref{cer} and \ref{defwqo}   imply the following result from \cite{s13}:

 \begin{proposition}\label{er2}
Let $k\geq3$ and let $\mathbb{A}$ be the quotient-poset of any initial segment of
$(ER(\mathcal{N});\leq_c)$ that contains the set of equivalence relations with components in $\mathcal{B}_k$. Then the first-order theory
of $\mathbb{A}$ is hereditarily undecidable.
 \end{proposition}

A similar undecidability result for the segments of Weihruch's degrees follows from the results in Section \ref{weired} and Theorem \ref{main2}. 
Another natural and important question is to study definability and decidability in the fragments of the quotient-poset of $(CBER;\leq_B)$. Unfortunately, we are not aware of any result in this direction.

\section{Hierarchies in quasi-Polish spaces}\label{opens}

In this section we discuss hierarchies of sets and $k$-partitions in quasi-Polish spaces mentioned in Section \ref{wadpol}. In particular, we provide  set-theoretic descriptions of these hierarchies, which give some new information even for the classical case of the Wadge hierarchy in Baire space.

\subsection{Hierarchies of sets}\label{hset}

Here we briefly discuss some notions of the hierarchy theory mentioned in the Introduction which provide a convenient language to discuss various concrete hierarchies.
The next definition of a hierarchy of sets was first proposed by J. Addison \cite{ad65}. 

\begin{definition}\label{d-in-hi0}
 Let $X$ be a set and $\eta$ be an ordinal.
\begin{enumerate}\itemsep-1mm

\item   An {\em
$\eta$-hierarchy of sets in $X$} is a sequence
$\{H_\alpha\}_{\alpha<\eta}$ of subsets of $P(X)$ such that
$H_\alpha\subseteq H_\beta\cap co$-$H_\beta$ for all
$\alpha<\beta<\eta$, where $co$-$H_\beta$ is the class of complements of sets in $H_\beta$.

\item  The classes $H_\alpha$ and
$co$-$H_\alpha\setminus H_\alpha$ are  {\em non-self-dual
levels of $\{H_\alpha\}$}, while the classes $(H_\alpha\cap
co$-$H_\alpha)\setminus(\bigcup_{\beta<\alpha}H_\beta\cap
co$-$H_\beta)$ are  {\em self-dual levels of $\{H_\alpha\}$}.

\item  The classes $H_\alpha\setminus co$-$H_\alpha$,
$co$-$H_\alpha\setminus H_\alpha$, and $(H_\alpha\cap
co$-$H_\alpha)\setminus(\bigcup_{\beta<\alpha}H_\beta\cap
co$-$H_\beta)$ are the {\em  components of $\{H_\alpha\}$}.

\item  $\{H_\alpha\}$ {\em does not collapse} if
$H_\alpha\not\subseteq co$-$H_\alpha$ for all $\alpha<\eta$.

\item  $\{H_\alpha\}$ is {\em non-trivial} if $H_\alpha\not\subseteq
co$-$H_\alpha$ for some $\alpha<\eta$.

\item   $\{H_\alpha\}$ {\em fits} a QO $\leq$ on $\bigcup_\alpha H_\alpha$ if every non-self-dual level is downward closed and has a largest (up to equivalence) element w.r.t. $\leq$.
\end{enumerate}

\end{definition}

Further definitions in this subsection were suggested in \cite{s92,s95}. The next one introduces some relations between hierarchies.

\begin{definition}\label{refin}
Let $\{H_\alpha\}$ and $\{G_\beta\}$ be hierarchies of sets in $X$.

\begin{enumerate}\itemsep-1mm

\item  $\{H_\alpha\}$ is  a {\em refinement of $\{G_\beta\}$ in a 
level $\beta$} if $\bigcup_{\gamma<\beta}(G_\gamma\cup
co$-$G_\gamma)\subseteq\bigcup_\alpha H_\alpha\subseteq (G_\beta\cap
co$-$G_\beta)$. Such a refinement is called exhaustive if
$\bigcup_\alpha H_\alpha= G_\beta\cap co$-$G_\beta$.

\item   $\{H_\alpha\}$ is a (global) {\em  refinement of $\{G_\beta\}$} if
for any $\beta$ there is an $\alpha$ with $H_\alpha=G_\beta$, and
$\bigcup_\alpha H_\alpha=\bigcup_\beta G_\beta$.

\item  A hierarchy is {\em discrete in a given level} if it has no
non-trivial refinements in this level. A hierarchy is (globally)
discrete if it is discrete in each level.

\item  $\{H_\alpha\}$ {\em is  an extension of $\{G_\beta\}$} if the sequence
$\{G_\beta\}$ is an initial segment of the sequence $\{H_\alpha\}$.

\item  $\{H_\alpha\}$ is {\em perfect in a  level $\beta$}
if $\bigcup_{\gamma<\beta}(H_\gamma\cup co$-$H_\gamma)= H_\beta\cap
co$-$H_\beta$.  A hierarchy is (globally) {\em perfect} if it is perfect
in all levels.
\end{enumerate}

\end{definition}

For instance, the transfinite Borel
hierarchy is an extension of the finite Borel hierarchy, the Borel
hierarchy is an exhaustive refinement of the Luzin hierarchy in
the first level (the Suslin theorem), the Hausdorff hierarchy over
any non-zero level of the Borel hierarchy is an exhaustive
refinement of the Borel hierarchy in the next level (the
Hausdorff-Kuratowski theorem), all the classical hierarchies in Baire space fit the Wadge reducibility, the Wadge hierarchy is perfect. 

Next we discuss a technical notion of a  base hierarchy (or just a base). 
By {\em $\alpha$-base  in $X$} (or just a base) we mean an $\alpha$-hierarchy of sets ${\mathcal
L}=\{{\mathcal L}_\beta\}_{\beta<\alpha}$ in $X$ such that each
level ${\mathcal L}_\beta$ is a lattice of sets containing $\emptyset, X$ as elements. The 1-base $({\mathcal L})$ is identified with
${\mathcal L}$. Note that any $(n+1)$-base $({\mathcal
L}_0,\ldots,{\mathcal L}_n)$ may be extended to the $\omega$-base
$\{{\mathcal L}_k\}_{k<\omega}$ (or even to a longer base) by
setting ${\mathcal L}_k$ to be the class of Boolean combinations of sets in ${\mathcal L}_n$ for all $k>n$. In the
sequel we deal mostly with 1-bases, $2$-bases, $\omega$-bases and
$\omega_1$-bases. A $\sigma$-base is a base every level of which is an upper $\sigma$-semilattice.

Let ${\mathcal L}$ be a 1-base.  Using the difference operators $D_\alpha$ from Section \ref{hquasi}, we can define the finitary version $\{D_n({\mathcal L})\}_{n<\omega}$ of the difference hierarchy over ${\mathcal L}$ with the usual inclusions of levels. If ${\mathcal L}$ is a $\sigma$-base, the infinitary version 
$\{{\mathcal L}_\alpha\}_{\alpha<\omega_1}$ will also have the usual properties. 
In particular, if  ${\mathcal L}$ is an $\omega$-$\sigma$-base (i.e., a $\sigma$-base which is an $\omega$-hierarchy), we have the inclusion $\bigcup_{\alpha<\omega_1}D_\alpha({\mathcal L}_n)\subseteq{\mathcal L}_{n+1}\cap co$-${\mathcal L}_{n+1}$ for each $n<\omega$. 

Thus, we have a natural  refinement of any $\omega$-base ${\mathcal L}$ in any non-zero level. Take for simplicity only the finitary refinements. We can continue the refinement process by adjoining new refinements in any non-discrete level obtained so far. This refinement process, studied in detail in \cite{s95}, ends (after collecting all the resulting levels together) with the {\em fine hierarchy over ${\mathcal L}$} of length $\varepsilon_0$. This hierarchy is a finitary abstract version of the Wadge hierarchy. Several such hierarchies over concrete bases have interesting properties \cite{s95,s08}. We do not give all details here because in the next subsection we discuss a more general case. The properties of such hierarchies strongly depend on some structural properties of classes of sets ${\mathcal A}\subseteq P(X)$ (see e.g.
\cite{ke95}). 

\begin{definition}\label{d-in-po3}
\begin{enumerate}\itemsep-1mm

\item  The class ${\mathcal A}$ has the {\em separation property}  if for
every two disjoint sets $A,B\in{\mathcal A}$ there is a set
$C\in{\mathcal A}\cap co$-${\mathcal A}$ with $A\subseteq
C\subseteq\overline{B}$. 

\item  The class ${\mathcal A}$ has the {\em reduction property} i.e. for
all $C_0, C_1\in{\mathcal A}$ there are disjoint
$C^\prime_0,C^\prime_1\in{\mathcal A}$ such that
$C^\prime_i\subseteq C_i$ for both $i<2$ and $C_0\cup
C_1=C^\prime_0\cup C^\prime_1$. The pair $(C^\prime_0,C^\prime_1)$
is called a reduct for the pair $(C_0,C_1)$.

\item  The class ${\mathcal A}$ has the {\em $\sigma$-reduction property}
if for each countable sequence $C_0, C_1,\ldots$ in ${\mathcal A}$
there is a countable sequence $C^\prime_0, C^\prime_1,\ldots$ in
${\mathcal A}$ (called a reduct of $C_0, C_1,\ldots$) such that
$C^\prime_i\cap C^\prime_j=\emptyset$ for all $i\not=j$ and
$\bigcup_{i<\omega}C^\prime_i=\bigcup_{i<\omega}C_i$.

\end{enumerate}

\end{definition}

It is well-known that if ${\mathcal A}$ has the
reduction property then the dual class $co$-${\mathcal A}$ has the
separation property, but not vice versa, and that
if ${\mathcal A}$ has the $\sigma$-reduction property then
${\mathcal A}$ has the reduction property but not vice versa.
Nevertheless, if ${\mathcal A}$ has the reduction property then for
any finite sequence $(C_0,\ldots,C_n)$ in ${\mathcal C}$ there is a
reduct $C^\prime_0,\ldots,C^\prime_n\in{\mathcal C}$ for
$(C_0,\ldots,C_n)$ which is defined similarly to the countable
reduct above.

The next properties of an $\omega$-base ${\mathcal L}$ imply good properties of the fine hierarchy over ${\mathcal L}$ (and if ${\mathcal L}$ is a $\omega$-base then also of the infinitary version of the fine hierarchy).

\begin{definition}\label{baseprop}
Let ${\mathcal L}$ be an $\omega$-base (resp.  $\sigma$-base) in $X$.
\begin{enumerate}\itemsep-1mm

\item  ${\mathcal L}$ is {\em  reducible} (resp. {\em $\sigma$-reducible}) if for every $n$ the level ${\mathcal L}_n$ has the
reduction (resp. $\sigma$-reduction) property.

\item  ${\mathcal L}$ is {\em  interpolable} (resp. {\em $\sigma$-interpolable}) if for each $n<\omega$  the class $co$-${\mathcal L}_{n+1}$ has the
separation property and ${\mathcal
L}_{n+1}\cap$co-${\mathcal L}_{n+1}$ coincides with the class of Boolean combinations of sets in ${\mathcal L}_n$ (resp. with $\bigcup_{\alpha<\omega_1}D_\alpha({\mathcal L}_n)$).

\end{enumerate}
\end{definition}

\subsection{Hierarchies of $k$-partitions}\label{hpart}

Here we extend hierarchies of sets from the previous subsection to hierarchies of $k$-partitions. 

The discussion in Section \ref{wlike} suggests that the levels of hierarchies of $k$-partitions should be ordered under inclusion in a more complicated way than for the hierarchies of sets. Accordingly, it is not natural to use ordinals to notate the levels of such hierarchies, as in Definition \ref{d-in-hi0}. In \cite{s12} we looked at the most general case of hierarchies whose levels are notated by an {\em arbitrary} poset $P$, let us call them $P$-hierarchies. Simple considerations show that only in the case where $P$ is WPO do we have reasonable behaviour in the components of a $P$-hierarchy (they should at least partition the sets covered by the $P$-hierarchy). Accordingly, we stick to hierarchies named by WQOs. For such hierarchies, reasonable extensions of Definitions \ref{d-in-hi0} and \ref{refin} were suggested in \cite{s12,s16}. 

Our experience with classifying $k$-partitions suggests to use the QO $\mathcal{T}_k(\omega)$ and its initial segments as notation system for the finitary versions of hierarchies of $k$-partitions, and the QO $\widetilde{\mathcal{T}}_k(\omega_1)$ and its initial segments as notation system for the infinitary versions.

First we consider the infinitary version of the difference hierarchy (DH) of $k$-partitions (the finitary version is obtained by sticking to the finite trees). The notation system for this hierarchy is $\widetilde{\mathcal{T}}_k$ (for the finitary version --- $\mathcal{T}_k$). Let ${\mathcal L}$ be a 1-base in $X$ which is a $\sigma$-base. Let $(T;c)\in\widetilde{\mathcal{T}}_k$, so $T\subseteq\omega^*$ is a well-founded tree and $c:T\to k$.

We
say that a $k$-partition $A:X\rightarrow k$ is {\em defined by a $T$-family
$\{B_\tau\}_{\tau\in T}$ of $\mathcal{L}$-sets} if $A_i=\bigcup_{\tau\in
T_i}\tilde{B}_\tau$ for each $i<k$ where
$\tilde{B}_\tau=B_\tau\setminus\bigcup\{B_\sigma\mid\sigma\sqsupset\tau\}$ and $T_i=c^{-1}(i)$. Note that we automatically have $\bigcup_\tau B_\tau=X$.  This definition is especially clear for the case when the family $\{B_\tau\}_{\tau\in T}$ is reduced (i.e., $B_{\tau i}\cap B_{\tau j}=\emptyset$ for all distinct $i,j$ with $\tau i,\tau j\in T$) because then $\{\tilde{B}_\tau\}_{\tau\in T}$ is a partition of $X$. Note that any reduced family with $\bigcup_\tau B_\tau=X$ defines a $k$-partition but this fails for the general families.

By the {\em difference hierarchy of $k$-partitions over ${\mathcal L}$} we mean the family $\{{\mathcal L}(T)\}_{T\in\widetilde{\mathcal{T}}_k}$ were ${\mathcal L}(T)$ is the set of $k$-partitions defined by
$T$-families of
$\mathcal{L}$-sets. We
mention some  properties from \cite{s16} (see also \cite{s12} where the
finitary version is considered).

\begin{proposition}\label{treedh1}
 \begin{enumerate}\itemsep-1mm
 \item  If ${\mathcal L}$ has the $\sigma$-reduction property then any level ${\mathcal L}(T)$ of the DH coincides with the set of $k$-partitions defined by the reduced $T$-families of $\mathcal{L}$-sets.
 \item  If $T\leq_hS$ then ${\mathcal
L}(T)\subseteq{\mathcal L}(S)$.
 \item Let $f$ be a function on $X$ such that $f^{-1}(A)\in\mathcal{L}$
for each $A\in\mathcal{L}$. Then $A\in{\mathcal L}(T)$ implies
$A\circ f\in{\mathcal L}(T)$.
 \item The DH of $2$-partitions over ${\mathcal L}$ coincides with the DH of sets over ${\mathcal L}$, in particular, $\{{\mathcal L}(T)\mid T\in\widetilde{\mathcal{T}}_2\}=\{D_\alpha({\mathcal L}),co$-$D_\alpha({\mathcal L})\mid \alpha<\omega_1\}$.
 \end{enumerate}
  \end{proposition}
  
If ${\mathcal L}$ is an $\omega$-base in $X$ which is a $\sigma$-base, the considerations above provide some basic information on the DHs of $k$-partitions over any level of ${\mathcal L}$. Obviously, for all $n$ and $T\in\widetilde{\mathcal{T}}_k$ the level ${\mathcal L}(T)$ is contained in the set of $k$-partitions with the components in ${\mathcal L}_{n+1}\cap co-{\mathcal L}_{n+1}$.

Now we extend these DHs of $k$-partitions to the fine hierarchy (FH) of $k$-partitions. Its levels are notated by the iterated $h$-QOs. In \cite{s16} we used the initial segment $\widetilde{\mathcal{T}}_k(\omega)$ as the notation system (for the finitary case $\mathcal{T}_k(\omega)$ was used in \cite{s12}). Simplifying notation, we stick to the initial segment $\widetilde{\mathcal{T}}_k(2)$, this also means sticking to the 2-base $({\mathcal L}_0,{\mathcal L}_1)$.  

Let $(T;c)\in\widetilde{\mathcal{T}}_k(2)$, so $T\subseteq\omega^*$ is a well-founded tree and $c:T\to \widetilde{\mathcal{T}}_k$, hence for any $\tau\in T$ we have a tree $(S^\tau;c^\tau)=c(T)$ with $c^\tau:S^\tau\to k$.
We
say that a $k$-partition $A:X\rightarrow k$ is {\em defined by families
$\{B_\tau\}_{\tau\in T}$ and $\{C_\sigma^\tau\}_{\tau\in T,\sigma\in S^\tau}$ of $\mathcal{L}$-sets} if $B_\tau\in\mathcal{L}_0,C_\sigma^\tau\in\mathcal{L}_1$, and $A_i=\bigcup\{\tilde{B}_\tau\cap \tilde{C}_\sigma^\tau\mid\tau\in
T,\sigma\in S^\tau_i\}$ for each $i<k$. Again, the definition is much clearer for the case when the families $\{B_\tau\}$ and $\{C_\sigma^\tau\}$ are reduced (for the second family this means $C^\tau_{\sigma i}\cap C^\tau_{\sigma j}=\emptyset$ for all distinct $i,j$ with $\sigma i,\sigma j\in S^\tau$, for each $\tau\in T$) because then $\{\tilde{B}_\tau\}_{\tau\in T}$ is a partition of $X$ and $\{\tilde{B}_\tau\cap \tilde{C}_\sigma^\tau\}_{\sigma\in S^\tau}$ is a partition of $\tilde{B}_\tau$ for each $\tau\in T$. Note that any such reduced family with $\bigcup_\tau B_\tau=X$ defines a $k$-partition but this fails for the general families.

By the {\em fine hierarchy of $k$-partitions over $({\mathcal L}_0,{\mathcal L}_1)$} we mean the family $\{{\mathcal L}(T)\}_{T\in\widetilde{\mathcal{T}}_k(2)}$ were ${\mathcal L}(T)$ is the set of $k$-partitions defined by
$T$-families of
$\mathcal{L}$-sets. We
mention some  properties from \cite{s16} (see also \cite{s12} where the
finitary version is considered). Note that these definitions show that the FH of $k$-partitions is in a sense an iterated version of the DHs.

\begin{proposition}\label{treedh1}
 \begin{enumerate}\itemsep-1mm
 \item  If ${\mathcal L}$ is $\sigma$-reducible then any level ${\mathcal L}(T)$ of the FH coincides with the set of $k$-partitions defined by the reduced $T$-families of $\mathcal{L}$-sets.
 \item  If $T\leq_hS$ then ${\mathcal
L}(T)\subseteq{\mathcal L}(S)$.
 \item Let $f$ be a function on $X$ such that $f^{-1}(A)\in\mathcal{L}$
for each $A\in\mathcal{L}$. Then $A\in{\mathcal L}(T)$ implies
$A\circ f\in{\mathcal L}(T)$.
 \item The FH of $2$-partitions over ${\mathcal L}$ coincides with the FH of sets over ${\mathcal L}$, in particular, $\{{\mathcal L}(T)\mid T\in\widetilde{\mathcal{T}}_2(2)\}$ coincides with the set of levels $\alpha<\omega_1^{\omega_1}$ of the FH of sets.
 \end{enumerate}
  \end{proposition}

\subsection{Hierarchies of sets in quasi-Polish spaces}\label{dhsets}

As we noticed in Section \ref{wadpol} it is not straightforward to extend Wadge hierarchy to quasi-Polish spaces using the Wadge reducibility in those spaces.

A natural way to do this is to apply the refinement process explained in Section \ref{hset} to the  $\omega_1$-base $\mathcal{L}_X=\{\bfSig^0_{1+\alpha}(X)\}$ in arbitrary quasi-Polish space $X$. Note that this base is interpolable. It is $\sigma$-reducible when $X$ is zero-dimensional but it is not reducible in general because the level $\bfSig^0_1(X)$ often does not have the reduction property (though the higher levels always have the $\sigma$-reduction property \cite{s13}).

At the first step of the process we obtain the classical Hausdorff hierarchies over each level. Further refinements may be done by the construction from the end of the previous subsection for $k=2$. In \cite{s16} this was done for the initial segment $T\in\widetilde{\mathcal{T}}_2(\omega)$ which is semi-linear-ordered with order type $\lambda=sup\{\omega_1,\omega_1^{\omega_1},\omega_1^{(\omega_1^{\omega_1})},\ldots\}$. By choosing the trees $T$ with the root label $0$, we obtain the increasing sequence of pointclasses $\{\bfSig_\alpha(X)\}_{\alpha<\lambda}$ which is a good candidate to be the (initial segment of)  Wadge hierarchy in $X$. 

From results in \cite{wad84,lo83} it is not hard to deduce that for $X=\calN$ these classes coincide with the corresponding levels of Wadge hierarchy in Section \ref{wadgedst}. This gives an alternative (to those in \cite{wad84,lo83}) set-theoretical  characterization of these levels. There is no doubt that  this definition maybe  extended to $T\in\widetilde{\mathcal{T}}_2(\omega_1)$ (yielding a set-theoretical  characterization of the levels of Wadge hierarchy within $\bf\Delta^0_\omega$) and to the higher levels.  

Some nice properties of the introduced classes may be obtained  using the admissible representations from the end of Section \ref{tspaces}. Let $\delta$ be a total admissible representation of the quasi-Polish space $X$. According to the  results in \cite{br,sr07}, $A\in\bfSig^{-1,\theta}_\alpha(X)$ iff $\delta^{-1}(A)\in\bfSig^{-1,\theta}_\alpha(\calN)$ for all $\alpha,\theta<\omega_1$, $\theta\geq1$ (in particular, $A\in\bfSig^0_\alpha(X)$ iff $\delta^{-1}(A)\in\bfSig^0_\alpha(\calN)$ for all $1\leq\alpha<\omega_1$). In \cite{s16} this was extended to the  fact that $A\in\bfSig_\alpha(X)$ iff $\delta^{-1}(A)\in\bfSig_\alpha(\calN)$, for all $\alpha<\lambda$. We do believe that this extends to the whole Wadge hierarchy.  

As suggested independently in \cite{s16,pe15}, one can also {\em define} the Wadge hierarchy $\{\bfSig_\alpha(X)\}_{\alpha<\nu}$ in $X$ by $\bfSig_\alpha(X)=\{A\subseteq X\mid \delta^{-1}(A)\in\bfSig_\alpha(\calN)\}$. One easily checks that the definition of $\bfSig_\alpha(X)$ does not depend on the choice of $\delta$, $\bigcup_{\alpha<\nu}\bfSig_\alpha(X)=\bf{B}(X)$,   ${\bfSig}_\alpha(X)\subseteq{\bfDelta}_\beta(X)$ for all $\alpha<\beta<\nu$, and any $\bfSig_\alpha(X)$ is downward closed under the Wadge reducibility on $X$. This definition is short but gives no real understanding of how the levels look like. Hence, also in this approach the set-theoretic characterization of levels is of principal interest.

\subsection{Hierarchies of $k$-partitions in quasi-Polish spaces}\label{kpart}

Here we extend the hierarchies of the previous subsection to $k$-partitions. Applying the general definitions of Section \ref{hpart} to the $\omega_1$-base $\mathcal{L}_X=\{\bfSig^0_{1+\alpha}(X)\}$ in a quasi-Polish space $X$ we obain the DHs of $k$-partitions $\{{\mathcal L}_\alpha(X,T)\}_{T\in\widetilde{\mathcal{T}}_k}$, $\alpha<\omega_1$, and the FH of $k$-partitions $\{{\mathcal L}(X,T)\}_{T\in\widetilde{\mathcal{T}}_k(2)}$.

As follows from \cite{s16}, the results from the previous subsection about the FH extend to $k$-partitions in the following sense: $A\in\mathcal{L}(X,T)$ iff $A\circ\delta\in\mathcal{L}(\calN,T)$, for each $T\in\widetilde{\mathcal{T}}_k(2)$. Similarly, for the DHs we have: $A\in {\mathcal L}_\alpha(X,T)$ iff $A\circ\delta\in {\mathcal L}_\alpha(\calN,T)$, for all $A\subseteq X$, $\alpha<\omega_1$, and $T\in\widetilde{\mathcal{T}}_k$. 

The FH of $k$-partitions $\{{\mathcal L}(\calN,T)\}_{T\in\widetilde{\mathcal{T}}_k(2)}$ is related to the Wadge reducibilty of $k$-partitions as follows. 
As we know from Section \ref{wpartbaire}, there is a function $\mu:\widetilde{{\mathcal T}}_k(2)\to{\mathbf \Delta}^0_3(k^\mathcal{N})$ 
inducing an isomorphism between the quotient-posets of the $\sigma$-join-irreducible elements in $(\bfDelta^0_3(k^\mathcal{N});\leq_W)$ and of 
$(\widetilde{{\mathcal T}}_k(2);\leq_h)$. 
For any $T\in\widetilde{{\mathcal T}}_k(2)$, we have $\mathcal{L}(\calN,T)=\{A\in\bfDelta^0_3(k^\mathcal{N})\mid A\leq_W\mu(F)\}$.

\section{Hierarchies in computability theory}\label{comput}

In this section we briefly discuss some hierarchies and reducibilities in
computability theory. They are
important because they provide  tools for classifying
many interesting decision problems in
logic and theoretical computer science. The idea to use reducibilities as a classification tool first appeared in computability theory and later it was borrowed by many other fields.

\subsection{Preliminaries}\label{prelcompu}

We assume that the reader is familiar with the main notions of
computability theory and simply recall some notation and not broadly
known definitions. For more details the reader may use any of the
many available books on the subject, e.g.
\cite{ro67,so87}.

If not specified otherwise, all functions are assumed in this
section to be functions on $\omega$, and all sets to be subsets of
$\omega$.  Thus, for an $n$-ary partial function $\phi$, we have
$dom(\phi)\subseteq\omega^n$ and $rng(\phi)\subseteq\omega$. Instead
of $(x_1,\ldots,x_n)\in dom(\phi)$ ($(x_1,\ldots,x_n)\not\in
dom(\phi)$) we sometimes write $\phi(x_1,\ldots,x_n)\downarrow$
(respectively, $\phi(x_1,\ldots,x_n)\uparrow$). We assume the reader
to be familiar with the computable partial  (c.p.) functions,
computable (total) functions  and computably enumerable (c.e.) sets.
For any $n>1$, there is a computable bijection $\lambda
x_1,\ldots,x_n.\langle x_1,\ldots,x_n\rangle$ (the Cantor coding
function) between $\omega^n$ and $\omega$. This fact reduces many
considerations to the case of unary functions and predicates.

By a {\em  numbering} we mean any
function $\nu$ with $dom(\nu)=\omega$, and by {\em  numbering of a
set $S$}
--- any numbering $\nu$ with $rng(\nu)=S$.  A numbering $\mu$ is {\em reducible} to a
numbering $\nu$ (in symbols $\mu\leq\nu$), if $\mu=\nu\circ f$ for
some computable function $f$, and $\mu$ is {\em equivalent} to
$\nu$ ($\mu\equiv\nu$), if $\mu\leq\nu$ and $\nu\leq\mu$. Relate to
any numberings $\mu,\nu$ and to any sequence of numberings
$\{\nu_k\}_{k<\omega}$ the numberings $\mu\oplus\nu$,  and
$\bigoplus_k\nu_k$, called respectively the {\em  join of $\mu$ and
$\nu$}  and the {\em  infinite join of $\nu_k (k<\omega)$}  defined as
follows:
 $$
(\mu\oplus\nu)(2n)=\mu n,\;(\mu\oplus\nu)(2n+1)=\nu n,\;
(\bigoplus_k\nu_k)\langle x,y\rangle=\nu_x(y).
 $$
 
Let $\{\varphi_n\}$ be the standard numbering of (unary)
c.p. functions on $\omega$. We assume  the reader to be acquainted with the computations
relative to a given set $A\subseteq\omega$ or a function
$\xi\in\omega^\omega$ (which in this situation are often called {\em
oracles}). E.g., such computations may be formally defined using
Turing machines with oracles. Enumerating all programs for such
machines we obtain numberings $\varphi^A$ ($\varphi^\xi$) of all
partial functions computable in $A$ (in $\xi$). 

A numbering $\nu:\omega\to S$ is {\em complete w.r.t. $a\in S$} if for every c.p. function $\psi$ on $\omega$ there is a total computable function $g$ such that $\nu(g(x))=\nu(\psi(x))$ in case $\psi(x)\downarrow$ and $\nu(g(x))=a$ otherwise.
For any set $S$ and any $a\in S$, define a
unary operation $p_a$ on $S^\omega$ as follows:  $[p_a(\nu)]n=a$ for
$\upsilon(n)\uparrow$ and $[p_a(\nu)]n=\nu\upsilon(n)$ for
$\upsilon(n)\downarrow$, where  $\upsilon$ is the universal p.c.
function $\upsilon(\langle n,x\rangle)=\varphi_n(x)$. These {\em
completion operations} were introduced in \cite{s82} as a variant of
similar operations from \cite{er68}. They are very relevant to fine
hierarchies, as the following particular case of results in
\cite{s82} demonstrates. The notions related to completeness are relativized to a given oracle $A$ in the obvious way.

\begin{theorem}\label{completeness}
 For every $2\leq k<\omega$, $(k^\omega;\leq,\oplus,p_0,\ldots,p_{k-1})$ is a
$dc$-semilattice.
 \end{theorem}

In case $k=2$ the QO $(k^\omega;\leq)$ coincides with the QO $(P(\omega);\leq_m)$, where $\leq_m$ is the $m$-reducibility, which is a popular structure of computability theory. Another important QO on $P(\omega)$ is the Turing reducibility.
Recall that $A$ is {\em Turing-reducible} ($T$-reducible) to
$B$ (in symbols, $A\leq_T B$) if $A$ is computable in $B$, i.e.
$A=\varphi^B_n$ for some $n$. The {\em Turing jump
operator} $A\mapsto A^\prime$ on $P(\omega)$ is defined by
$A^\prime=\{n\mid \varphi_n^A(n)\downarrow \}$. For any $n<\omega$,
define the $n$-th jump $A^{(n)}$ of $A$ by $A^{(0)}=A$ and
$A^{(n+1)}=(A^{(n)})^\prime$.

The arithmetical hierarchy $\mathcal{L}=\{\Sigma^0_{n+1}\}_{n<\omega}$ is an $\omega$-base in $\omega$. It is reducible but not interpolable, it does not collapse and fits the $m$-reducibility. Moreover, $\emptyset^{(n+1)}$ is $m$-complete in $\Sigma^0_{n+1}$ for each $n<\omega$ (see \cite{ro67}). In the next two subsections we consider the refining process for this base, concentrating for simplicity on the finitary hierarchies. Let $\leq^A$ (resp. $\leq^n$) denote the reducibility of numberings by functions computable in $A$ (resp. in $\emptyset^{(n)}$).

\subsection{Difference hierarchies of sets and $k$-partitions}\label{dhcompu}

Let $\{\Sigma^{-1,n}_m\}_{m<\omega}$ be  the DH over $\Sigma^0_{n+1}$. These hierarchies and their transfinite extensions over the Kleene ordinal notation system where thoroughly investigated in  \cite{er68,s83}. We recall here only their characterizations in terms of suitable jump operators. Since we will consider several such operators, we give a general notion \cite{s92}. 

By a {\em jump operator} we mean a unary operation
$J$ on $P(\omega)$ such that $A\oplus \overline{A}\leq_m J(A)$ and
$J(A)$ is a complete numbering w.r.t. $0$ uniformly in  $A$ (uniformity in, say, second condition means the existence of a computable sequence $\{g_e\}$ of total computable functions such that, for all $A,e,x$ we have: $\nu(g_e(x))=\nu(\varphi^A_e(x))$ in case $\varphi_e(x)\downarrow$ and $\nu(g(x))=0$ otherwise, where $\nu=J(A):\omega\to\{0,1\}$). From the
properties of complete numberings it follows that actually we
have $A\oplus \overline{A}<_m J(A)$. It is clear that for any set
$A$ complete w.r.t. 0 the sequence $\{J^{n}(A)\}$ of iterates  of $J$ starting from the
set $A$ is strictly increasing w.r.t. the $m$-reducibility, and the corresponding principal ideals form an $\omega$-hierarchy denoted as $(J,A)$. 

As mentioned above, if $TJ$ is the Turing jump then
$(TJ,\emptyset)$ is the arithmetical hierarchy. As observed in
\cite{er68}, the operation
$mJ(A)=\upsilon^{-1}(A\oplus\overline{A})$, where $\upsilon\langle
n,x\rangle=\varphi_n(x)$ is the universal c.p. function, is  a jump
operator  called the {\em m-jump}. By \cite{er68}, $(mJ,\emptyset)$
coincides with the Ershov's hierarchy $\{\Sigma^{-1,0}_m\}_{m<\omega}$. Similarly, if we take in place
of $mJ$ the $m$-jump relativized to $\emptyset^{(n)}$, for each
$n<\omega$, we obtain the  DH $\{\Sigma^{-1,n}_m\}_{m<\omega}$ over $\Sigma^0_{n+1}$
\cite{s83}.

Now consider the DH $\{\Sigma^0_{n+1}(T)\}_{T\in\mathcal{T}_k}$ of $k$-partitions over $\Sigma^0_{n+1}$. This hierarchy can also be characterized in terms of natural operations on $k^\omega$. Namely, let $p^n_i$, $i<k$, be the relativization of the operation $p_i$ from the end of the previous subsection to $\emptyset^{(n)}$. Define the  function $\mu:\mathcal{T}_k\to k^\omega$ just as in Section \ref{wpartbaire}, only for the finite forests and the finitary joins of $k$-partitions. Let $\mu^n$ be defined similarly but with $p^n_i$ instead of $p_i$. Then from (the relativization of) Theorem \ref{completeness} we obtain:

\begin{theorem}\label{embegp}
 For all $2\leq k<\omega$ and $n<\omega$, the function $\mu^n$ induces an embedding of the quotient-poset of $(\mathcal{F}_k;\leq_h)$ into that of $(k^\omega;\leq^n)$ (as well of their functional expansions to the signature $\{\oplus,p_0,\ldots,p_{k-1}\}$). Moreover, $\Sigma^0_{n+1}(T)=\{A\mid A\leq\mu^n(T)\}=\{A\mid A\leq^n\mu^n(T)\}$ for each $T\in\mathcal{T}_k$.
 \end{theorem}

\subsection{Fine  hierarchies of sets and $k$-partitions}\label{finecompu}

By the preceding subsection, the  DHs over $\mathcal{L}=\{\Sigma^0_{n+1}\}_{n<\omega}$ may be characterized in
terms of suitable jump operations. Is there a similar
characterization for the FHs? The answer is positive, and
actually the FH of sets was first discovered in \cite{s83} in this
way. 

Since the jump-characterization is non-trivial and yields
additional information on the FH, we provide some
details. Which jump operations to use? Of course, at least   the
$m$-jumps $J^n_m$ relativized to $\emptyset^{(n)}$, for all
$n<\omega$. By the preceding subsection, $(J^n_m,\emptyset)$ is the
difference hierarchy over $\Sigma^0_{n+1}$. A wider class of
$\omega$-hierarchies is constructed by considering the sets generated
from the empty set by all the operations $J^n_m(n<\omega)$, see
\cite{s83}. It is not hard to check that in this way we obtain a
non-collapsing hierarchy with order type $\omega^\omega$. This
already shows that these jump operations do not yield the whole FH but only its small fragment. 

In order to find a sufficient class of jump operations, we defined in \cite{s83} an
operation $r:S^\omega\times S^\omega\times k^\omega\rightarrow
S^\omega$ (where $S$ is a set and $2\leq k<\omega$) that
includes the jump operations  above, the Turing jump and many
others. We set $r(\mu,\nu,f)=\bigoplus_np^f_{\nu(n)}(\mu)$. Then
$r(\mu,\lambda x.a,f)\equiv p^f_a(\mu)$  for all $a\in S$, hence $r$
generalizes the operations of completion from Subsection
\ref{prelcompu}. Note that for $S=k=2$ the operation $r$ is a
ternary operation on sets satisfying $r(\omega,\emptyset,A)\equiv
A^\prime$, hence $r$ generalizes also the Turing jump. It induces
also several other jump operators. Namely, for any sets $B$ and $C$,
if $B$ is complete w.r.t. 0 then $A\mapsto r(A\oplus\overline{A},
B,C)$ is a jump operator. This follows from  the definition and the
property that if $\nu$ is $f$-complete w.r.t. $a$ then so is also
the numbering $r(\mu,\nu,f)$. The last property together with other
properties of $r$ generalizing the properties of the completion
operations were established in \cite{s83}. These properties play a
central role in the algebraic proof of the result below that
classifies elements of the subalgebra generated by the operations
$r,\;\bar{}$ and $\oplus$ from $\emptyset$ within $2^\omega$. As a
corollary, we obtain the jump-characterization of the FH of sets over $\mathcal{L}$. For details and much of additional information see e.g. \cite{s95,s12}.

We conclude this subsection with the brief discussion of the  FH $\{{\mathcal L}(T)\}_{T\in\mathcal{T}_k(\omega)}$ of $k$-partitions over $\mathcal{L}$. We define by induction on $m$ the sequence $\{\rho^n_m\}_n$ of functions $\rho^n_m:T\in\mathcal{T}_k(m+1)\to k^\omega$ as follows. Let $\rho^n_0:=\mu^n$ (where $\mu$ is defined before Theorem \ref{embegp}) and suppose by induction that $\rho^n_m$, $n\geq0$, are defined.

Let $(T;c)\in\mathcal{T}_k(m+2)$, so $c:T\to\mathcal{T}_k(m+1)$. Relate to any node $\sigma\in T$ the $k$-partition $\rho_T(\sigma)$ by induction on the rank $rk(\sigma)$ of $\sigma$ in $(T;\sqsupseteq)$ as follows: if $rk(\sigma)=0$ then $\rho_T(\sigma):=\rho^{n+1}_m(Q)$ where $Q=c(\sigma)\in\mathcal{T}_k(m+1)$; otherwise, $\rho_T(\sigma):=r(\bigoplus\{\rho_T(\sigma n)\mid n<\omega,\sigma n\in T\},\rho^{n+1}_m(Q), \emptyset^{(n)})$. 
We set $\rho^n_{m+1}(T):=\rho_T(\varepsilon)$. 

From the properties of $r$ in \cite{s83,s95} and Proposition \ref{dcq} it follows by induction on $m$ that $\rho^n_m$ induces an embedding of the quotient-poset of $(\mathcal{T}_k(m+1);\leq_h)$ into that of $(k^\omega;\leq^n)$, and $\rho^n_{m+1}$ extends $\rho^n_m$ modulo $\equiv^n$, Therefore, $\rho^n:=\bigcup_m\rho^n_m$ induces an embedding of the quotient-poset of $(\mathcal{T}_k(\omega);\leq_h)$ into that of $(k^\omega;\leq^n)$. 
Set $\rho:=\rho^0$ and extend $\rho$ to $\mathcal{T}^\sqcup_k(\omega)$ by $\rho(F):=\bigoplus\{\rho_T(n)\mid n<\omega, (n)\in T\}$ where $T:=\{\varepsilon\}\cup F$.

\begin{theorem}\label{embegp1}
 For each $2\leq k<\omega$, the function $\rho$ induces an embedding of the quotient-poset of $(\mathcal{T}^\sqcup_k(\omega);\leq_h)$ into that of $(k^\omega;\leq)$ (as well as of their  functional expansions). Moreover, $\mathcal{L}(T)=\{A\mid A\leq\rho(T)\}$ for each $T\in\mathcal{T}_k(\omega)$.
 \end{theorem}

\subsection{Natural degrees}\label{natural}

As is well known, the degree structures in computability theory are extremely rich and complicated, including the structures of many-one and Turing degrees. Some of these degrees (e.g., those obtained by iterating the Turing or the $m$-jumps starting from the empty set) are ``natural'' in the sense that they are equivalent to a lot of sets appearing in mathematical practice (outside the computability theory). It turns out that in fact only a small number of degrees are ``natural'' in this sense (e.g., no ``natural'' non-computable set strictly below $\emptyset^\prime$ under Turing reducibility is known).

A main idea of \cite{s83} was to find in the rich structure of the $m$-degrees of (hyper-)arithmetical sets a natural easy substructure that contains $m$-degrees of all sets which appear naturally in mathematics. In this search we tried  to expand the structure of $m$-degrees with natural jump operations and then look at the degrees generated from the empty set, as explained above. The result was the discovery of the FH of this section which was later characterized set-theoretically \cite{s92,s95} as the abstract finitary version of the Wadge hierarchy.
Moreover, it was described \cite{s92} how to obtain the Wadge hierarchy from the FH using the uniform relativization and taking the ``limit'' on the oracles. 

In parallel,  it was formally proved in \cite{st82,ss88,be88} (using  game-theoretic techniques) that the ``natural'' Turing degrees are, essentially, the iterates of the Turing jump through the transfinite.

Recently \cite{km16}, a similar result was achieved for the ``natural'' $m$-degrees, under a precise notion of ``naturalness'' based on the uniform relativizations.
Namely, a function $f : \calN\to P(\omega)$ is {\em uniformly $(\leq_T ,\leq_m)$-preserving}  if, for every $X, Y \in\calN$,
$X \leq_T Y$ implies $f(X) \leq_m f(Y )$ uniformly, i.e.,
 there is a computable function $u$ on $\omega$ such that, for all $X, Y \in\calN$ and $e\in\omega$, the condition
``$X \leq_T Y$ via $e$'' implies that $f(X) \leq_m f(Y )$ via $u(e)$. The uniformly $(\equiv_T ,\equiv_m)$-preserving functions are defined in the same way. It is easy to see that the $m$-degrees of sets complete in the levels of the FH are natural in the sense of  \cite{km16}.

In \cite{km16} it was shown that the degree structure of the uniformly $(\equiv_T ,\equiv_m)$-order preserving functions under a natural QO of ``many-one reducibility on a cone'' is isomorphic to the structure of Wadge degrees. Moreover, the result holds for $Q^\omega$ in place of $2^\omega=P(\omega)$ in the definition above, where $Q$ is an arbitrary BQO. The proof heavily uses a generalization of Theorem \ref{emsbqo}. This is another demonstration of the interplay between computability theory and DST.

\section{Hierarchies in automata theory}\label{auto}

In this section we discuss some hierarchies and reducibilities
arising in automata theory. Automata theory is an important part of
computer science with many deep applications. In fact, many results
of this extensive field became already a part of the information
technology being realized in most of the existing hardware and
software systems. At the same time, automata theory remains an area
of active research, with many  open problems. The
theory is naturally divided in two parts devoted to the study of
finite and infinite behavior of computing devices. A positive feature of
this field is that many important decision problems concerning deterministic finite automata (dfa's)
are decidable. Accordingly, much effort is devoted to finding the optimal
decision algorithms and to the complexity issues.

Investigation of the infinite behavior of computing devices is of
great interest for computer science because many hardware and
software concurrent systems (like processors or operating systems)
may not terminate. In many cases, the infinite behavior of a device
is captured by the notion of $\omega$-language recognized by the
device. The most basic notion of this field is that of regular
$\omega$-languages, i.e. $\omega$-languages recognized by finite
automata. Regular $\omega$-languages play an important role in the theory and technology of specification and verification of finite state systems.

Regular $\omega$-languages were introduced by J.R. B\"{u}chi in the
1960s and studied by many people including B.A. Trakhtenbrot, R.
McNaughton and M.O. Rabin. The subject quickly developed into a rich
topic with several deep applications. Much information and
references on the subject may be found  e.g. in
\cite{tb70,th90,th96,wag79,pp04}. We assume the reader to be familiar with the standard notions and
facts of automata theory which may be found e.g. in
\cite{tb70,pp04}.

\subsection{Preliminaries}\label{premauf}

If not stated otherwise, $A$ denotes some finite
alphabet with at least two letters. Let $A^\ast$ and $A^+$ be the
sets of finite (respectively, of finite non-empty) words over $A$.
Sets of words are called languages. We mainly use
the logical approach to the theory of regular languages. This is the
reason why we mostly deal with subsets of $A^+$ (they correspond to
the non-empty structures, the empty structure is excluded because
dealing with it in logic is not usual). With suitable changes
analogs of the results below hold also for the subsets of $A^\ast$.

By an {\em automaton} (over $A$) we mean a triple ${\mathcal
M}=(Q,A,f)$ consisting  of a finite non-empty set $Q$ of states, the
input alphabet $A$ and a transition function $f:Q\times A\rightarrow
Q$. The transition function is naturally extended to the function
$f:Q\times A^*\rightarrow Q$ defined by induction
$f(q,\varepsilon)=q$ and $f(q,u\cdot x)=f(f(q,u),x)$, where
$\varepsilon$ is the empty word, $u\in A^*$ and $x\in A$. A {\em
word acceptor} is a triple $({\mathcal M},i,F)$ consisting of an
automaton ${\mathcal M}$, an initial state $i$ of ${\mathcal M}$ and
a set of final states $F\subseteq Q$. Such an acceptor recognizes
the language $L({\mathcal M},i,F)=\{u\in A^*\mid f(i,u)\in F\}$.
Languages recognized by such acceptors are called {\em regular}.

Relate to any alphabet $A=\{a,\ldots\}$ the signatures
$\varrho=\{\leq,Q_a,\ldots\}$ and $\sigma=\{\leq,Q_a,\ldots,\bot,
\top,p,s\}$, where $\leq$ is a binary relation symbol, $Q_a$ (for
any $a\in A$) is a unary relation symbol, $\bot$ and $\top$ are
constant symbols, and $p,s$  are unary function symbols. A word
$u=u_0\ldots u_n\in A^+$ may be considered as a structure ${\bf
u}=(\{0,\ldots,n\};<,Q_a,\ldots)$ of signature $\sigma$, where $<$
has its usual meaning, $Q_a(a\in A)$ are unary predicates on
$\{0,\ldots,n\}$ defined by $Q_a(i)\leftrightarrow u_i=a$, the
symbols $\bot$ and $\top$ denote the least and the greatest
elements, while $p$ and $s$ are respectively the predecessor and
successor functions on $\{0,\ldots,n\}$ (with $p(0)=0$ and
$s(n)=n$).

For a sentence $\phi$ of $\sigma$, set $L_\phi=\{u\in A^+\mid{\bf
u}\models\phi\}$. Sentences $\phi,\psi$ are treated as equivalent
when $L_\phi=L_\psi$. A language is {\em $FO_\sigma$-axiomatizable}
if it is of the form $L_\phi$ for some first-order sentence $\phi$
of signature $\sigma$. Similar notions apply to other signatures in
place of $\sigma$. It is well-known (see e.g.
\cite{str94,pp04}) that the class of
$FO_\sigma$-definable languages (as well as the class of
$FO_\varrho$-definable languages) coincides with the important class
of {\em regular aperiodic languages} which are also known as {\em
star-free languages}.

By {\em initial automaton} (over $A$) we mean a tuple $(Q,A,f,i)$
consisting  of a dfa $(Q,A,f)$ and an initial state $i\in Q$. Similarly to the function
$f:Q\times A^*\rightarrow Q$, we may define the function $f:Q\times
A^\omega\rightarrow Q^\omega$ by $f(q,\xi)(n)=f(q,\xi\upharpoonright n)$. Relate
to any initial automaton ${\mathcal M}$ the set of cycles
$C_{\mathcal M}=\{f_{\mathcal M}(\xi)\mid\xi\in A^\omega\}$ where
$f_{\mathcal M}(\xi)$ is the set of states which occur infinitely
often in the sequence $f(i,\xi)\in Q^\omega$. 

A {\em Muller acceptor} has the form $({\mathcal M},{\mathcal F})$
where ${\mathcal M}$ is an initial automaton and ${\mathcal
F}\subseteq C_{\mathcal M}$; it recognizes the set $ L({\mathcal
M},{\mathcal F})=\{\xi\in A^\omega\mid f_{\mathcal
M}(\xi)\in{\mathcal F}\}$. It is well known that Muller acceptors
recognize exactly the {\em regular $\omega$-languages} called also
just regular sets. The class ${\mathcal R}$ of all regular
$\omega$-languages is a proper subclass  of ${\mathbf\Delta}^0_3(A^\omega)$.

\subsection{Well quasiorders and regular languages}\label{wqoreg}

A basic fact of automata theory (Myhill-Nerode theorem) states that a language $L\subseteq A^*$ is regular iff it is closed w.r.t. some congruence of finite index on $A^*$ (recall that a congruence is an equivalence relation $\equiv$ such that $u\equiv v$ implies $xuy\equiv xvy$, for all $x,y\in A^*$). In \cite{eh83} the following version of Myhill-Nerode theorem was established:

\begin{theorem}\label{regwqo}
A language $L\subseteq A^*$ is regular iff it is upward closed w.r.t. some monotone WQO on $A^*$ 
(where a QO $\leq$ on $A^*$ is {\em monotone} if $u\leq v$ implies  $xuy\leq xvy$, for all $x,y\in A^*$).
\end{theorem}

Note that any congruence $\equiv$ of finite index on $A^*$ is a monotone WQO on $A^*$ such that the quotient-poset of $(A^*;\equiv)$ is a finite antichain. As observed in Proposition 6.3.1 of \cite{lv99}, an equivalence on $A^*$ is a congruence of finite index iff it is a monotone WQO.

Associate with any monotone WQO $\leq$ on $A^*$ the class $\mathcal{L}_\leq$  of upward closed sets in $(A^*;\leq)$. Then $\mathcal{L}_\leq$ is a lattice of regular sets. Clearly, $\mathcal{L}_\leq$ is closed under the complement iff $\leq$ is a congruence of finite index. In the literature one can find many examples of monotone QOs $\leq$ for which $\mathcal{L}_\leq$ is a finite lattice not closed under the complement (in particular, such examples arise from one-sided Ehrenfeucht-Fra\"iss\'e games, see e.g. \cite{s09}).

Are there other interesting examples of monotone WQOs? An important example is given by the subword relation $\leq^*$ from Section \ref{wqo}; in this case $\mathcal{L}_{\leq^*}$ is infinite. As observed independently in \cite{gs01} and \cite{s01}, $\mathcal{L}_{\leq^*}$ coincides with $\Sigma_1^\sigma$, hence there is a relation to the logical approach to automata theory (see the next subsection for additional details). There are other natural examples, for instance for each $k<\omega$ the following relations $\leq_k$ on non-empty words studied e.g. in \cite{st85,gs01,s01}: $u\leq_kv$, if $u=v\in A^{\leq k}$ or $u,v\in
A^{>k}$, $p_k(u)=p_k(v)$, $s_k(u)=s_k(v)$, and there is a
$k$-embedding $f:u\rightarrow v$. Here $p_k(u)$ (resp. $s_k(u)$) is the
prefix (resp., suffix) of $u$ of length $k$, and the $k$-embedding
$f$ is a monotone injective function from $\{0.\ldots,|u|-1\}$ to
$\{0.\ldots,|v|-1\}$ such that $u(i)\cdots u(i+k)=v(f(i))\cdots
v(f(i)+k)$ for all $i<|u|-k$. Note that the relation $\leq_0$ coincides with the subword relation.

It would be of interest to have more examples of natural monotone WQOs or maybe even a characterization of a wide class of monotone WQOs, in order to understand which classes of regular languages may be obtained in this way.  Theorem \ref{regwqo} turned out useful in the study of rewriting systems, serving as a tool to prove regularity of languages obtained by such systems. Many interesting facts on this may be found in  \cite{lv99} and references therein. This interesting direction has a strong semigroup-theoretic flavour.

Another development of Theorem \ref{regwqo} was initiated in \cite{miz} where some analogues of this theorem for infinite words were found. A QO $
\preceq$ on $A^\omega$ is a {\em periodic extension} of a QO $\leq$ on $A^*$ if   $\forall i<\omega(u_i\leq v_i)$ implies $u_0u_1\cdots\preceq v_0v_1\cdots$ and $\forall p\in A^\omega\exists u,v\in A^*(p\preceq uv^\omega\wedge uv^\omega\preceq p)$. Clearly, every periodic extension of a monotone WQO on $A^*$ is WQO on $A^\omega$. For instance, the subword relation on infinite words is a periodic extension of the subword relation on finite words and is therefore WQO. A basic fact in \cite{miz} is the following characterization of regular $\omega$-languages.

\begin{theorem}\label{regwqo1}
An $\omega$-language $L\subseteq A^\omega$ is regular iff it is upward closed w.r.t. some periodic extension of a monotone WQO on $A^*$.
\end{theorem}

Relate to any monotone WQO $\leq$ on $A^*$ the class $\mathcal{L}^\omega_\leq$  of upward closed sets in $(A^\omega;\preceq)$, for some periodic extension $\preceq$ of $\leq$. Then $\mathcal{L}^\omega_\leq$ is a class of regular $\omega$-languages. To our knowledge, almost nothing is known on which classes of regular $\omega$-languages are obtained in this way. It seems natural to explore possible relationships of such classes to the ``logical'' hierarchies of  regular $\omega$-languages which are important but are much less understood than the logical hierarchies of   languages in the next subsection.  For some information on logical hierarchies of  regular languages see e.g. \cite{pp04,dk09} and references therein.

\subsection{Hierarchies of regular languages}\label{basauf}

We denote by $\Sigma^\sigma_n$  the class of languages that can be
axiomatized by a $\Sigma^0_n$-sentence of signature $\sigma$. The classes
$\Sigma^\rho_n$  are defined analogously with respect to
$\rho$. There is a level-wise correspondence of
these classes to the well-known concatenation hierarchies of
automata theory (see e.g. \cite{th96,str94,pp04}).

The $\omega$-bases $\mathcal{L}^\sigma=\{\Sigma^\sigma_n\}$ and $\mathcal{L}^\rho=\{\Sigma^\rho_n\}$ do not collapse, and they are neither
reducible nor interpolable (see e.g. \cite{s09} and references therein). 
In \cite{sw05} a natural reducibility $\leq_{qf\rho}$ by quantifier-free formulas of
signature $\rho$ was introduced and studied. This reducibility fits the hierarchy $\mathcal{L}^\rho$.  

One can of course consider the refinements of the hierarchies $\mathcal{L}^\sigma$ and $\mathcal{L}^\rho$. Among these, the difference hierarchies of sets were studied in detail, see e.g. \cite{st85,gs01,s01,s09,gss16}. Note that many variants of the mentioned hierarchies and reducibilities on regular languages were also considered in the literature (say, for other signatures or other logics in place of the first-order logic). 

The main problems about the mentioned hierarchies concern decidability of the corresponding classes of languages or relations between languages (if the languages are given, say, by  recognizing automata). Many  such decidability problems turn out to be complicated, in particular the decidability of only lower levels of the mentioned hierarchies is currently known. 

The relation of this theme to WQOs was not investigated systematically, though the relation to some WQOs was used in proving decidability of levels of the DHs over $\Sigma^\sigma_1$ (and also for some other natural bases, see  \cite{gs01,s01,s09}). Such algorithms are based on the characterization of $\Sigma^\sigma_1$ in terms of the subword partial order mentioned in the previous subsection. Applicability of WQO-theory to higher levels remains unclear.

Concerning the quantifier-free reducibilities, some interesting structural results were obtained in \cite{sw05,s09}. The relation $\leq_{qf\rho}$ is not WQO on the regular languages but it is open whether $\leq_{qf\rho}$ is WQO on the regular aperiodic languages (or at least on some reasonable subclasses).

The mentioned ``logical'' hierarchies of regular languages may be defined a similar way  for the regular $\omega$-languages, and it is known that this extension brings many new aspects, in particular one has to deal with the topological issues (see e.g. \cite{dk09} and references therein). The relation to WQO-theory is not clear.

\subsection{Hierarchies of $\omega$-regular languages and $k$-Partitions}\label{partaui}

On the class ${\mathcal R}$ of regular $\omega$-languages there is a natural
2-base ${\mathcal L}=({\mathcal R}\cap\mathbf{\Sigma}^0_1,{\mathcal R}\cap\mathbf{\Sigma}^0_2)$. As shown in \cite{s98}, this base is reducible and interpolable. There is also a natural reducibility $\leq_{DA}$ that fits this hierarchy (namely, the reducibility by the so-called deterministic asynchronous finite transducers \cite{wag79}, i.e., by dfa's with output which may print a word at any step). One can also consider the Wadge reducibility, which is denoted in \cite{wag79} by $\leq_{CA}$; it does not fit this hierarchy.

In \cite{wag79} K. Wagner gave in a sense the finest possible
topological classification of regular $\omega$-languages which subsumes several hierarchies considered before him. Among his main results are the following:

\begin{enumerate}\itemsep-1mm
 \item  The QO $({\mathcal R};\leq_{CA})$ is  semi-well-ordered
with the order type $\omega^\omega$.
 
 \item  The $CA$-reducibility coincides on ${\mathcal R}$ with
 the $DA$-reducibility.
 
 \item   Every level of the
hierarchy formed by the principal ideals of $({\mathcal R};\leq_{DA})$ is decidable.
\end{enumerate}

  In
\cite{s98} the Wagner hierarchy  was related to the Wadge
hierarchy and to the author's fine hierarchy, namely it is just the FH of sets over ${\mathcal L}$.

Here we briefly discuss the extension of the Wagner hierarchy to the $\omega$-regular $k$-partitions, the class of which is denoted by $\mathcal{R}_k$. Recall from \cite{s11} that a {\em Muller
$k$-acceptor}  is a pair $({\mathcal A},c)$ where
${\mathcal A}$ is an automaton and $c: C_{\mathcal A}\rightarrow k$  a
$k$-partition of $C_{\mathcal A}$. Such a $k$-acceptor recognizes the
$k$-partition $L({\mathcal A},c)=c\circ f_{\mathcal A}$ where $f_{\mathcal
A}:A^\omega\rightarrow C_{\mathcal A}$ is the function defined  in Section \ref{premauf} As shown in \cite{s11}, a $k$-partition $L:A^\omega\rightarrow k$ is regular iff it is
recognized by a Muller $k$-acceptor. Below we also use the 2-base ${\mathcal M}=(\mathbf{\Sigma}^0_1,\mathbf{\Sigma}^0_2)$. The main results in \cite{s11} maybe summarized as follows:

\begin{theorem}\label{main-fh}
\begin{enumerate}\itemsep-1mm
 \item The quotient-posets of $({\mathcal{T}^\sqcup_k(2);\leq_h})$,
$(\mathcal{R}_k;\leq_{CA})$ and
$(\mathcal{R}_k;\leq_{DA})$ are isomorphic.
 \item  The relations $\leq_{CA}$ and $\leq_{DA}$ coincide on $\mathcal{R}_k$.
 \item  Every level $\mathcal{L}(T)$  of the FH of $k$-partitions over $\mathcal{L}$ is decidable.
 \item For each $T\in\mathcal{T}_k(2)$,
 $\mathcal{L}(T)=\mathcal{R}_k\cap\mathcal{M}(T)$.
 \end{enumerate}
\end{theorem}

Item (2) is obtained by applying the B\"uchi-Landweber theorem on infinite regular games \cite{bl69}.
The proof of item (1) is similar to that in Section \ref{wpartbaire}, but first we have to define  the corresponding operations on $k^{A^\omega}$, where $A=\{0,1,\ldots\}$ is a finite alphabet. This needs some coding because here we work with the compact Cantor space $A^\omega$ while in Section \ref{wpartbaire}  with the Baire space where the coding is easier.

For all $i<k$ and $A\in k^{A^\omega}$, define the
$k$-partition $p_i(A)$ as follows:
$[p_i(A)](\xi)=i$, if 
$\xi\not\in 0^\ast1X^\omega$, otherwise
$[p_i(A)](\xi)=A(\eta)$ where $\xi=0^n1\eta.$

Next we define unary operations $q_0,\ldots,q_{k-1}$ on
$k^{A^\omega}$.  To simplify notation, we do this only for the
particular case $A=\{0,1\}$ (the general  case is treated
similarly). Define a $DA$-function $f:3^\omega\to 2^\omega$ by
$f(x_0x_1\cdots)=\tilde{x}_0\tilde{x}_1\cdots$ where
$x_0,x_1\ldots<3$ and
$\tilde{0}=110000,\tilde{1}=110100,\tilde{2}=110010$ (in the same
way we may define $f:3^\ast\to 2^\ast$). Obviously,
$f(3^\omega)\in\mathcal{R}\cap\mathbf{\Pi}^0_1(A^\omega)$ and there is a
$DA$-function $f_1:2^\omega\to 3^\omega$ such that $f_1\circ
f=id_{3^\omega}$. For all $i<k$ and $k$-partitions $A$ of
$X^\omega$, define the $k$-partition $q_i(A)$ as follows:
 $[q_i(A)](\xi)= i$, if
$\xi\not\in f(3^\omega)\vee\forall p\exists n\geq p(\xi[n,n+5]=\tilde{2})$,
$[q_i(A)](\xi)=A(f_1(\xi))$, if $\xi\in f(2^\omega)$, and
$[q_i(A)](\xi)=A(\eta)$ in the other cases, where $\xi=f(\sigma3\eta)$
 for some $\sigma\in 3^\omega$ and $\eta\in 2^\omega$. 

Finally, we define the binary operation  $+$ on $k^{A^\omega}$ as
follows. Define a $DA$-function $g:X^\omega\to X^\omega$ by
$g(x_0x_1\cdots)=x_00x_10\cdots$ where $x_0,x_1,\ldots\in X$ (in the same
way we may define  $g:X^\ast\to X^\ast$). Obviously,
$g(A^\omega)\in\mathcal{R}\cap\mathbf{\Pi}^0_1(A^\omega)$ and there is a
$DA$-function $g_1:A^\omega\to X^\omega$ such that $g_1\circ
g=id_{X^\omega}$. For all $k$-partitions $A,B$ of  $X^\omega$, we
set:
 $[A+B](\xi)= A(g_1(\xi))$ if 
$\xi\in g(X^\omega)$, otherwise
 $[A+B](\xi)=B(\eta)$, where $\xi=g(\sigma)i\eta$
for some $\sigma\in A^\omega,i\in A\setminus\{0\}$ and $\eta\in
A^\omega$.  

The operations $p_i,q_i,+$  have the same properties
as the corresponding operations in Section \ref{wpartbaire} (only this time we have no infinite disjoint union, so we speak e.g. about $dc$-semilatices instead of the $dc\sigma$-semilatices). Therefore, defining the functions $\mu,\nu,\rho$ just as in Section \ref{wpartbaire} (but for finite trees $T$) we get that  $\rho:\mathcal{T}^\sqcup_k(2)\to k^{A^\omega}$ induces the embedding of the quotient-poset of $(\mathcal{T}^\sqcup_k(2);\leq_h)$ into those of $(\mathcal{R}_k;\leq_{CA})$ and
$(\mathcal{R}_k;\leq_{DA})$. Moreover, one easily checks that
for any $T\in\mathcal{T}_k(2)$, $\rho(T)$ is $CA$-complete in
$\mathcal{M}(T)$ and $DA$-complete in $\mathcal{L}(T)$.  

That this embedding is in fact an isomorphism, and that items (3), (4) hold, follows from analysing some invariants of the Muller $k$-acceptors ${\mathcal A}$ based on the QOs $\leq_0$ and  $\leq_1$ on the set of cycles $C_{\mathcal A}$ which extends a standard technique in the Wagner hierarchy.

\section{Conclusion}\label{con}

When this paper was under review (as a chapter of a Handbook on WQOs), preprint \cite{km17} appeared which contains important results on Wadge degrees. It gives a characterization of the quotient-poset of  $(Q^*;\leq^*)$  from Theorem \ref{emsbqo} for every countable BQO $Q$. Unifying notation, we denote $(Q^*;\leq^*)$ as $(\mathbf{B}(Q^\mathcal{N});\leq_W)$ and call elements of $Q^\mathcal{N}$ $Q$-partitions of $\mathcal{N}$. In fact, the authors of \cite{km17} also characterize the quotient-poset of $(\mathbf{\Delta}^0_{1+\alpha}(Q^\mathcal{N});\leq_W)$ for each $\alpha<\omega_1$ where $\mathbf{\Delta}^0_{1+\alpha}(Q^\mathcal{N})$ consists of all $A:\mathcal{N}\to Q$ such that $A^{-1}(q)\in\mathbf{\Delta}^0_{1+\alpha}(\mathcal{N})$ for every $q\in Q$.

As in Section \ref{wpartbaire}, the characterization uses a suitable iteration $\widetilde{\mathcal{T}}_{\alpha}(Q)$ of the introduced in \cite{s16} operator  $\widetilde{\mathcal{T}}$ on the class of all BQOs starting from $Q$, and an extension $\Omega:\widetilde{\mathcal{T}}^\sqcup_{\alpha}(Q)\to \mathbf{\Sigma}^0_{1+\alpha}(Q^\mathcal{N})$ of our embedding $\mu$. Using an induction on BQO $(\mathbf{\Sigma}^0_{1+\alpha}(Q^\mathcal{N});\leq_W)$ one can show that for every $A\in\mathbf{\Sigma}^0_{1+\alpha}(Q^\mathcal{N})$ there is $F\in\widetilde{\mathcal{T}}^\sqcup_{\alpha}(Q)$ with $\Omega(F)\equiv_WA$; this yields the desired isomorphism of quotient-posets. Thus, the idea and the scheme of proof is the same as in \cite{s16,s17} but the proof of surjectivity of $\Omega$ in \cite{km17} is  quite different from our proof of particular cases. The proof in \cite{km17} uses a nice extension of the notion of conciliatory set \cite{du01} to $Q$-partitions and a deep relation of this field to some basic facts about Turing degrees. 

We conclude this survey with collecting some open questions which seem of interest to the discussed topic:

\begin{enumerate}\itemsep-1mm

\item What are the  maximal order types of the concrete BPOs mentioned in the paper (except those which are semi-well-ordered or are already known). 

\item Characterize the maximal order types of computable WPOs of rank $\omega$ (or any other infinite computable ordinal).

\item Characterize the degree spectra of countable WPOs. In particular, is it true that for any given countable graph there is a countable WPO with the same degree spectrum?

\item Associate with any WPO $P$ the function $f_P:rk(P)\to\omega$ by: $f_P(\alpha)$ is the cardinality of $\{x\in P\mid rk_P(x)=\alpha\}$. Is there a computable WPO $P$ such that $f_P$ is not  computable? 

\item Characterize the finite posets $Q$ for which the first-order theory of the quotient-poset of $\mathcal{T}^\sqcup_Q$ is decidable.

\item Is there a finite poset $Q$ of width $\geq3$ such that the automorphism groups of $Q$ and of the quotient-poset of $\mathcal{T}^\sqcup_Q$ are not isomorphic?

\item Extend the above-mentioned characterizations of the initial segments of $Q$-partitions of Baire space ($Q$ is BQO) beyond the Borel $Q$-partitions.
\medskip

{\bf Acknowledgement.} I am grateful to Leibniz-Zentrum f\"ur Informatik for accepting and taking care of Dagstuhl Seminars 08271, 11411, 15392, 16031 which were important for promoting the topic of this paper.

\end{enumerate}

%\section{Conclusion}\label{con}

%%%%%%%%%%%%%%%%%%%%%%%%%%%%%%%%%%%%%%%%%%%%%%%%%%%%%%%%%
%
%%%%%%%%%%%%%%%%%%%%%%%%%%%%%%%%%%%%%%%%%%%%%%%%%%%%%%%%%

%%%%%%%%%%%%%%%%%%%%%%%%%%%%%%%%%%%%%%%%%%%%%%%%%%%%%%%


\begin{thebibliography}{asaaa67}

\bibitem{ab04} P. A. Abdulla et al. Using forward reachability analysis for verification of lossy channel systems? In: Form. Methods Sys. Des. 25.1 (2004), pp. 39--65.

\bibitem{ad59} J.W. Addison. Separation principles in the
hierarchies of classical and effective set theory. {\em Fund.
Math.}, 46 (1959), 123--135.

\bibitem{ad62} J.W. Addison. Some problems in herarchy theory. {\em Recursive Function
Theory}, Proc. of Symp. in Pure Math., AMS, 5 (1962), 123--130.

\bibitem{ad62a} J.W. Addison. The theory of hierarchies. {\em Logic, Methodology and Philosophy of
Science}, Proc. of 1960 Int. Congress, Stanford, Palo Alto, 1962,
26--37.

\bibitem{ad65} J.W. Addison. The method of alternating chains.
In: {\em The theory of models}, Amsterdam, North Holland, 1965,
p.1--16.

\bibitem{aj} S.  Abramsky S., A.  Jung. Domain theory. In:  Handbook of
Logic in Computer Science, v. 3, Oxford, 1994, 1--168.

\bibitem{ak00} C.J. Ash, J. Knight. Computable Structures and the Hyperarithmetical Hierarchy. Elsevier Science, Amsterdam (2000).

\bibitem{ak00a} S. Adams, A.S. Kechris. Linear algebraic groups and countable Borel equivalence relations. Journal of AMS,  13, No 4,  909--943.

\bibitem{an06}
A. Andretta. More on Wadge determinacy, Annals of Pure and Applied Logic, 144,
2006, no.~1-3, 2--32.

\bibitem{am03}
A. Andretta, D.A. Martin. Borel-Wadge degrees, Fund. Math.
177, 2003, no.~2, 175--192.

%\bibitem{ap03} M. Aschenbrenner, W.Y. Pong. Orderings of Monomial Ideals. arXiv:math/0305384 [math.LO], 2003.

\bibitem{bl14} A.C. Block. Operations on a Wadge-type hierarchy of ordinal-valued functions. Master’s thesis, Universiteit van Amsterdam, 2014.

\bibitem{be88} H. Becker. A characterization of jump operators. Journal of Symbolic Logic, 53(3):708–-728, 1988.

%\bibitem{bg12}  Becher V.,  Grigorieff, S.:
%Borel and Hausdorff hierarchies in topological spaces of Choquet games and their effectivization. Mathematical Structures in Computer Science 25(7): 1490--1519 (2015).

%\bibitem{bg12a}  Becher V.,  Grigorieff, S.:
%Wadge hardness in Scott spaces and its effectivization. Mathematical Structures in Computer Science 25(7): 1520--1545 (2015).

\bibitem{bl69} J.R. B\"{u}chi and L.H. Landweber. Solving sequential
conditions by finite-state strategies. {\em Trans. Amer. Math.
Soc.,} 138 (1969), p.295--311.

%\bibitem{bra05}  Brattka V.: Effective Borel measurability
%and reducibility of functions,  Mathematical Logic Quarterly, 51:1
%(2005), 19--44.

\bibitem{bg09} V. Brattka, G. Gherardi. Weihrauch degrees,
omniscience principles and weak computability.
http://arxive.org/abs 0905.4679 (2009)

\bibitem{bg09a} V. Brattka, G. Gherardi. Effective choice and
boundedness principles in computable analysis.
http://arxive.org/abs 0905.4685 (2009).

\bibitem{ca13} R. Carroy. A quasiorder on continuous functions. Journal of Symbolic Logic,
    78, No 2 (2013), 633--648. 

\bibitem{ce91} D. Cenzer. Polynomial-time versus recursive models. Annals of Pure and Applied Logic, 54, No 1, 1991  17--58.

%\bibitem{dg09} V.Diekert, M. Kufleitner. Fragments of first-order logic over infinite words. ArXiv:0906.2995v2[cs.FL] 2 Oct 2009. 

\bibitem {br} %[Br13]
M. de Brecht.
Quasi-Polish spaces,
\emph{Annals of pure and applied logic}, \textbf{164}, (2013), 356--381.

\bibitem{df98} R.G. Downey, M.R. Fellows. Parameterized Complexity. Springer, New York, 1998.

\bibitem{dk09} V. Diekert, M. Kufleitner.
Fragments of First-Order Logic over Infinite Words. Theory Comput. Syst. 48(3): 486--516 (2011).

\bibitem{jp77} D.H.J. de Jongh, R. Parikh> Well-partial orderings and hierarchies. Nederl. Akad. Wetensch. Proc. Ser. A 80 = Indag. Math. 39(3), 195--207 (1977).

\bibitem{dp94} B.A. Davey and H.A. Priestley. {\em Introduction to
Lattices and Order}. Cambridge, 1994.

\bibitem{du01} J. Duparc. Wadge hierarchy and Veblen hierarchy, part I. Journal of Symbolic Logic, 66, No 1 (2001), 56--86.

\bibitem{lv99} A. de Luca, S. Varricchio. Finiteness and Regularity
in Semigroups and Formal Languages. Springer, Berlin, 1999.

\bibitem{eh83} A. Ehrenfeucht, D. Hausser, and G. Rozenberg. On regularity of context-free
languages. Theoretical Computer Science, 27:311-–332, 1983.

\bibitem{en89} %[En89]
R. Engelking. 
General Topology,
\emph{Heldermann, Berlin} (1989).

\bibitem{engelen} F. van Engelen. Homogeneous zero-dimensional absolute Borel sets.  CWI Tracts, Amsterdam, 1986.

\bibitem{ems87} F. van Engelen, A. Miller, J. Steel. Rigid Borel
sets and better quasiorder theory.  Contemporary mathematics,
65 (1987), 199--222.

\bibitem{er68} Yu.L. Ershov. On a hierarchy of sets 1,2,3 (in Russian).
 Algebra i Logika, 7,  No 4 (1968), 15--47

\bibitem{eg99} Yu.L. Ershov, S.S.  Goncharov. {\em Constructive
Models}. Novosibirsk, Scientific Book, 1999 (in Russian, there is an
English Translation).

\bibitem{eltt}
Yu. L. Ershov, I.A. Lavrov, A.D.  Taimanov, M.A. Taitslin.
Elementary theories. Uspechi Mat. Nauk {\bf 20} N 4 (1965) 37--108.
(in Russian)

%\bibitem{er80} Yu.L. Ershov. {\em Decidability problems and
%constructive models}. Moscow, Nauka, 1980 (in Russian).

%\bibitem{g03} %[GH80] 
%Giertz, G., Hoffmann, K.H., Keimel, K., Lawson, J.D., Mislove, M.W., Scott, D.S.: 
%A compendium of Continuous Lattices,
%\emph{Springer, Berlin}, 1980.

\bibitem{gao09} Su Gao. {\em Invariant Descriptive Set Theory}.
Pure and Applied Mathematics, A Series of Monographs and
Textbooks, 293. Taylor $\&$ Francis Group, 2009.

%\bibitem{gl10} J. Goubault-Larrecq. Noetherian spaces in verification. ICALP (2) 2010: 2--21.

\bibitem{gs01} C. Gla\ss er amd H. Schmitz. The Boolean structure of
dot-depth one. {\em Journal of Automata, Languages and
Combinatorics}, 6 (2001), 437--452.

\bibitem{gss16} C. Glasser, H. Schmitz, V. Selivanov. Efficient algorithms for membership in boolean hierarchies of regular languages. Theoretical Computer Science, 646, issue C, 86-108 (2016), doi: 10.1016/j.tcs.2016.07.017. ISSN 0304-3975.

%\bibitem{hem06}  Hemmerling, A.: The Hausdorff-Ershov hierarchy in
%Euclidean spaces.  Archive for Mathematical Logic, 45 (2006),
%323--350.

\bibitem{he93} P. Hertling. Topologische Komplexit\"atsgrade von
Funktionen mit endlichem Bild.  Informatik-Berichte 152, 34 pages,
Fernuniversit\"at Hagen, December 1993.

\bibitem{he96}
P.~Hertling. {\em Unstetigkeitsgrade von Funktionen in der
effektiven   Analysis}. PhD thesis, Fachbereich Informatik,
FernUniversit\"{a}t Hagen, 1996.

\bibitem{hs14} P. Hertling and V. Selivanov. Complexity issues for preorders on finite labeled forests. «Logic, Computation, Hierarchies», Eds. Vasco Brattka, Hannes Diener, and Dieter Spreen, Ontos Publishing, de Gruiter, Boston-Berlin, 2014, 165-190.

\bibitem{hw94} P. Hertling, K. Weihrauch.
Levels of degeneracy and exact lowercomplexity bounds for geometric algorithms.
Proc. of the 6th Canadian Conf. on Computational Geometry Saskatoon (1994)
237-24

\bibitem{hi52} G. Higman. Ordering by divisibility in abstract algebras. Proc. Lond. Math. Soc. 2(3), 326--336 (1952).

\bibitem{hir} D.R. Hirschfeldt. Slicing the Truth - On the Computable and Reverse Mathematics of Combinatorial Principles. Lecture Notes Series, Institute for Mathematical Sciences, National University of Singapore: Volume 28, World Scientific, 2014. 

\bibitem{ho93} W. Hodges. Model Theory. Cambridge University Press, 1993.

\bibitem{hurwal}
W. Hurewicz\ and\ H. Wallman, {\it Dimension Theory}, Princeton Mathematical Series, v. 4, Princeton Univ. Press, Princeton, NJ, 1948.

\bibitem{ik10} D. Ikegami. Games in Set Theory and Logic, PhD Thesis, University of Amsterdam, 2010.

\bibitem{sc10} D. Ikegami, P. Schlicht, H. Tanaka. Continuous reducibility for the real line, preprint, submitted, 2012.

%\bibitem{jr82}J.E. Jayne,\ and\ C.A. Rogers, First level Borel functions and
%isomorphisms, J. Math. Pures Appl. (9) {\bf 61} (1982), no.~2,177--205.

%\bibitem{JR79a}
%J. E. Jayne\ and\ C. A. Rogers, Borel isomorphisms at the first level. I, Mathematika {\bf 26} (1979), no.~1, 125--156.

%\bibitem{JR79b}
%J. E. Jayne\ and\ C. A. Rogers, Borel isomorphisms at the first level. II, Mathematika {\bf 26} (1979), no.~2, 157--179.

%\bibitem{JR82}
%J. E. Jayne\ and\ C. A. Rogers, First level Borel functions and isomorphisms, J. Math. Pures Appl. (9) {\bf 61} (1982), no.~2, 177--205.

\bibitem{jm09} J. Jezek, R. McKenzie. Definability in substructure orderings, iv: finite
lattices. Algebra universalis, 61(3--4):301--312, 2009.

\bibitem{jm09a} J. Jezek, R. McKenzie.  Definability in substructure orderings, i: fifinite
semilattices. Algebra universalis, 61(1):59--75, 2009.

\bibitem{jm09b} J. Jezek, R. McKenzie.  Definability in substructure orderings, iii:
finite distributive lattices. Algebra universalis, 61(3-4):283--300, 2009.

\bibitem{jm10} J. Jezek, R. McKenzie.  Definability in substructure orderings, ii:
finite ordered sets. Order, 27(2):115--145, 2010.

\bibitem{ku15} A. Kunos. Definability in the embeddability ordering of finite directed graphs.
Order, 32(1):117--133, 2015.

\bibitem{kr60} J.B. Kruskal. Well-quasi-ordering, the tree theorem, and Vazsonyi’s conjecture. Trans. Am.
Math. Soc. 95, 210--225 (1960).

%\bibitem{kacmotsem}
%M. Ka\v{c}ena, L. Motto Ros, B. Semmes, Some observations on "A
% new proof of a theorem of Jayne and Rogers", accepted for publication on Real Analysis Exchange, 2012.
 
\bibitem{kns17}   P. Karandikar, M. Niewerth, Ph. Schnoebelen. On the state complexity of closures and interiors of regular languages with subwords and superwords. Theoret. Comput. Sci. 610 (2016), pp. 91--107.

\bibitem{ksc15}   P. Karandikar, Ph. Schnoebelen. Decidability in the logic of subsequences and supersequences. In: Proc. FST\&TCS 2015, LIPIcs 45 Leibniz-Zentrum f\"ur Informatik, 2015, pp. 84--97.

\bibitem{ksc15a}    P. Karandikar, Ph. Schnoebelen. Generalized Post embedding problems. In: Theory of Computing Systems 56.4 (2015), pp. 697--716.

\bibitem{ksc16}  P. Karandikar Ph. Schnoebelen. The height of piecewise-testable languages with applications in logical complexity. In: Proc. CSL 2016, LIPIcs62. Leibniz-Zentrum f\"ur Informatik, 2016, 37:1--37:22. 

\bibitem{ks06}   O.V. Kudinov and V.L. Selivanov.
Undecidability in the homomorphic quasiorder of finite labeled
forests. Proc. CiE 2006, {\em Lecture Notes in Computer Science}, v.
3988, Berlin: Springer, 2006, 289--296.

\bibitem{ks07}  O.V. Kudinov, V.L. Selivanov. Definability in the homomorphic quasiorder of finite labeled forests. Proc. of CiE-2007, Lecture Notes in Computer Science, v. 4497. Berlin: Springer, 2007, 436--445. 

\bibitem{ks07a} O.V. Kudinov, V.L. Selivanov. Undecidability in the Homomorphic Quasiorder of Finite Labelled Forests. Journal of Logic and Computation, 17 (2007), 113--1151.

\bibitem{ks09}  O.V. Kudinov, V.L. Selivanov. A Gandy theorem for abstract structures and applications to firstorder definability. Proc. CiE-2009, LNCS 5635, Springer, Berlin, 290--299, 2009. 

\bibitem{ks09a} O. Kudinov, V. Selivanov. Definability in the infix order on words. Proc. of DLT-2009 (V. Diekert and D. Nowotka, eds.), Lecture Notes in Computer Science, v. 5583. Berlin: Springer, 2009, 454--465.

\bibitem{ksy10} O.V. Kudinov, V.L. Selivanov, L.V. Yartseva. Definability in the subwordx order. Proc. of CiE-2010 (F. Ferreira, B. Loewe, E. Mayordomo and L.M. Gomes, eds.), Lecture Notes in Computer Science, v. 6158. Berlin: Springer, 2010, 246--255.

\bibitem{ksz09} O.V. Kudinov, V.L. Selivanov, A.V. Zhukov. Definability in the h-quasiorder of labeled forests. Annals of Pure and Applied Logic, 159(3), 318--332, 2009.

\bibitem{ku06} D. Kuske. Theories of orders on the set of words, Theoretical Informatics and Applications 40 (2006), 53--74. 

\bibitem{kur}
K. Kuratowski, Sur une g\'en\'eralisation de la notion d'hom\'eomorphie, Fund. Math, {\bf 22} (1934), 206--220.

\bibitem{km67} K. Kuratowski and A. Mostowski.
{\em Set Theory}. North Holland, 1967.

\bibitem{km16} T. Kihara, A.  Montalban. The uniform Martin's conjecture for many-one degrees. ArXiv: 1608.05065 v1 [Math.LO] 17 Aug 2016. (accepted by Transactions of AMS).

\bibitem{km17} T. Kihara, A.  Montalban. On the structure of the Wadge degrees of BQO-valued Borel functions.  ArXiv: 1705.07802 v1 [Math.LO] 22 May 2017. (accepted by Transactions of AMS).

\bibitem{ke95} %[Ke95]
Kechris, A.S.:
Classical Descriptive Set Theory,
\emph{Springer, New York}, (1995).

\bibitem{la71}  Laver, R.: On Fra\"iss\'e’s order type conjecture. Ann. of Math. 93(2), 89–-111 (1971).

\bibitem{la78} R. Laver, Better-quasi-orderings and a class of trees, in Studies in Foundations and Combinatorics,
Adv. in Math. Supplementary Series 1, Academic Press, New York, 1978, pp.
31–48.

\bibitem{lo83} A. Louveau. Some results in the Wadge hierarchy
of Borel sets. {\em Lec. Notes in Math.}, No 1019 (1983), p.28--55.

\bibitem{ma16} A.S. Marks. The universality of polynomial time Turing
equivalence. ArXiv:1601.033431 [math LO] 13 Jan 2016.

\bibitem{mass} A.S. Marks, T. Slaman, J. Steel. Martin’s conjecture, arithmetic equivalence and countable Borel equivalence relations.

\bibitem{mon07} A. Montalban. Computable linearizations of well-partial-orderings. Order (2007) 24:39–-48,
DOI 10.1007/s11083-007-9058-0.

\bibitem{mo09} Y.N. Moschovakis. {\em Descriptive  Set  Theory},
North Holland, Amsterdam, 2009.

\bibitem{ros09}
L. Motto Ros, Borel-amenable reducibilities for sets of reals, J. Symbolic Logic {\bf 74} (2009), no.~1, 27--49.

%\bibitem{motbaire}
%L. Motto Ros, Baire reductions and good Borel reducibilities, J. Symbolic Logic {\bf 75} (2010), no.~1, 323--345.

%\bibitem{motsuperamenable}
%L. Motto Ros, Beyond Borel-amenability: scales and superamenable reducibilities, Ann. Pure Appl. Logic {\bf 161} (2010), no.~7, 829--836.

%\bibitem{motsem}
%L. Motto Ros\ and\ B. Semmes, A new proof of a theorem of Jayne and Rogers, Real Anal. Exchange {\bf 35} (2010), no.~1, 195--203.

\bibitem{mss12} 	L. Motto Ros, P. Schlicht and V. Selivanov. Wadge-like reducibilities on arbitrary quasi-Polish spaces. Mathematical structures in computer science,   v. 25 (2015), special issue 8, pp. 1705--1754, doi:10.1017/S0960129513000339.

\bibitem{ma93} A. Marcone, Foundations of bqo theory and subsystems of second order arithmetic, Ph.d.
thesis, The Pennsylvania State University, 1993.

\bibitem{ma} A. Marcone, Foundations of bqo theory, Trans. Amer. Math. Soc., 345, No 2 (1994),  641--660.

%\bibitem{mil} E. C. Milner, Basic wqo- and bqo-theory, in [8], pp. 487–502.

\bibitem{mw85} R. Mansfield, G. Weitkamp. Recursive aspects of descriptive set theory. Oxford
University Press, New York, 1985. 

\bibitem{nw65a} C. St. J. A. Nash-Williams, On well-quasi-ordering infinite trees, Proc. Cambridge Philos.
Soc. 61 (1965), 697–720.

\bibitem{nw65} C.St.J.A. Nash-Williams. On well-quasi-ordering transfinite sequences. Proc. Camb. Philos. Soc.
61, 33–-39 (1965).

\bibitem{nw65} C. St. J. A. Nash-Williams. On well-quasi-ordering infinite trees. Proc. Cambridge Philos.
Soc., 61: 697–-720, 1965.

\bibitem{nw68} C. St. J. A. Nash-Williams, On better-quasi-ordering transfinite sequences, Proc. Cambridge
Philos. Soc. 64 (1968), 273–290.

\bibitem{ni00} A. Nies. Definability in the c.e. degrees:
questions and results. {\em Contemporary Mathematics}, 257 (2000),
207--213.

\bibitem{miz} M. Ogawa. Well quasiorders and regular $\omega$-languages. Theoretical Computer Science, 324, No 1 (2004), 55--60.

\bibitem{pe15} Y. Pequignot. A Wadge hierarchy for second countable spaces.
Archive for Mathematical Logic 54.5-6 (2015), pp. 659–683. doi:
10.1007/s00153-015-0434-y.

\bibitem{pp04} D. Perrin and J.-E. Pin. {\em Infinite Words}. v. 141 of
{\em Pure and Applied Mathematics}, Elsevier, 2004.

%\bibitem{qu46} W. V. Quine. Concatenation as a basis for arithmetic.  J. Symb. Logic, 11(4), 1946, 105--114. 

\bibitem{ro67}H.  Rogers, jr. {\em  Theory of Recursive Functions and
Effective Computability}. McGraw-Hill, New York, 1967.

\bibitem{rs04} N. Robertson, P.D. Seymour. Graph minors. XX. Wagner’s conjecture. J. Comb. Theory Ser. B
92(2), 325–-357 (2004).

\bibitem{rt16}  R. Ramanujam. R.S. Thinniyam. Definability in first order theories of graph
orderings. In Logical Foundations of Computer Science, pages 331--348. Springer,
2016.

\bibitem{sc79} D. Schmidt. Well-partial orderings and their maximal order types. Habilitationsschrift,
University of Heidelberg (1979).

\bibitem{so87} R.I. Soare. Recursively enumerable sets and degrees. Perspectives in Mathematical Logic. A
study of computable functions and computably generated sets. Springer, Berlin (1987).

\bibitem{s82} V.L. Selivanov, V.L. On the structure of degrees of generalized
index sets. {\em Algebra and Logic}, 21,  (1982), 316--330.

\bibitem{s83} V.L. Selivanov. Hierarchies  of
hyperarithmetical  sets and functions. {\em Algebra and Logic,} 22
(1983), p.473--491.

\bibitem{s88} V.L. Selivanov. On algorithmic complexity of algebraic systems. Mathematical Notes, 44,
No 5--6 (1988) p.944--950.

\bibitem{s92} V.L. Selivanov.
{\em Hierarchies, Numberings, Index Sets.} Handwritten notes,
1992, 300 pp.

\bibitem{s95} 	V.L. Selivanov. Fine hierarchies and Boolean terms. The Journal of Symbolic Logic, 60, No 1 (1995), 289--317.

\bibitem{s98} V.L. Selivanov. Fine hierarchy of regular
$\omega$-languages. {\em Theoretical Computer Science}, 191
(1998), 37--59.

\bibitem{s01} V.L. Selivanov. A logical
approach to decidability of hierarchies of regular star-free
languages. {\em Lecture Notes in Computer Science}, v. 2010.
Berlin, Springer, 2001, 539--550.

\bibitem{s04} V.L. Selivanov. Boolean hierarchies of partitions
over reducible bases. {\em Algebra and Logic}, 43, N 1 (2004),
44--61.

\bibitem{s05} V.L. Selivanov. Variations on the Wadge
reducibility. {\em Siberian Advances in
Mathematics}, 15, N 3 (2005), 44--80.

%\bibitem{s07a} V.L. Selivanov. Classifying omega-regular partitions. Preproceedings of
%LATA-2007, Universitat Rovira i Virgili Report Series, 35/07,
%529--540.

\bibitem{s06}  %[Se06] 
V. Selivanov. 
Towards a descriptive set theory for domain-like structures,
\emph{Theoretical Computer Science}, \textbf{365}, 2006, 258--282.

\bibitem{s07} V.L. Selivanov. The quotient algebra of labeled
forests modulo h-equivalence.
{\em Algebra and Logic}, 46, N 2 (2007), 120--133.

\bibitem{s07a} V.~L. Selivanov. Hierarchies of
$\bfDelta^0_2$-measurable $k$-partitions.  Math. Logic
Quarterly, 53 (2007), 446--461.

\bibitem{s08} V.~L. Selivanov. Fine hierarchies and $m$-reducibilities in
theoretical computer science. {\em  Theoretical Computer Science},
405 (2008), 116--163.

%\bibitem{s08a} V.~L. Selivanov. Fine hierarchy of regular aperiodic $\omega$-languages. {\em
%International Journal of Foundations of Computer Science}, 19, No
%3 (2008) 649--675.

\bibitem{s09} V.~L. Selivanov. Hierarchies and reducibilities on regular languages related to
modulo counting. {\em RAIRO Theoretical Informatics and
Applications}, 41 (2009), 95--132.

\bibitem{s10}  V.L. Selivanov.  On the Wadge reducibility of k-partitions. Journal of Logic and
Algebraic Programming, 79, No 1, 2010, 92--102.  PII:
S1567-8326(09)00024-1 DOI: 10.1016/ j.jlap.2009. 02.008

\bibitem{s11}  V.L. Selivanov.
A fine nierarchy of  $\omega$-regular $k$-partitions. B. L\"owe et.al. (Eds.):
CiE 2011, LNCS 6735, pp. 260--269. Springer, Heidelberg (2011).

\bibitem{s12} V.~L. Selivanov.
Fine hierarchies via Priestley duality. Annals of Pure and Applied
Logic, 163 (2012) 1075--1107.

%\bibitem{s03} V.L. Selivanov.   Wadge degrees of $\omega$-languages of
%deterministic Turing machines.  Theoretical Informatics
%and Applications, 37 (2003), 67--83.

%\bibitem{s04a} V.L. Selivanov. Difference hierarchy in
%$\varphi$-spaces. {\em Algebra and Logic}, 43, N 4 (2004),
%238--248.

\bibitem{s13} %[Se13]
V. Selivanov. Total representations,
\emph{Logical Methods in Computer Science}, \textbf{9(2)}, 2013, 1--30.
DOI: 10.2168/LMCS-9(2:5)2013.

%\bibitem{s15} 	V.L. Selivanov: Towards the Effective Descriptive Set Theory. CiE 2015 (Eds. Arnold Beckmann, Victor Mitrana, Mariya Ivanova Soskova). Lecture Notes in Computer Science 9136, Springer 2015, ISBN 978-3-319-20027-9): 324--333.

\bibitem{s16} Selivanov, V.L.: Towards a descriptive theory of cb0-spaces. Mathematical Structures in Computer Science. v. 28 (2017), issue 8, 1553--1580. DOI: http://dx.doi.org/10.1017/S0960129516000177. Earlier version in: ArXiv: 1406.3942v1 [Math.GN] 16 June 2014.

\bibitem{s17} 	V.L. Selivanov.  Extending Wadge theory to k-partitions. J. Kari, F. Manea and Ion Petre (eds.) CiE 2017, LNCS 10307, 387--399, Berlin, Springer.


\bibitem{sc11}
P. Schlicht, Continuous reducibility for Polish spaces, 2012, submitted.

\bibitem{si87} Logic and Combinatorics, edited by S. G. Simpson, Contemporary Mathematics 65, American
Mathematical Society, Providence, RI, 1987.

%\bibitem{sch:ext} %[Sch02] 
%Schr\"oder, M.:
%Extended admissibility, 
%\emph{Theoretical Computer Science}, \textbf{284}, (2002), 519--538.

\bibitem{si85} S.G. Simpson. Bqo-theory and Fra\"iss\'e conjecture. Chapter 9 of \cite{mw85}, 1985.

\bibitem{ss88} T.A. Slaman and J. R. Steel. Definable functions on degrees. In Cabal Seminar
81–-85, volume 1333 of Lecture Notes in Math., pages 37-–55. Springer, Berlin, 1988.

\bibitem{sr07}
J. Saint Raymond. Preservation of the Borel class under countable-compact-
covering mappings. Topology and its Applications 154 (2007), 1714–-1725.

\bibitem{st85} J. Stern. Characterizations of some classes of
regular events. {\em Theoretical Computer Science}, 35 (1985),
163--176.

\bibitem{str94}  H. Straubing. {\em Finite automata, formal logic and circuit
complexity}. Birkh\"auser, Boston, 1994.

\bibitem{st82} J.R. Steel. A classification of jump operators. J. Symbolic Logic, 47(2):347–358, 1982.

\bibitem{ste80} J. Steel. Determinateness and the separation property.
{\em J. Symbol. Logic,} 45 (1980), p.143--146.

\bibitem{sw05} V.L. Selivanov and K.W. Wagner. A reducibility for the dot-depth
hierarchy. {\em Theoretical Computer Science},
345, N 2-3 (2005), 448--472.

\bibitem{tb70} B.A. Trakhtenbrot and J.M. Barzdin. {\em Finite
automata. Behavior and Synthesis}. Mir, Moscow, 1970 (Russian,
English translation: North Holland, Amsterdam, 1973).

\bibitem{th90} W. Thomas. Automata on infinite objects.
{\em Handbook of Theor. Computer Science}, v. B (1990), 133--191.

\bibitem{th96} W. Thomas. Languages, automata and logic.
{\em Handbook of Formal Language theory}, v. B (1996), 133--191.

\bibitem{tmr53} A. Tarski, A. Mostowski, and J. Robinson.
{\em Undecidable Theories}. North Holland, Amsterdam, 1953.

\bibitem{vw76} R.  Van Wesep. Wadge  degrees  and  descriptive set theory.
 Lecture Notes in Mathematics, 689 (1976), p. 151--170.

\bibitem{wad84} W. Wadge.  Reducibility and Determinateness in the
Baire Space. PhD thesis, University of California, Berkely, 1984.

\bibitem{wag79} K. Wagner. On $\omega$-regular sets.
{\em Inform. and Control}, 43 (1979),  123--177.

\bibitem{w92} K. Weihrauch. The degrees of discontinuity of some translators
between representations of the real numbers.
Technical Report TR-92-050, International Computer Science Institute,
Berkeley, 1992.

\bibitem{wei00} K. Weihrauch
Computable Analysis. Springer Verlag, Berlin (2000)

\bibitem{wi12}  A. Wires. Definability in the substructure ordering of simple graphs. 2012. http://www.math.uwaterloo.ca/ awires/SimpleGraphs.pdf.

\bibitem{wi16}  A. Wires. Definability in the substructure ordering of simple graphs. Annals of Combinatorics, 20(1):139--176, 2016.





\end{thebibliography}
\end{document}